\AtBeginDocument{\date{30 March 2015}}

\documentclass[prd,preprintnumbers,nofootinbib,twocolumn]{revtex4-1}
\pdfoutput=1
\usepackage{graphicx}
\usepackage{amsmath}
\usepackage{amssymb}
\usepackage[bookmarksopen,bookmarksnumbered]{hyperref}
\usepackage{color}
\usepackage{units}
\usepackage{rotating}

\allowdisplaybreaks

\def\xleft{\mathopen{}\left}

\DeclareRobustCommand*\diff[2][]{%
   \mathop{
     \mathrm{d}^{#1}
     \mskip-0.2\thinmuskip
    #2}\nolimits
}

\newcommand\3[1]{\boldsymbol{#1}}

\newcommand{\T}[1]{\boldsymbol{#1}_{\text{T}}}
\newcommand{\Tsc}[1]{#1_{\text{T}}}

\def\asy{\mathop{\text{asy}}}

\newcommand\BIBvol[1]{\textbf{#1}}

\newcommand{\bmax}{b_{\rm max}}
\newcommand\bstar{\3{b}_*}
\newcommand\bstarsc{b_*}
\newcommand\mubstar{\mu_{\bstarsc}}
\newcommand\muQ{\mu_Q}

\newcommand\pORm[2]{\genfrac{}{}{0pt}{2}{+#1}{-#2}}

\begin{document}


\title{Understanding the large-distance behavior of
  transverse-momentum-dependent parton densities and the Collins-Soper
  evolution kernel} 
  
\preprint{YITP-WSB-14-45}  
\preprint{JLAB-THY-15-2001}  

\author{John Collins}
\email{jcc8@psu.edu}
\affiliation{%
  104 Davey Lab., Penn State University, University Park PA 16802, USA
}
\author{Ted Rogers}
\email{tedconantrogers@gmail.com}
\affiliation{C.N.~Yang Institute for Theoretical Physics,  Stony Brook University, 
  Stony Brook NY 11794, USA}
\affiliation{Department of Physics, Southern Methodist University, Dallas, Texas 75275, USA}
\affiliation{Department of Physics, Old Dominion University, Norfolk, VA 23529, USA}
\affiliation{Theory Center, Jefferson Lab, 12000 Jefferson Avenue, Newport News, VA 23606, USA}

\begin{abstract}
  There is considerable controversy about the size and importance of
  nonperturbative contributions to the evolution of 
  transverse-momentum-dependent (TMD) parton distribution functions.  Standard
  fits to relatively high-energy Drell-Yan data give
  evolution that when taken to lower $Q$ is too rapid to be consistent
  with recent data in semi-inclusive 
  deeply inelastic scattering.  Some authors provide very different
  forms for TMD evolution, even arguing that nonperturbative
  contributions at large transverse distance $\Tsc{b}$ are not needed
  or are irrelevant. Here, we systematically analyze the issues, both
  perturbative and nonperturbative.  We make a motivated proposal for
  the parameterization of the nonperturbative part of the TMD
  evolution kernel that could give consistency: with the variety of
  apparently conflicting data, with theoretical perturbative
  calculations where they are applicable, and with general theoretical
  nonperturbative constraints on correlation functions at large
  distances.  We propose and use a scheme- and scale-independent
  function $A(\Tsc{b})$ that gives a tool to compare and diagnose
  different proposals for TMD evolution.  
  We also advocate for phenomenological 
  studies of $A(\Tsc{b})$ as a probe of TMD evolution. The results are important
  generally for applications of TMD factorization. In particular, they
  are important to making predictions for proposed polarized Drell-Yan
  experiments to measure the Sivers function.
\end{abstract}

\maketitle

\section{Introduction}
\label{sec:intro}

Factorization involving transverse-momentum-dependent (TMD) parton
densities and fragmentation functions is important to understanding a
variety of hard-scattering reactions in QCD.  The domain of utility of
such ``TMD factorization'' is where there is a relevant measured
transverse momentum, $\Tsc{q}$, that is much less than the large scale
$Q$ of the hard scattering.

Although there has been substantial success (e.g.,
\cite{Landry:2002ix, Konychev:2005iy, Qiu:2000hf, Qiu:2002mu,
  Fai:2003zc}) in fitting data with TMD factorization in the standard
framework of Collins, Soper and Sterman (CSS) \cite{Collins:1981uk,
  Collins:1981uw, Collins:1984kg, Collins:2011qcdbook}, a number of
recent papers --- e.g., Refs.\ \cite{Becher:2010tm,
  Echevarria:2012pw, Sun:2013dya, 
  Sun:2013hua} --- disagree with the way this has been done, as we will
explain in more detail in Sec.\ \ref{sec:assess}.
There appear to be substantially different predictions
for
lower-energy experiments, and there even appear to be inconsistencies in how
the phenomenology of TMD factorization is to be implemented.

These
problems particularly impact proposals for experiments
\cite{Barone:2005pu, Liu:2010kb, Lutz:2009ff}
for the Drell-Yan process with a transversely polarized hadron.
The proposed experiments
are designed (among other
things) to test the predicted reversal of sign \cite{Collins:2002kn}
of the Sivers function \cite{Sivers:1990cc, Sivers:1991fh} between
semi-inclusive deeply-inelastic scattering (SIDIS) and the Drell-Yan process.
Predictions for polarized Drell-Yan use the
Sivers functions measured in SIDIS at relatively low
$Q$ (e.g., $Q\sim\unit[\sqrt{2.4}]{GeV}$ at HERMES
\cite{Airapetian:2009ae, Airapetian:2010ds}).  
However, to make the
predictions one needs to use correct evolution of the TMD functions to
the higher $Q$ of the polarized Drell-Yan experiments.  The same
applies when one wants to analyze data on the Collins function
both from the HERMES and COMPASS experiments and
from $e^+e^-$ annihilation from the Belle \cite{Abe:2005zx,
  Seidl:2008xc} and BABAR \cite{BABAR:2013yha} experiments.

The main issue concerns the evolution of the TMD
parton densities (from which arises a dilution of the
fractional Sivers asymmetry and other similar quantities).  In the CSS
framework, the evolution is governed by the function
called\footnote{Essentially the same function is notated
$-2 D$ by
  Echevarr\'\i a et al.\ \cite{GarciaEchevarria:2011rb}, and
  $-F_{q\bar{q}}$ by Becher and Neubert \cite{Becher:2010tm}.}
$\tilde{K}(\Tsc{b})$ \cite{Collins:1981uk, Collins:1981uw,
  Collins:1984kg, Collins:2011qcdbook}, with $\T{b}$ being a
transverse position variable.  The results of standard fits \cite{Landry:2002ix,
  Konychev:2005iy} of the nonperturbative part of $\tilde{K}$ to
unpolarized Drell-Yan data at $Q\gtrsim\unit[9]{GeV}$ give what Sun
and Yuan \cite{Sun:2013dya,Sun:2013hua} argue to be a too rapid
evolution of TMD functions at the lower values of $Q$ relevant for
currently available measurements of the Sivers function.  However,
Aybat et al.\ \cite{Aybat:2011ta} find that the rather rapid evolution
given by the fit of Brock-Landry-Nadolsky-Yuan (BLNY) \cite{Landry:2002ix} does
agree with the change in the Sivers asymmetry between the values of
$Q$ for the HERMES \cite{Airapetian:2009ae} and preliminary COMPASS
\cite{Bradamante:2011xu} data.  More recently, Aidala et al.\
\cite{Aidala:2014hva} make a very detailed examination of the HERMES
and COMPASS data.  They agree that there is indeed a discrepancy between this
data and the predictions based on the earlier Drell-Yan data, but
argue that the low- and high-energy data are mostly probing different
regions of transverse position; thus the discrepancy concerns the
extrapolation of a parametrization of nonperturbative physics
outside the region where it was fitted.  

Another issue is that there
is more than one formulation of TMD factorization, and that these do
not always appear to be compatible.  There is the original CSS
formulation in Refs.\ \cite{Collins:1981uk, Collins:1981uw,
  Collins:1984kg} and its new version in Ref.\
\cite{Collins:2011qcdbook}.  There is the related but different
formulation by Ji, Ma and Yuan \cite{Ji:2004wu}.  There are at least
three not obviously identical formulations in the Soft-Collinear Effective Theory (SCET) framework: by
Becher and Neubert \cite{Becher:2010tm}, by Echevarr\'\i a, Idilbi,
and Scimemi \cite{GarciaEchevarria:2011rb} (EIS), and by Mantry and Petriello
\cite{Mantry:2009qz, Mantry:2010bi}.  Furthermore, although Sun and
Yuan \cite{Sun:2013dya, Sun:2013hua} explicitly base their work on the
CSS method, the actual formula they used to fit data differs
substantially from those used in the original fits to Drell-Yan data.
Sun and Yuan's formula represents a certain approximation that is essentially
identical to one given by Boer in Ref.\ \cite[Eq.\
(144)]{Boer:2008fr}.  The different formulae give rather different
results, as seen in a recent paper by Boer \cite{Boer:2013zca}.

To understand the origin of the proliferation in methods, it is helpful
to note that there is often a clash of motivations for applying TMD
factorization in specific phenomenological situations.

One increasingly prominent 
motivation is to study the intrinsic transverse motion related to
nonperturbative binding and hadronic physics.  For this it is
desirable to have a cross section where the effects of
intrinsic transverse momentum do not become washed out by perturbative
gluon radiation.  
For these applications, $Q$ is taken to be large
enough for an overall TMD factorization theorem to be valid,
while also being small enough that perturbative
radiation does not greatly obscure interesting nonperturbative
dynamics associated with the hadron wavefunction.\footnote{A standard example is
  the Sivers function, where associated with intrinsic
  nonperturbative transverse momentum is a transverse single-spin
  asymmetry with a process-dependent sign.}  The measurement of TMD
parton densities and fragmentation functions then gives a useful probe of
nonperturbative transverse momentum dynamics inside a hadron. Examples of recent work where 
this is the driving motivation include 
Refs.~\cite{Avakian:2009jt,Courtoy:2008dn,Lorce:2011zta,Burkardt:2012sd,Anselmino:2012re,Signori:2013mda,Echevarria:2014xaa}.

By contrast, at high-energy hadron colliders, one is often interested
in studying purely perturbative phenomena in cross sections with large
transverse momentum, or in utilizing perturbative QCD calculations as
part of searches for new physics beyond the Standard Model.  TMD
factorization
contains a useful way of resumming large logarithms in
perturbative calculations. Examples of recent work where this second 
motivation is the primary focus include Refs.~\cite{Becher:2010tm,Bozzi:2010xn,Banfi:2011dx,Guzzi:2013aja,Meade:2014fca}.
In these applications, sensitivity to
nonperturbative transverse momentum phenomena could understandably be
seen as a nuisance and a source of uncertainty, while these same
phenomena are the main objects of study in the situation described in the
previous paragraph.

Our perspective is that the two motivations described above are best 
seen as complementary rather than conflicting. Both the perturbative calculability 
of the small $\Tsc{b}$ dependence and the universality of the large $\Tsc{b}$ 
dependence are important features of the TMD factorization 
theorem, and predictive power is optimized when both are fully exploited. 
Verifying the universality of the large $\Tsc{b}$ behavior 
can provide valuable insight into aspects of the nonperturbative 
dynamics relevant to studies of hadron structure. Moreover, even for measurements at relatively large $Q$, where 
sensitivity to the nonperturbative region of $\Tsc{b}$ is expected to be small,
nonperturbative components are potentially necessary if 
very high precision calculations are desired.\footnote{One example is the $W$ mass where the 
large $\Tsc{b}$ behavior is an important source of uncertainty  in precision experimental constraints. 
See, for example, page 57 of~\cite{LopesdeSa:2013zga}. See also 
a discussion of related issues in Ref.~\cite{Nadolsky:2004vt}.} 
Thus, the TMD factorization theorem, with its accommodation of both types of behavior, simultaneously addresses 
both of the motivations described above within a single formalism.

One of the issues that now becomes prominent is the nature of the
evolution of the TMD functions between different energies.  Notably, 
there is disagreement on whether there are nonperturbative
contributions to TMD evolution and how significant they are.

A nonperturbative contribution to the 
evolution might appear troublesome at first glance. nonperturbative 
inputs in perturbative calculations are sources of uncertainty.  
However, the Collins-Soper kernel, including its nonperturbative components, has 
what we call ``strong'' universality. 
Namely, it is process independent, but it is also insensitive to
the types of external hadrons involved, any polarization dependence, the flavors of the quarks,  and the 
scale $Q$. (The only dependence is on the color representation of the colliding parton --- there is, in principle, 
a different nonperturbative evolution for quark and gluon TMD parton distribution functions (pdfs).) 
In this sense, it has a much greater degree of universality than, say, collinear pdfs, which, while 
independent of the process, are dependent upon the nature and identity of parent hadron, including the polarization, 
and on the flavor and polarization of the quark. Once the nonperturbative TMD \emph{evolution} is known, it can 
be used in all situations where a TMD factorization theorem is valid. 
Verifying this strongly universal nature of the nonperturbative evolution is an important test 
of the TMD-factorization theorem itself. Moreover, its relationship to matrix elements of the fundamental 
field operators is known --- it is essentially a nonperturbative contribution to the 
vacuum expectation value of a certain kind of Wilson loop. 
(This follows immediately from the definitions of the 
TMD pdfs; see, for example,~\cite[Eqs.~(13.47, 13.48)]{Collins:2011qcdbook}.)
Therefore, determining its influence on 
cross sections provides a probe of fundamental nonperturbative quark-gluon dynamics, which 
can then be compared with nonperturbative approaches to understanding QCD matrix elements. 
Such nonperturbative methods could include, for example, lattice based calculations~\cite{Musch:2010ka,Musch:2011er,Ji:2014hxa}.

Now, the TMD
functions are normally used in their Fourier transformed versions, in
terms of a transverse position variable $\T{b}$.
CSS took it as
obvious that a large enough value of $\Tsc{b}$ is in the region of
nonperturbative physics in QCD.  They therefore argued that one must
provide some kind of cutoff on perturbative calculations and insert a
parametrization, to be fit to data, to handle nonperturbative
contributions at large $\Tsc{b}$.  Such fits were done in Refs.\
\cite{Landry:2002ix, Konychev:2005iy}.  
But other papers have either avoided \cite{Becher:2010tm, Sun:2013dya,
  Sun:2013hua} the use of a nonperturbative contribution to the
function $\tilde{K}(\Tsc{b})$ that controls evolution, or have argued
\cite{Echevarria:2012pw} that the nonperturbative contribution is not
needed until well beyond the region of $\Tsc{b}$ that was regarded as
nonperturbative in the earlier fits.
Furthermore, Kulesza, Sterman, and Vogelsang
\cite{Kulesza:2002rh} have presented a method in which the Landau pole
manifested in perturbative calculations in $\Tsc{b}$-space is avoided
by a contour deformation, and the nonperturbative behavior associated
with large $\Tsc{b}$ is inferred from power law behavior obtained in
perturbation theory and extrapolated to large $\Tsc{b}$. In 
Refs.~\cite{Korchemsky:1994is,Tafat:2001in}, the sizes of power corrections 
were estimated on the basis of renormalon studies.

In this paper, we give a detailed analysis of the issues, and propose
methods and solutions for resolving the disagreements.

First, we survey a sample of the different ways solutions to the evolution 
equations can be written. This will illustrate how different but essentially 
equivalent styles of presentation may emphasize particular specific features while preserving the 
broad, underlying predictive power of the TMD factorization formalism. For instance, 
we will note the connection between a TMD parton model language and 
the standard presentation of the results of the CSS formalism. 

Then, as a diagnostic tool, we define a master function\footnote{%
  The function in fact equals the one that CSS \cite{Collins:1984kg}
  called $A$.  But we now give a more general definition, not tied to
  a particular version of the CSS formalism in the perturbative
  region.  In the CSS formalism our definition gives $A =
  - \partial\tilde{K}/\partial\ln \Tsc{b}^2$.  Contrary to CSS, we
  now treat $A$ as a function of $\Tsc{b}$ instead of
  $\alpha_s(1/\Tsc{b})$, as is more appropriate when examining its
  behavior beyond perturbation theory.
} 
$A(\Tsc{b})$ that can be used to quantify the evolution of
the shape of TMD functions,  separately from the normalization.
This function is scheme- and
renormalization-group-independent, and 
the function is predicted by QCD to be
independent of
all other
kinematic variables (e.g., $Q$) and of parton and hadron flavor.
Because $A$ is scheme-independent, a
comparison of $A$ between different formalisms and approximations can be used as a
diagnostic to determine where in $\Tsc{b}$ the methods disagree.
This then indicates which experiments would give sensitivity to the
differences. 
Since $A$ is a scheme-independent property of QCD and is
a function of only one variable, we propose that calculations and fits for
TMD evolution can be presented as including a determination of $A$ in particular
regions of $\Tsc{b}$, and that it would be useful to obtain a global
fit for $A$.  The result would be a universal function that controls
the evolution with
energy
of the shape
of all the many TMD functions (for color 
triplet quarks).  As a first step in such an analysis, we use the
master function $A$ to compare evolution in various formalisms and
from different analyses of data.

Finally, we examine phenomenological and theoretical issues 
concerning the
nonperturbative large-$\Tsc{b}$ behavior of the TMD pdfs and of
$\tilde{K}$. 
We will argue that the standard Gaussian parametrizations of
the TMD functions give the wrong limiting behavior at asymptotically
large $\Tsc{b}$.
Instead, we agree with Schweitzer et al.\ \cite{Schweitzer:2012hh}
that TMD parton densities and fragmentation functions should decay
exponentially at large $\Tsc{b}$, with a decay length corresponding to
the mass of the lowest relevant state.

As regards the CSS evolution kernel $\tilde{K}$, the same argument
suggests that it goes to a constant at large $\Tsc{b}$ (and hence that
our master function $A$ goes to zero).  This is also suggested
phenomenologically by the apparent slowness of TMD evolution at low
$Q$, where phenomena are dominated by the nonperturbative range of
large $\Tsc{b}$.  
We will suggest possible parametrizations that can
reconcile our proposed large $\Tsc{b}$ asymptote with previous fits
that use a quadratic $\Tsc{b}^2$ dependence for $\tilde{K}$.  These
quadratic parametrizations correspond to a $Q$-dependent width for
the standard Gaussian ansatz for TMD functions.  The quadratic form
for $\tilde{K}$ (and hence $A$) can only be valid over a limited range
of moderately large $\Tsc{b}$.  One should not continue the
$\Tsc{b}^2$ form to the larger values of $\Tsc{b}$ that are important
for processes at low $Q$.  The result is then that the evolution of
TMD pdfs is much weaker at low $Q$ than would otherwise happen.

A related proposal for the nonperturbative form was given by Collins
and Soper \cite{Collins:1985xx}.  Their form was logarithmic instead
of quadratic, and the particular formula was designed to provide a
better match to the perturbative part of $\tilde{K}$ in the
extrapolation to large $\Tsc{b}$.

A number of the ideas in this paper have been discussed and presented
at various conferences and workshops, so they are already becoming
current (e.g., \cite{Su:2014wpa}).  The particular
aim of the present paper is to give a detailed, systematic, and
unified account of the issues.

The outline of this paper is as follows: First, in Sec.\
\ref{sec:CSS}, we review the CSS formalism for TMD factorization in
the form given in Ref.\ \cite{Collins:2011qcdbook}, together with an
analysis of various forms of solution of the evolution equations.  In
Sec.\ \ref{sec:univ} we review the universality properties of the
functions in the formalism.  Then in Sec.\ \ref{sec:master}, we
motivate and define the master function $A$.  In Sec.\
\ref{sec:assess}, we assess various approximations for $\tilde{K}$ and
show the phenomenological difficulties referred to earlier.  In Sec.\
\ref{sec:large-b}, we propose new forms for parametrizing the
large-$\Tsc{b}$ region.

\section{CSS formalism}
\label{sec:CSS}

In this section we review TMD factorization for the
Drell-Yan cross section in the version of the CSS formalism recently
derived in Ref.\ \cite{Collins:2011qcdbook}, where can be found
justifications of assertions in this section that are otherwise
unreferenced.  (The results presented are for the case that all quarks
are light.  In the presence of heavy quarks, as in real QCD, the
modifications have not been completely worked out, to the best of our
knowledge. See, however, Ref.~\cite{Nadolsky:2002jr} for work in this direction in 
the specific case of heavy quark production.)

It should be emphasized that the CSS formalism in
Ref.~\cite{Collins:2011qcdbook} is, in fact, equivalent to the earlier
formalism of~\cite{Collins:1981uk,Collins:1981uw, Collins:1984kg}, but
with a different organization of the factors that is intended to be
more convenient, and with improved operator definitions of the
TMD functions.  This implies, in particular, that high order
calculations for quantities like anomalous dimensions performed for
the original version of the CSS formalism can be carried over to the
new version, after allowing for an effective scheme change. 
Thus, 
use of the updated formalisms does not necessarily imply an increased degree of 
complexity, relative to previous work, for doing calculations.

Results of the same structure apply to other processes of
interest, e.g., semi-inclusive deep-inelastic scattering 
(SIDIS)~\cite{Meng:1991da,Meng:1995yn,Nadolsky:1999kb} and
inclusive production of two hadrons in $e^+e^-$
annihilation~\cite{Collins:1981uk}. 
It will be sufficient to use Drell-Yan scattering as our main
example.

Much of the presentation in this section follows previous treatments.
The reader should be warned that some steps may require familiarity 
with more complete derivations such as can be found in 
Ref.~\cite{Collins:1981uk,Collins:1981uw, Collins:1984kg, Collins:2011qcdbook}.
We will emphasize aspects that are particularly relevant to
comparison with data, to the predictive power of the formalism, and to
the comparison with other formalisms and approximations.  In
particular, we will give a careful account of the nature of quantities
that receive nonperturbative contributions, notably the function
$\tilde{K}$ that controls the evolution of TMD densities.

\subsection{The fundamental equations}
\label{sec:CSS.eqs}

The Drell-Yan process is the production of a high-mass lepton-pair in a
high-energy collision of two hadrons, $A+B\to l+\bar{l}+X$, with the
restriction to production of the lepton pair through a virtual photon
and/or $Z$ boson\footnote{The modifications for other related
  processes, e.g., $W$ production are, of course, elementary.} in the
lowest order in electroweak interactions.

Kinematic variables are defined as follows: The momenta of the
incoming hadrons are $p_A$ and $p_B$, the momentum of the lepton pair
is $q$, polar angles $\theta$ and $\phi$ in the Collins-Soper frame
\cite{Collins:1977iv} are used to specify the directions of the lepton
momenta, and the element of solid angle for the lepton direction is
$\diff{\Omega}$.  Certain auxiliary variables are convenient for the
factorization formalism.  To define these, we use light-front
coordinates in the overall center-of-mass frame, with $A$ and $B$ 
moving in the $+z$ and $-z$ directions.  Then $q^{\pm}=(q^0\pm
q^3)/\sqrt{2}$, and we let the mass, rapidity, and transverse momentum
of the lepton pair be $Q=\sqrt{q^2}$, $y=\frac{1}{2}\ln(q^+/q^-)$, and
$\T{q}$.  Longitudinal momentum fractions are defined by $x_A =
Qe^y/\sqrt{s}$ and $x_B = Qe^{-y}/\sqrt{s}$, where $\sqrt{s}$ is the
overall center-of-mass energy.  Factorization applies up to power
suppressed corrections when $s$ and $Q$ are made large with fixed
$x_A$ and $x_B$.

\subsubsection{Factorization}

The factorization formula is:
\begin{widetext}
\begin{align}
\label{eq:kt.fact}
  \frac{ \diff{\sigma} }{ \diff[4]{q}\diff{\Omega} } 
  ={}&
    \frac{2}{s} \sum_j
    \frac{ \diff{\hat{\sigma}_{j\bar{\jmath}}}(Q,\mu,\alpha_s(\mu)) }{ \diff{\Omega} }
    \int \diff[2]{\T{b}}
    ~ e^{i\T{q}\cdot \T{b} }
    ~ \tilde{f}_{j/A}(x_A,\T{b};\zeta_A,\mu) 
    ~ \tilde{f}_{\bar{\jmath}/B}(x_B,\T{b};\zeta_B,\mu)
\nonumber\\
    & + \mbox{polarization terms} 
      + \mbox{high-$\Tsc{q}$ term ($Y$)}
      + \mbox{power-suppressed}.
\end{align}
\end{widetext}
Here, $\tilde{f}_{j/H}(x,\T{b};\zeta,\mu)$ is the TMD density of a
quark of flavor $j$ in hadron $H$, but Fourier transformed to
transverse coordinate space.  A suitable definition is given in Ref.\
\cite[Sec.\ 13.15]{Collins:2011qcdbook}, with the change of direction
of the Wilson lines appropriate to the Drell-Yan process \cite[Sec.\
14.15]{Collins:2011qcdbook}.  The sum over $j$ is over all flavors of
quark and antiquark.  The hard-scattering coefficient is
$\diff{\hat{\sigma}_{j\bar{\jmath}}}(Q,\mu,\alpha_s(\mu))/\diff{\Omega}$,
normalized like a cross section, for production of the lepton pair
from a quark-antiquark collision, with renormalization scale $\mu$.
When the renormalization scale is taken to be of 
order $Q$, the hard
scattering is perturbatively calculable because of the asymptotic
freedom of QCD.  The parton densities have two scale arguments: one is
the renormalization scale $\mu$, and the other is a scale $\zeta$ that
is used to parametrize 
how the effects of soft-gluon radiation are partitioned between
the two TMD parton densities.  The two $\zeta$ scales obey
$\zeta_A\zeta_B=Q^4$.

It is convenient to define a contribution to the $\T{b}$-space
integrand by
\begin{align}
\label{eq:Wj.def}
\tilde{W}_j(\T{b};Q) & \equiv   
     Q^2 \frac{ \diff{\hat{\sigma}_{j\bar{\jmath}}}(Q,\mu,\alpha_s(\mu)) }{ \diff{\Omega} } \nonumber \\ 
  & \; \times \tilde{f}_{j/A}(x_A,\T{b};\zeta_A,\mu) \,
  \tilde{f}_{\bar{\jmath}/B}(x_B,\T{b};\zeta_B,\mu) \, .
\end{align}
Note that this quantity includes the hard part, and that the
normalization differs from that of a similar quantity in 
Ref.~\cite{Collins:1984kg}.  The overall factor of $Q^2$ in
Eq.~\eqref{eq:Wj.def} is to make $\tilde{W}_j$ dimensionless.
It is also convenient to define a summed $\tilde{W}$ whose Fourier
transform corresponds to the cross section itself:
\begin{equation}
\label{eq:W.def}
  \tilde{W} = \sum_j \tilde{W}_j \, .
\end{equation}
Then the first line of Eq.~\eqref{eq:kt.fact} is
\begin{equation}
\label{eq:Wj.sigma}
\frac{2}{s Q^2}\sum_j  \int \diff[2]{\T{b}} e^{i\T{q}\cdot \T{b} }
   \tilde{W}_j(\T{b};Q) \, ,
\end{equation}
i.e.,
\begin{equation}
\label{eq:W.sigma}
 \frac{2}{s Q^2} \int \diff[2]{\T{b}} e^{i\T{q}\cdot \T{b} }
   \tilde{W}(\T{b};Q) \, . 
\end{equation}

The hard scattering is computed from graphs for the massless on-shell
quark-antiquark-to-lepton-pair reaction with subtractions for
collinear and soft regions appropriate to the definition of the
TMD parton densities that are used.  The leptons' angular distribution
is computed by using the valid approximation that in the hard
scattering the incoming quark $j$ and antiquark $\bar{\jmath}$ move in
exactly the $+z$ and $-z$ directions in the Collins-Soper frame.

The cross section, with its angular distribution, can be expressed in
terms of a hadronic tensor $W^{\mu\nu}(q,p_A,p_B)$ and corresponding
scalar structure functions ($W_1$ etc) in the standard way
\cite{Lam:1978pu, Mirkes:1992hu, Ralston:1979ys, Donohue:1980tnI}:
Given that only annihilation through an electroweak boson is involved,
the cross section is written in terms of the product of $W^{\mu\nu}$
and a lowest-order leptonic factor.  The hadronic tensor is decomposed
in terms of standard basis tensors times scalar structure functions.
There is a corresponding decomposition of the cross section in terms
of basis functions for the dependence on the lepton polar angles, 
including spin dependence~\cite{Balazs:1995nz,Balazs:1997xd,Arnold:2008kf}.  By
matching the basis function expansion for the full cross section and
the hard-scattering cross section (or some equivalent method), one
obtains from Eq.\ (\ref{eq:kt.fact}) corresponding TMD factorization
formulae for the structure functions.

The first line in Eq.\ (\ref{eq:kt.fact}) gives the cross section when
$\Tsc{q} \ll Q$, the hadrons are unpolarized, and quark polarization
is ignored.  In that case the TMD parton densities are independent of
the azimuthal angle of $\T{b}$.  When hadron and quark polarization
are taken into account,
further similar terms arise; these can be
characterized in terms of suitable polarization-dependent TMD
densities \cite{Bacchetta:2006tn, Mulders:1996dh}.  For example, there
is the Sivers function \cite{Sivers:1990cc, Sivers:1991fh}, which
gives the azimuthal 
dependence of the TMD density of unpolarized quarks in a transversely
polarized proton.

Treatment of the polarization-dependent terms involves a mostly
straightforward generalization of the CSS method: See, for example,
Refs.\ \cite{Boer:2001he, Aybat:2011ge, Boer:2013zca}.

A further addition to the formula is present, because the first part
of Eq.\ (\ref{eq:kt.fact}), including the polarization-dependent
terms, gives a valid approximation to the cross section only when
$\Tsc{q}\ll Q$.  At large $\Tsc{q}$, ordinary collinear factorization
is valid.  Therefore a correction term is added so that the total is
correct to the leading power of $Q$ for any value of $\Tsc{q}$; the
correction term is called $Y$ by CSS \cite{Collins:1981uk,
  Collins:1984kg}.  

\subsubsection{Summary of subsidiary results}
\label{sec:subsidiary}

Most of the predictive power of TMD factorization comes not from the
factorization formula (\ref{eq:kt.fact}) alone, but from its
combination with further results, as follows.

First, there are evolution equations (Sec.~\ref{eq:evolequations}, Eqs.~(\ref{eq:CSS.evol}--\ref{eq:RG.H}))
for the $\zeta$ and $\mu$ dependence of
the factors.  These equations enable the parton densities to be
written in terms of parton densities at fixed scales.  The parton
densities are universal across processes\footnote{Apart from the
  predicted sign-reversal of $T$-odd functions, like the Sivers
  function, between Drell-Yan and SIDIS.}.

The second source of predictive power is from perturbative
calculations of the kernels of the evolution equations.  These include
anomalous dimensions and the universal function $\tilde{K}(\Tsc{b},\mu)$
that controls $\zeta$ dependence.  With the aid of a
renormalization-group transformation, $\tilde{K}(\Tsc{b},\mu)$ can be
calculated perturbatively when $\Tsc{b}$ is not too large.
(See Sec.~\ref{eq:evolequations}, Eqs.~(\ref{eq:gammaK}--\ref{eq:gammaj}) below.)

The universality of the various nonperturbative functions, as
summarized in Sec.\ \ref{sec:univ} below, gives further predictions.  This
especially concerns the function $\tilde{K}(\Tsc{b},\mu)$, which is
nonperturbative for large $\Tsc{b}$.

Another source of predictions is that after evolution is applied to set
$\zeta$ and $\mu^2$ to be of 
order $Q^2$, the hard scattering, $\diff{\hat{\sigma}_{j\bar{\jmath}}}(Q,\mu,\alpha_s(\mu))/\diff{\Omega} $, 
is perturbative; that is, it can be
expanded in powers of the effective coupling $\alpha_s(Q)$ at a high
scale, without the logarithmic enhancements of coefficients that would
otherwise occur.

Finally, there is a kind of operator product
expansion (OPE) for the TMD parton densities at small $\Tsc{b}$. (See Sec.~\ref{sec:OPE.small-b}.) In a
theory, such as a superrenormalizable nongauge theory, where the elementary parton model is valid, the value of a
coordinate-space TMD parton density at zero $\Tsc{b}$ equals the
integral over all transverse momentum of the momentum space TMD
function, by elementary properties of Fourier transforms.  The result
is also the corresponding integrated parton density.  See Ref.\
\cite[Sec.\ 6.8]{Collins:2011qcdbook} for details.

But in renormalizable theories and especially QCD, there is a strong
enough singularity at $\Tsc{b}\to0$, that such results must be
modified (e.g., \cite[Ch.\ 13]{Collins:2011qcdbook}) and the
appropriate modification is the OPE at small $\Tsc{b}$.  This enables
the TMD functions at small $\Tsc{b}$ to be expressed in terms of
ordinary integrated parton densities and perturbatively calculable
coefficient functions.  The coefficient functions in this OPE are
currently known\footnote{Catani et al.\ \cite{Catani:2012qa} also give
  the results for a number of high-order calculations
  for a version of CSS resummation. Since there may be a scheme change
  compared with the formalism that we and the authors of
  \cite{Gehrmann:2012ze,Gehrmann:2014yya} used, it remains to check
  consistency of the different calculations.}
to order $\alpha_s^2$
\cite{Gehrmann:2012ze,Gehrmann:2014yya}.  Intuitively, the OPE can be
characterized by saying that when a momentum-space TMD density is
integrated over transverse momenta up to order $Q$, the result is the
integrated parton density at scale $Q$ plus perturbative corrections
of order $\alpha_s(Q)$.  

Naturally, Dokshitzer--Gribov--Lipatov--Altarelli--Parisi (DGLAP)
evolution also enters 
here, so that the OPE plus DGLAP evolution gives the TMD parton
densities at small $\Tsc{b}$ in terms of ordinary parton densities at
a fixed scale.

It should be added that even without any of these subsidiary results,
the factorization in Eq.~\eqref{eq:Wj.def} alone provides predictions, since as regards the
dependence on the longitudinal momentum fraction parameters $x_A$ and
$x_B$, the cross section is a function jointly of both variables.  But
each parton density depends only on one of these variables.

\subsubsection{Evolution equations}
\label{eq:evolequations}

The CSS evolution equation for the $\zeta$ dependence of the TMD
parton densities is
\begin{equation}
\label{eq:CSS.evol}
  \frac{ \partial \ln \tilde{f}_{f/H}(x,\Tsc{b}; \zeta; \mu) }
       { \partial \ln \sqrt{\zeta} }
  = 
  \tilde{K}(\Tsc{b};\mu).
\end{equation}
The kernel $\tilde{K}$ is independent of the flavor and spin of the
quark, of the nature of the hadron target, and of the momentum
fraction $x$.  It is also the same for fragmentation functions as well
as parton densities, and is the same between the versions of parton
densities for Drell-Yan and the SIDIS processes, and for all the different
polarized parton densities.  A different kernel does appear in gluon
densities, since $\tilde{K}$ depends on the color representation
carried by the parton.  Note that both the parton densities and the
kernel $\tilde{K}$ have
contributions from the infra-red or long-distance
domain,\footnote{For $\tilde{K}$, infra-red contributions are
  power-suppressed at small $\Tsc{b}$ but not at large $\Tsc{b}$.} and
hence these functions depend on quark masses, as
well as on the coupling $\alpha_s(\mu)$; but we have not indicated this
dependence explicitly.

The renormalization group (RG) equation for the kernel is
\begin{equation}
\label{eq:RG.K}
  \frac{ \diff{\tilde{K}(\Tsc{b};\mu)} }{ \diff{\ln \mu } }
  = -\gamma_K\xleft(\alpha_s(\mu)\right), 
\end{equation}
and for the parton densities, the RG equation is  
\begin{equation}
\label{eq:RG.TMD.pdf}
  \frac{ \diff{ \ln \tilde{f}_{j/H}(x,\Tsc{b};\zeta;\mu) }}
       { \diff{\ln \mu} }
    = \gamma_j( \alpha_s(\mu); 1 )
      - \frac12 \gamma_K(\alpha_s(\mu)) \ln \frac{ \zeta }{ \mu^2 },
\end{equation}
in the notation of Ref.\ \cite{Collins:2011qcdbook}.  The RG
coefficient $\gamma_j$ is specific to quark $j$.  However the relevant
calculations are the same for all flavors of spin-$\frac12$ quark. The
$\zeta$ dependence on the right-hand side is determined from the fact that
differentiation of a parton density with respect to $\mu$ commutes with
differentiation with respect to $\zeta$.  (An alternative notation for the
whole of the right-hand side of (\ref{eq:RG.TMD.pdf}) is 
$\gamma_j( \alpha_s(\mu); \zeta/\mu^2 )$.)

An RG equation for the hard scattering follows from the RG invariance
of physical cross sections:
\begin{multline}
\label{eq:RG.H}
  \frac{ \diff{} }
       { \diff{\ln \mu} }
      \ln \left[
      \frac{ \diff{\hat{\sigma}_{j\bar{\jmath}}}(Q,\mu,\alpha_s(\mu)) }{ \diff{\Omega} }
      \right]
\\
    = 
      -2 \gamma_j( \alpha_s(\mu); 1 )
      + \gamma_K(\alpha_s(\mu)) \ln \frac{ Q^2 }{ \mu^2 }.
\end{multline}

In our calculations, we will need the one-loop values for the above
quantities, and the two-loop value of $\gamma_K$:\footnote{See
  \cite{Collins:2011qcdbook} for one-loop calculations of
  $\gamma_K$ from its definition.  The value to three-loop order was
  found by Moch, Vermaseren, and Vogt \cite{Moch:2005id}; they compute
  a quantity they call $A$, which is our $\gamma_K/2$ --- see their
  Eq.\ (2.4). Their value was
  recently confirmed by Grozin et al.\ \cite{Grozin:2014hna}. }
\begin{align}
  \gamma_K(\alpha_s(\mu)) & =  2C_F \frac{\alpha_s(\mu)}{\pi} 
\nonumber\\ & \hspace*{-1cm}
           + \left( \frac{\alpha_s(\mu)}{\pi} \right)^2
             C_F \left[ C_A \left(\frac{67}{18}-\frac{\pi^2}{6} \right)
                       -\frac{10}{9}T_Fn_f  
                \right] 
\nonumber\\ & \hspace*{-1cm}
   + O(\alpha_s(\mu)^3),
\label{eq:gammaK}
\\
   \tilde{K}(\Tsc{b};\mu) 
  = {}& - \frac{\alpha_s(\mu)}{\pi} C_F
       \left[ \ln\frac{\Tsc{b}^2\mu^2}{4} + 2\gamma_E \right]
\nonumber \\ & 
      + O(\alpha_s(\mu)^2),
\label{eq:K}
\\
  \gamma_j(\alpha_s(\mu);1)  = {}&
  \frac{3 C_F}{2} \frac{\alpha_s(\mu)}{\pi}
  + O(\alpha_s(\mu)^2).
\label{eq:gammaj}
\end{align}

\subsubsection{Small-$\Tsc{b}$ expansion}
\label{sec:OPE.small-b}

At small $\Tsc{b}$, the unpolarized TMD parton densities can be
expressed in terms of the corresponding integrated parton densities,
$f_{k/H}(x;\mu)$, by a kind of OPE:
\begin{align}
\label{eq:TMD.OPE}
  \tilde{f}_{j/H}(x,\Tsc{b};\zeta;\mu) 
  = {}& \sum_k \int_{x-}^{1+} \frac{ \diff{\xi} }{ \xi }
       \,\tilde{C}_{j/k}\xleft( x/\xi,\Tsc{b};\zeta,\mu,\alpha_s(\mu) \right)
\nonumber\\
&\times
        f_{k/H}(\xi;\mu)
~+~ O\xleft[(m\Tsc{b})^p \right].
\end{align}
Here, the sum is over all flavors $k$ of parton: quarks, antiquarks,
and gluons.  When $\Tsc{b}$ is small, the coefficient functions,
$\tilde{C}_{j/k}$, can be usefully expanded in perturbation theory,
provided that $\sqrt{\zeta}$ and $\mu$ are comparable to $1/\Tsc{b}$,
so that large logarithms involving these parameters are not present.
Corrections to the OPE are suppressed by a power of $\Tsc{b}$, as
indicated by the last term in (\ref{eq:TMD.OPE}).

The lowest order coefficient is effectively unity:
\begin{equation}
  \label{eq:C.LO}
  \tilde{C}_{j/k}\xleft( x/\xi,\Tsc{b};\zeta,\mu,\alpha_s(\mu) \right)
  = \delta_{jk} \, \delta(\xi/x-1) + O(\alpha_s).
\end{equation}
See Ref.\ \cite{Gehrmann:2012ze,Gehrmann:2014yya} for the coefficient
functions to order $\alpha_s^2$.

An OPE of the same form as (\ref{eq:TMD.OPE}) applies also to the
helicity and transversity densities (but generally with coefficient
functions that differ beyond lowest order).  The proofs work the same
way.  In each case one can characterize the OPE in the same way as in
last part of Sec.\ \ref{sec:subsidiary}: It formulates the
QCD corrections to the parton-model idea that the integral of a TMD
density over all transverse momenta is the corresponding integrated
density.

As for the other (polarization-dependent) TMD densities, like the
Sivers function, generalizations of Eq.\ (\ref{eq:TMD.OPE}) apply ---
see e.g., \cite[Eq.\ (9)]{Kang:2011mr}.
They relate these other TMD densities to more involved quantities
associated with matrix elements of higher-twist operators (e.g., the
Qiu-Sterman function \cite{Qiu:1990xxa,Qiu:1990xy}).  As such, they are
less useful, because, if nothing else, the Qiu-Sterman function is
much less well measured than conventional unpolarized parton
densities.  Thus it will often be useful not to apply an OPE to the
Sivers function, etc.

\subsection{Solutions}

The evolution equations can be used to reformulate the factorization
formula in such a way that:
\begin{itemize}
\item Universality properties are exhibited.  In particular,
functions with nonperturbative content, like parton densities, are at a single fixed scale.
\item Perturbatively calculated quantities have no large logarithms in
  their expansions in powers of $\alpha_s$.
\end{itemize}
We present useful solutions in several forms.  The different forms
can be used to
emphasize different aspects of the physics.

As a starting point, we can use the original factorization formula
(\ref{eq:kt.fact}) with $\zeta$ set to $Q^2$, and $\mu$ proportional
to $Q$:
\begin{widetext}
\begin{align}
\label{eq:kt.fact.Q}
  \frac{ \diff{\sigma} }{ \diff[4]{q}\diff{\Omega} } 
  ={}&
    \frac{2}{s} \sum_j
    \frac{ \diff{\hat{\sigma}_{j\bar{\jmath}}}(Q,\muQ,\alpha_s(\muQ)) }{ \diff{\Omega} }
    \int \diff[2]{\T{b}}
    ~ e^{i\T{q}\cdot \T{b} }
    ~ \tilde{f}_{j/A}(x_A,\T{b};Q^2,\muQ) 
    ~ \tilde{f}_{\bar{\jmath}/B}(x_B,\T{b};Q^2,\muQ)
\nonumber\\
    & + \mbox{polarization terms} 
      + \mbox{high-$\Tsc{q}$ term ($Y$)}
      + \mbox{power-suppressed}.
\end{align}
\end{widetext}
Here $\muQ = C_2 Q$, and, as usual, the constant $C_2$ can be chosen
with the aim of optimizing the accuracy of perturbation theory for
$\diff{\hat{\sigma}}$.  This formula exhibits a parton-model form with
a perturbatively calculable hard scattering, while characteristic QCD
effects are hidden inside the $Q$ dependence of the parton densities.

A corresponding equation to (\ref{eq:kt.fact.Q}) is used in ordinary
collinear factorization: There one has a perturbatively calculable
hard scattering convoluted with parton densities at a scale
proportional to $Q$.  Then one uses DGLAP evolution to express the
$Q$-dependent collinear parton densities in terms of those at a fixed
reference scale.  We apply the same strategy to TMD factorization.
Differences in the implementation arise from three sources: (1) DGLAP
evolution results in a complicated convolution to relate collinear
parton densities at different scales; this convolution is hidden
inside numerical computer codes for its implementation.  For the TMD
case, the solutions to the evolution equations are simple enough to be
exhibited explicitly.  (2) There are two scale arguments $\zeta$ and $\mu$
in the TMD case; this just reflects two sources of scale dependence as
implemented in the technical definition of the TMD parton densities.
(3) There is an extra variable $\T{b}$ in some of the functions, and
extra steps in the analysis are used to treat this.

In the rest of this section,
we will first evolve the TMD parton densities and $\tilde{K}$ so that
factorization is presented in terms of these functions at fixed values
of their scale arguments.  This displays the universality properties
of parton densities and of the CSS kernel $\tilde{K}$, and is
especially suitable when these functions are in a nonperturbative
region.  It also exhibits in a
rather direct manner the nature of the evolution with $Q$ and the
strong predictions that can be obtained 
even without the use of perturbation theory to compute TMD parton
densities and $\tilde{K}(\Tsc{b})$ at small $\Tsc{b}$.  We will also
exhibit a modified form of this solution which more clearly displays the fact that
measurements to give an unambiguous determination of
$\tilde{K}(\Tsc{b})$ require changes of the center-of-mass energy
$\sqrt{s}$, not merely changes in $Q$ at fixed $\sqrt{s}$.

Next
we exploit RG transformations to allow perturbative
expansions without logarithms, when appropriate, and we 
arrange for explicitly defined separate
functions that contain the nonperturbative parts.
The result is a version
of a formula given by CSS in Ref.\ \cite{Collins:1984kg}.
It
utilizes perturbative information both for $\tilde{K}$ and for the OPE
for the TMD densities, and therefore makes maximum use of the
predictive power of perturbative calculations.  But the structure of
the formula, (\ref{eq:soln.2}) below, does not make clear the fact
that it uses TMD parton densities in a form close to that of the
parton model.  

We will also present a 
modified form of this solution that uses perturbative
information only for $\tilde{K}$, and effectively treats the TMD
densities as if they are functions to be obtained from data.  The
reason for this is that it is common that
low-energy data have been
fitted with TMD parton densities in a pure parton-model 
formalism~\cite{Anselmino:1995tv,Anselmino:2004nk,Anselmino:2005ea,
Collins:2005ie,Collins:2005rq,Anselmino:2005an,Anselmino:2009st,
Schweitzer:2010tt,Anselmino:2013vqa,Anselmino:2013rya,Signori:2013mda}.
Normally a Gaussian dependence on $\Tsc{b}$ is used in such fits; they
are intended to take account of nonperturbative contributions to the
TMD functions.  This is an approximation that ignores the predicted
small-$\Tsc{b}$ dependence in real QCD.  It is, of course, important to
know how such TMD functions evolve.

The multiplicity of different forms of solution of the evolution
equations gives a potentially confusing set of different ways of using
TMD factorization phenomenologically.  In fact, the different forms of 
solution correspond to
different methods that have appeared in the literature (typically in
conjunction with approximations).  What should be clear from the order
of presentation in this paper is that they are all related to a single
unifying factorization framework.

\begin{widetext}
\subsubsection{Fixed scales}

The solution with the parton densities and the CSS kernel at fixed
scales is:
\begin{align}
\label{eq:soln.1}
  \frac{ \diff{\sigma} }{ \diff[4]{q}\diff{\Omega} } 
  ={}&  \frac{2}{s} \sum_j
    \frac{ \diff{\hat{\sigma}_{j\bar{\jmath}}}(Q,\muQ,\alpha_s(\muQ)) }{ \diff{\Omega} }
    \int \diff[2]{\T{b}}
    e^{i\T{q}\cdot \T{b} }
    \tilde{f}_{j/A}\big( x_A, \T{b}; Q_0^2,\mu_0 \bigr)
    \,
    \tilde{f}_{\bar{\jmath}/B}\big( x_B, \T{b}; Q_0^2,\mu_0 \bigr)
\nonumber\\&
  \,\times
  \left( \frac{ Q^2 }{ Q_0^2 } \right)^{\tilde{K}(\Tsc{b};\mu_0) }
  \exp\left\{   
           \int_{\mu_0}^{\muQ}  \frac{ \diff{\mu'} }{ \mu' }
           \biggl[ 2 \gamma_j(\alpha_s(\mu'); 1) 
                 - \ln\frac{Q^2}{ (\mu')^2 } \gamma_K(\alpha_s(\mu'))
           \biggr]
  \right\}
\nonumber\\&
+ \mbox{polarized terms}
+ \mbox{large $\Tsc{q}$ correction, $Y$}
+ \mbox{p.s.c.}
\end{align}
Here $Q_0$ and $\mu_0$ are chosen fixed reference scales.  They have
exactly the same status as a similar parameter that is used in DGLAP
evolution of ordinary parton densities (e.g., \cite{Owens:2012bv}),
and that is often denoted by $Q_0$.  In both cases, there is a
functional form of parton densities at the fixed reference scale (or
scales), and evolution has been used to obtain $Q$-dependent parton
densities, as used in Eq.\ (\ref{eq:kt.fact.Q}).  The solution
(\ref{eq:soln.1}) can be obtained from (\ref{eq:kt.fact.Q}) by first
applying the RG equation (\ref{eq:RG.TMD.pdf}) for TMD densities and
then the CS equation (\ref{eq:CSS.evol}), to express the $Q$-dependent
densities in terms of the densities at the scales used in
(\ref{eq:soln.1}).  It will generally be convenient to set
$\mu_0=\mu_{Q_0}=C_2Q_0$. 
In Eq.~\eqref{eq:soln.1} (and in 
similar later equations) ``$\mbox{p.s.c.}$'' is a short-hand 
for ``power suppressed corrections''.

Although the reference scale $\mu_0$ is in principle arbitrary, it
should in practice be chosen large enough to be treated as being in
the perturbative region.  This allows finite-order perturbative
calculations of the anomalous dimensions, $\gamma_j$ and $\gamma_K$ to
be appropriate for all the values in the integral over $\mu'$.  Given
the choice $\mu_0=C_2Q_0$, it will generally be sensible to choose
$Q_0$ to be near the lower end of the range of $Q$ for the data to
which one applies factorization.  This is exactly the same as with
typical implementations of DGLAP evolution.

In the hard scattering, we have preserved $\muQ =C_2 Q$, so that it
has no large logarithms in its perturbative coefficients, and can
therefore be effectively calculated by low order perturbation theory
in powers of $\alpha_s(\mu_Q)$.  

It is convenient to notate the evolution factor on the second line of
(\ref{eq:soln.1}) as
$e^{-S(\Tsc{b},Q,Q_0,\mu_0)}$, where
\begin{equation}
  \label{eq:S.def}
  S(\Tsc{b},Q,Q_0,\mu_0)
  =
  - \tilde{K}(\Tsc{b};\mu_0) \ln \frac{ Q^2 }{ Q_0^2 }
  + \int_{\mu_0}^{\muQ}  \frac{ \diff{\mu'} }{ \mu' }
           \biggl[ - 2 \gamma_j(\alpha_s(\mu'); 1) 
                   + \ln\frac{Q^2}{ (\mu')^2 } \gamma_K(\alpha_s(\mu'))
           \biggr].
\end{equation}

The two equations (\ref{eq:kt.fact.Q}) and (\ref{eq:soln.1}) exhibit a
very close relationship to a TMD parton model formula.  Eq.\
(\ref{eq:kt.fact.Q}) is of a parton model form except that: (a) the
hard part has perturbative higher order corrections, (b) the TMD
parton densities are scale dependent, and (c) there is a $Y$-term.
But when the cross section is expressed in terms of TMD parton
densities at the reference scales, we find in Eq.\ (\ref{eq:soln.1}) 
a factor $e^{-S}$ that gives the important effects of gluon radiation.

Furthermore, the solution (\ref{eq:soln.1}) exhibits universality properties of the
TMD densities that are the same as in the parton model.  That is,
under all circumstances
where a TMD factorization theorem holds,
the same TMD densities, functions of $x$ and
$\T{b}$, are used, up to possible factors of $-1$ for 
$T$-odd functions.  Both the perturbatively calculable hard scattering
and the factor involving perturbatively calculable anomalous
dimensions only affect the normalization of the cross section, but not
its shape as a function of $\Tsc{q}$.  The remaining factor
$(Q^2/Q_0^2)^{\tilde{K}(\Tsc{b};\mu_0)} =\exp\xleft[
\tilde{K}(\Tsc{b};\mu_0) \ln(Q^2/Q_0^2) \right]$ gives a $Q$-dependent
change in shape of the distribution, a very characteristic effect of
gluonic emission in a gauge theory.

As is well known, a minor modification to universality arises because
the appropriate TMD parton densities differ between processes of the
Drell-Yan type and those of the SIDIS type.  The operators in the
definition of the TMD densities use oppositely directed Wilson lines
in the two cases.  Most TMD densities are
numerically unchanged, but $T$-odd
densities, like the Sivers function, change sign
\cite{Collins:2002kn}.

It is important to recall that the derivation of Eq.~\eqref{eq:soln.1}
depends on the TMD factorization and evolution equations, and that
these in turn depend on properties of the particular definitions used
for the TMD parton densities; see Ref.~\cite[Eq.~(13.106,
13.108)]{Collins:2011qcdbook} and the discussions leading up to this
definition.  (These remarks apply equally to the original CSS
derivations \cite{Collins:1981uk, Collins:1981uw, Collins:1984kg}.)

\subsubsection{Measuring CSS evolution}

Suppose we temporarily ignored the perturbative information 
about $\Tsc{b}$-dependence 
available at small $\Tsc{b}$ for the CSS kernel and for the TMD densities.  Then
one could determine $\tilde{K}$ from a limited set of data with
variable $Q$ at fixed $x_A$ and $x_B$ (up to errors associated with
the unknown power-suppressed corrections).  The evolution of the cross
section at every other value of 
$x_A$, $x_B$ and $Q$ would then
be determined.  The TMD densities can be determined from data from an
experiment at one value of $s$ (aside from the issue of flavor
dependence that can be best analyzed with data from other processes
like SIDIS).

However, if one examines data in a single experiment, i.e., at fixed $s$,
the $x$ dependence of the TMD functions is confounded with the $Q$
dependence of the factor involving $\tilde{K}$, since $Q^2=x_Ax_Bs$.
To exhibit the fact that $\tilde{K}$ can only be determined by varying
$s$ with $x_A$ and $x_B$ fixed, it may therefore be convenient to use
CSS evolution to change the choice of the $\zeta$-argument of the TMD
densities so as to make the corresponding $\tilde{K}$-dependent factor
a function of $s$ instead of $Q$.  For this purpose, an appropriate
solution is
\begin{align}
\label{eq:soln.1a}
  \frac{ \diff{\sigma} }{ \diff[4]{q}\diff{\Omega} } 
  ={}&  \frac{2}{s} \sum_j
    \frac{ \diff{\hat{\sigma}_{j\bar{\jmath}}}(Q,\muQ,\alpha_s(\muQ)) }{ \diff{\Omega} }
    \int \diff[2]{\T{b}}
    e^{i\T{q}\cdot \T{b} }
    \tilde{f}_{j/A}\big( x_A, \T{b}; x_A^2Q_0^2,\mu_0 \bigr)
    \,
    \tilde{f}_{\bar{\jmath}/B}\big( x_B, \T{b}; x_B^2Q_0^2,\mu_0 \bigr)
\nonumber\\&
  \,\times
  \left( \frac{ s }{ Q_0^2 } \right)^{\tilde{K}(\Tsc{b};\mu_0) }
  \exp\left\{   
           \int_{\mu_0}^{\muQ}  \frac{ \diff{\mu'} }{ \mu' }
           \biggl[ 2 \gamma_j(\alpha_s(\mu'); 1) 
                 - \ln\frac{Q^2}{ (\mu')^2 } \gamma_K(\alpha_s(\mu'))
           \biggr]
  \right\}
\nonumber\\&
+ \mbox{polarized terms}
+ \mbox{large $\Tsc{q}$ correction, $Y$}
+ \mbox{p.s.c.}
\end{align}
\end{widetext}
One might choose $Q_0^2=s_0$, where $s_0$ is the value of $s$ for some
particularly important set of data.  Given fixed values of $Q_0$ and
$\mu_0$, the functions $\tilde{f}_{j/H}\big( x, \T{b}; x^2Q_0^2,\mu_0
\bigr)$ are just like differently defined TMD densities.  That is, as
regards the kinematic variables $x$ and $\T{b}$, they are functions of
the same two variables as the TMD functions in Eq.\ (\ref{eq:soln.1}).
They have a definite relation to the versions of the densities used in
the previous solution.

Which is the most appropriate form to use in practice is not so clear.
Furthermore, the simple use of these solutions to fit data ignores two
sources of perturbatively accessible information for $\Tsc{b}$
dependence: The calculation of $\tilde{K}$ and of the coefficient
functions in the OPE (\ref{eq:TMD.OPE}).  So in Sec.\
\ref{sec:max.pert.soln} we will give another form of solution,
(\ref{eq:soln.2}) below, that is more suitable for using perturbative
calculations in combination with fits in the nonperturbative domain.

However, that solution obscures an important unifying property that is
clearly visible in Eqs.\ (\ref{eq:soln.1}) and (\ref{eq:soln.1a}).
This is, quite simply, the existence of the QCD entities that are the
TMD densities and the CSS kernel $\tilde{K}$.  When using
(\ref{eq:soln.2}), we can regard the purpose of the associated
perturbative calculations and of the fitting as providing useful
estimates for the TMD densities and for the function $\tilde{K}$ as
they appear in Eqs.\ (\ref{eq:soln.1}) and (\ref{eq:soln.1a}).  One
can imagine the result of a global fit of TMD factorization being
presented as tables of the TMD functions and of $\tilde{K}(\Tsc{b})$,
just as with global fits of integrated parton densities.
The fitting process
could use the complicated formula
(\ref{eq:soln.2}), but the application of the results to predict cross
sections could use a simpler formula, e.g., the original factorization
formula (\ref{eq:kt.fact}) or one of the fixed scale formulae
(\ref{eq:soln.1}) and (\ref{eq:soln.1a}).  
If results of a global fit were presented for evolved $Q$-dependent
TMD densities (as is routine for collinear parton densities), then
users would not even have to use (\ref{eq:soln.1}),
(\ref{eq:soln.1a}), or (\ref{eq:soln.2}); they could just use the
simple parton-model-like form (\ref{eq:kt.fact.Q}). 
(See Ref.~\cite{Hautmann:2014kza} for 
recent efforts in this direction.)

\emph{Note:} With the form of solution given in Eqs.\
(\ref{eq:soln.1}) and (\ref{eq:soln.1a}), it is dangerous to estimate
the kernel $\tilde{K}(\Tsc{b};\mu_0)$ by a simple fixed order
perturbative expansion with the renormalization scale set to $\mu_0$.
This is because there is an integral over all $\T{b}$, and inevitably
there will be large logarithmic corrections from higher order terms.
Thus a low-order perturbative expansion of $\tilde{K}(\Tsc{b};\mu_0)$
can give inaccurate results, especially if $Q/Q_0$ differs
substantially from unity.  Of course, the integral also extends into a
clearly nonperturbative region of large $\Tsc{b}$.

\subsubsection{Solution optimized for perturbative calculations}
\label{sec:max.pert.soln}

We now present a solution that corresponds to one presented by CSS.
The aim is to allow the maximum use of perturbative calculations.
First, the small-$\Tsc{b}$ expansion is applied to the TMD parton
densities.  Then, to allow the effective use of fixed-order
perturbative calculations, the evolution equations are applied so that
in the functions $\tilde{K}$ and $\tilde{C}_{j/f}$, $\mu$ and
$\sqrt{\zeta}$ are of order $1/\Tsc{b}$.  Thus, when $\Tsc{b}$ is
sufficiently much smaller than $1/\Lambda_{\rm QCD}$, these functions
are given by their expansions in powers of a small coupling,
$\alpha_s(1/\Tsc{b})$, and no large logarithms of $\mu\Tsc{b}$, etc,
are present.

Note, however, that perturbatively calculated functions may appear in
an exponent---as for $\tilde{K}$ and the anomalous dimensions.  Thus
any errors in a perturbative calculation of such a function can be
magnified by a large logarithm.

In any case, perturbative calculations are not applicable at large
enough values of $\Tsc{b}$, which, given our knowledge of QCD, is a
nonperturbative region.  An indication of where the nonperturbative
region is likely to be quantitatively important is given by an analysis by
Schweitzer, Strikman, and Weiss \cite{Schweitzer:2012hh}.  Using a
chiral effective theory, they found that there are two relevant
nonperturbative distance scales: a chiral scale $\unit[0.3]{fm} =
\unit[1.5]{GeV^{-1}}$ and a confinement scale $\unit[1]{fm} =
\unit[5]{GeV^{-1}}$.  At large $\Tsc{b}$, they find that a TMD density
behaves like an exponential $e^{-\Tsc{b}/l}$ times a power of
$\Tsc{b}$, with $l$ being a characteristic scale.  The chiral and
confinement scales manifest themselves in the large-$\Tsc{b}$
dependence of the sea and valence quark densities.

To get maximum predictive information, one should therefore combine
the use of perturbative calculations at small $\Tsc{b}$ with fits to
data to measure $\tilde{K}$ and the TMD densities at large $\Tsc{b}$.
Undoubtedly, fits to data will eventually be supplemented by further constraints
from nonperturbative calculations like those of Ref.\
\cite{Schweitzer:2012hh} from chiral models, and those of Ref.\
\cite{Musch:2011er} from lattice gauge theory.

Since the integral in Eq.\ (\ref{eq:kt.fact}) extends from $\Tsc{b}=0$
to $\Tsc{b}=\infty$, one cannot avoid using parton densities and
$\tilde{K}$ in the nonperturbative large-$\Tsc{b}$ region\footnote{An
  important, but separate, practical question is whether or not the
  integrand in Eq.\ (\ref{eq:kt.fact}) is large enough in the
  nonperturbative region for the details of the nonperturbative
  parametrization to matter in the context of particular
  calculations. }.  Therefore it is necessary to combine
nonperturbative information with perturbative calculations.  CSS
\cite{Collins:1984kg} provided a prescription\footnote{The CSS
  prescription is not the only possibility.  See Refs.\
  \cite{Qiu:2002mu,Qiu:2000ga} for one alternative.
  In addition Bozzi et al.\ \cite[Eq.\ (17)]{Bozzi:2010xn}, motivated
  by \cite{Catani:1992ua}, proposed a modification to improve the
  behavior of the formalism at small $\Tsc{b}$.  This involves the
  replacement of $\Tsc{b}^2$ by $\Tsc{b}^2+4e^{-2\gamma_E}/Q^2$ in the
  parton densities and evolution factors in (\ref{eq:kt.fact.Q}),
  (\ref{eq:soln.1}), and our other solutions, together with a
  consequent change in $Y$, as computed in \cite[App.\
  B]{Bozzi:2010xn}.  This is probably a generally useful
  prescription.
}  for doing
this; we will call their method the ``$\bstarsc$ method'', after the
name of a variable defined by CSS.

They first defined quantities with a smooth upper cutoff on
transverse distance, at a chosen value $\bmax$, with the use of
the following function of $\T{b}$:
\begin{align}
\label{eq:bstar}
  \bstar = \frac{ \T{b} }{ \sqrt{ 1 + \Tsc{b}^2/\bmax^2} }.
\end{align}

What are called the perturbative parts of the TMD densities and of
$\tilde{K}$ were defined by replacing $\T{b}$ by $\bstar$.  Then the
nonperturbative\footnote{``nonperturbative'' is somewhat of a
  misnomer.  If $\bmax$ were chosen to be excessively small, the
  values of the ``nonperturbative'' parts near $\Tsc{b}=\bmax$
  could be reliably estimated perturbatively.} parts were defined as
whatever is left over.  This idea is implemented with the aid of
functions $g_{j/H}(x,\Tsc{b};\bmax)$ and $g_K(\Tsc{b};\bmax)$ defined
by
\begin{equation}
  \label{eq:gK.def}
  g_K(\Tsc{b};\bmax) 
  = -\tilde{K}(\Tsc{b},\mu) + \tilde{K}(\bstarsc,\mu)
\end{equation}
and
\begin{multline}
  \label{eq:gjH.def}
  e^{-g_{j/H}(x,\T{b};\bmax)}
\\
  = \frac{ \tilde{f}_{j/H}\big( x, \T{b}; \zeta, \mu \bigr) }
         { \tilde{f}_{j/H}\big( x, \bstar; \zeta, \mu \bigr) }
    e^{g_K(\Tsc{b};\bmax) \ln(\sqrt{\zeta}/Q_0)}.
\end{multline}
Here $Q_0$ is a chosen reference scale, that simply determines how
much of the TMD density is in $e^{-g_{j/H}}$ and how much is put into
the exponential of $g_K$ times a logarithm that appears in Eq.\
(\ref{eq:gjH.def}) and in (\ref{eq:soln.2}) below.
As indicated by our notation, we will choose $Q_0$ here to have the
same value as in Eq.\ (\ref{eq:soln.1}). 
We treat $g_{j/H}$ and $g_K$ as needing to be fit to data.

Both of $g_{j/H}$ and $g_K$ vanish approximately\footnote{As we will
  see in Sec.\ \ref{sec:small-b}, the existence of 
  perturbatively controlled logarithmic singular behavior of
  $\tilde{K}$ and $\tilde{f}$ at small $\Tsc{b}$ implies that
  $g_{j/H}$ and $g_K$ are not exactly quadratic at small $\Tsc{b}$.}
like $\Tsc{b}^2$ at small
$\Tsc{b}$, from their definition, and become significant when
$\Tsc{b}$ approaches $\bmax$ and beyond.

Both functions are independent of both $\mu$ and $\zeta$.  This is
because there is an exact cancellation in the terms obtained by
applying the CSS and RG equations to the quantities on the right of
Eqs.\ (\ref{eq:gK.def}) and (\ref{eq:gjH.def}).  The functions do
depend, however, on the choice of the value of $\bmax$ and on
the particular CSS prescription for segregating nonperturbative
information.  It is the full TMD parton densities and the function
$\tilde{K}$ that are independent of $\bmax$ and of the use of
the $\bstarsc$ prescription of CSS.

As regards the possible flavor and $x$ dependence of $g_K$ and
$g_{j/H}$, this follows from that of the corresponding parent
functions, i.e., $\tilde{K}$ and the TMD parton densities.  Since
$\tilde{K}$ is independent of quark flavor, hadron flavor, and parton
$x$, so is $g_K$.  But the TMD parton densities can depend on quark and hadron
flavor and on $x$, so the same is true of the $g_{j/H}$ functions.

Given these definitions, the evolution equations and the
small-$\Tsc{b}$ expansion can be used to write the factorization
formula as
\begin{widetext}
\begin{align}
\label{eq:soln.2}
  \frac{ \diff{\sigma} }{ \diff[4]{q}\diff{\Omega} } 
  ={}&  \frac{2}{s}   \sum_{j,j_A,j_B}
        \frac{ \diff{\hat{\sigma}_{j\bar{\jmath}}}(Q,\mu_Q,\alpha_s(\mu_Q)) }{ \diff{\Omega} }
        \int \frac{ \diff[2]{\T{b}} }{ (2\pi)^2 }  e^{i\T{q}\cdot \T{b} }
\nonumber\\& \times
  e^{-g_{j/A}(x_A,\Tsc{b};\bmax) }
  \int_{x_A}^1 \frac{ \diff{\hat{x}_A} }{ \hat{x}_A }
       f_{j_A/A}(\hat{x}_A;\mubstar) 
  ~ \tilde{C}_{j/j_A}\xleft( \frac{x_A}{\hat{x}_A},\bstarsc;
  \mubstar^2, \mubstar, \alpha_s(\mubstar) \right)
\nonumber\\& \times
  e^{ -g_{\bar{\jmath}/B}(x_B,\Tsc{b};\bmax)}
  \int_{x_B}^1 \frac{ \diff{\hat{x}_B} }{ \hat{x}_B }
       f_{j_B/B}(\hat{x}_B;\mubstar) 
  ~ \tilde{C}_{\bar{\jmath}/j_B}\xleft( \frac{x_B}{\hat{x}_B},\bstarsc; \mubstar^2, \mubstar, \alpha_s(\mubstar) \right)
\nonumber\\& \times  
  \left( \frac{ Q^2 }{ Q_0^2 } \right) ^ {- g_K(\Tsc{b};\bmax)}
  \left( \frac{ Q^2 }{ \mubstar^2 } \right) ^ {\tilde{K}(\bstarsc;\mubstar)}
  \exp\xleft\{ 
       \int_{\mubstar}^{\muQ}  \frac{ \diff{\mu'} }{ \mu' }
          \left[ 2 \gamma_j(\alpha_s(\mu'); 1) 
                 - \ln\frac{Q^2}{ (\mu')^2 } \gamma_K(\alpha_s(\mu'))
          \right]
  \right\}
\nonumber\\&
+ \mbox{polarized terms}
+ \mbox{large-$\Tsc{q}$ correction, $Y$}
+ \mbox{p.s.c.}
\end{align}
Here $\mubstar$ is chosen to allow perturbative calculations of
$\bstarsc$-dependent quantities without large logarithms:
\begin{equation}
  \label{eq:mub.defn}
  \mubstar = C_1 / \bstarsc,
\end{equation}
where $C_1$ is a numerical constant typically chosen to be $C_1 = 2
e^{-\gamma_{\rm E}}$. 

\subsubsection{Fixed scale densities with perturbative organization of
CSS evolution}
\label{sec:CSS.plus.fixed.scale}

As mentioned in Sec.\ \ref{sec:OPE.small-b}, an OPE of the form of
(\ref{eq:TMD.OPE}) applies not only to the unpolarized TMD densities,
but also to the helicity and transversity densities, i.e., to those
TMD densities that have a corresponding integrated density (of the
``twist-2'' kind).  Thus the form of solution in the first part of
Eq.\ (\ref{eq:soln.2}) applies also to the terms with the helicity and
transversity densities~\cite{Bacchetta:2013pqa}.

But for the remaining terms, e.g., those with a Sivers function, the
appropriate OPE is of a different form, involving twist-3
distributions like the Qiu-Sterman function.  Therefore for these terms, it
is useful to have a version of Eq.\ (\ref{eq:soln.2}) that does not
apply the OPE to the TMD densities.  Even when the simple OPEs can be
applied, it can still be useful to treat the TMD functions at a fixed
scale as being measured from data.  In this new form of the solution, 
we retain the TMD parton densities as in Eq.\
(\ref{eq:soln.1}), but we do RG improvement on $\tilde{K}$ to use its
perturbative expansion optimally; we also set $\mu_0=\mu_{Q_0}=C_2Q_0$, so
that we arrange for the evolution factor to be unity at $Q=Q_0$.  The
result is
\begin{align}
\label{eq:soln.2a}
  \frac{ \diff{\sigma} }{ \diff[4]{q}\diff{\Omega} } 
  ={}&  \frac{2}{s} \sum_j
    \frac{ \diff{\hat{\sigma}_{j\bar{\jmath}}}(Q,\muQ,\alpha_s(\muQ)) }{ \diff{\Omega} }
    \int \diff[2]{\T{b}}
    e^{i\T{q}\cdot \T{b} }
    \tilde{f}_{j/A}\big( x_A, \T{b}; Q_0^2,\mu_{Q_0} \bigr)
    \,
    \tilde{f}_{\bar{\jmath}/B}\big( x_B, \T{b}; Q_0^2,\mu_{Q_0} \bigr)
\nonumber\\&
  ~\times
  \exp\left\{ 
       \left[ - g_K(\Tsc{b};\bmax) + \tilde{K}(\bstarsc;\mubstar)
           - \int_{\mubstar}^{\mu_{Q_0}}  \frac{ \diff{\mu'} }{ \mu' }
                 \gamma_K(\alpha_s(\mu'))
       \right]
       \ln\frac{Q^2}{Q_0^2}
      \right\}
\nonumber\\&
  ~\times
  \exp\left\{   
           \int_{\mu_{Q_0}}^{\muQ}  \frac{ \diff{\mu'} }{ \mu' }
             \left[ 2 \gamma_j(\alpha_s(\mu'); 1) 
                 - \ln\frac{Q^2}{ (\mu')^2 } \gamma_K(\alpha_s(\mu'))
             \right]
  \right\}
\nonumber\\&
+ \mbox{polarized terms}
+ \mbox{large $\Tsc{q}$ correction, $Y$}
+ \mbox{p.s.c.}
\end{align}
The exponentials are unity when $Q=Q_0$.
That allows the fitting of parton densities as in the parton model at
this scale (with small perturbative corrections from the hard
scattering).  Then the exponentials show how to do evolution to other
energies in terms of perturbative quantities without logarithms, and a
single nonperturbative function.
The above solution is related to a form of solution given in
Ref.~\cite{Aidala:2014hva}. 

In both of Eqs.\ (\ref{eq:soln.2}) and (\ref{eq:soln.2a}), the
strategy in organizing the solution was to arrange that quantities
to be calculated perturbatively are used in a region where
the coupling is in a perturbative region and that there are no large
logarithms in the expansions of these quantities.  (The logarithms in
ordinary fixed coupling expansions have all moved into the explicit
logarithms and into the integrals over $\mu'$.)  Thus the general size
of the errors due to truncation of perturbation expansions can be
quantified.  

Quantities that are not to be calculated perturbatively in a given 
form of the solution
(e.g., $ \tilde{f}_{j/A}\big( x_A, \T{b}; Q_0^2,\mu_{Q_0} \bigr)$ in 
Eq.~\eqref{eq:soln.2a}) are assumed to be fitted or treated using 
nonperturbative methods.

\end{widetext}

\subsubsection{Choice of parameters}

There are a number of arbitrary parameters in the solutions
(\ref{eq:soln.1}), (\ref{eq:soln.1a}), (\ref{eq:soln.2}) and
(\ref{eq:soln.2a}), notably $\bmax$, $\mu_0$ and $Q_0$.  Their
occurrence might appear to reduce the predictive power of the
formalism.  However, these appearances are misleading.

The trickiest case is that of $\bmax$.  It is sometimes said
that $\bmax$ is a parameter to be fitted to data, e.g.,
\cite[Sec.\ I]{Echevarria:2012pw}.  But, as can be seen from the
definitions and derivation summarized above, this is not the case.
Equation (\ref{eq:soln.2}) is true independently of the choice of
$\bmax$.  When $\bmax$ is changed, the so-called
nonperturbative functions, $g_K$ and $g_{j/H}$, defined in (\ref{eq:gK.def}) and
(\ref{eq:gjH.def}), change their form.  In fits
to data, the results should be equivalent, provided that the
parametrizations used are flexible enough.

However, if limited fixed parametrizations are used for $g_K$ and
$g_{j/H}$, they may work better with one value of $\bmax$ than
another.  Also, if an excessively small value of $\bmax$ is
used, much of the fitting of the functions will be devoted to
recovering their dependence on $\Tsc{b}$ in a region where
perturbative calculations would be adequate.  We will see symptoms of
this later.

The status of the other parameters $\mu_0$ and $Q_0$ is easier to
explain.  These are
just like the scale used to specify a measured
value of the strong coupling, or the scale at which initial parton
densities are parametrized for DGLAP evolution in global fits to
ordinary integrated parton densities.  For example, it is commonly
chosen to report the value of the strong coupling at the scale of the
mass of the $Z$ boson.  The choice $\mu=m_Z$ is essentially arbitrary
(provided that it is in a perturbative region).  No matter what scale
is chosen, there is one number (a value of the coupling) that needs to
be reported as the result of a fit to data.  Once a particular scale
is chosen, the meaning of the coupling's numerical value is fixed.  If
someone prefers a different choice of scale, then the original
numerical coupling $\alpha_s(m_Z)$ may be transformed unambiguously to
a value at the new scale, without any gain or loss of predictive
power.  
The same remarks apply to our scales $Q_0$ and $\mu_0$.
The only exception could be due to the
influence of errors caused by truncation of perturbation series, which
is not an issue of principle.

\section{Universality properties}
\label{sec:univ}

The issues that motivated this paper concerned the fitting of TMD
parton densities and their evolution in one collection of data and the
use of the results to predict other experimental data.  However, in a
more comprehensive view of TMD factorization, there are properties of
the factors that concern their different kinds of universality.  So in
this section, we review the universality properties.  Proofs of the
statements made are to be found in Ref.\ \cite{Collins:2011qcdbook}
and elsewhere.
The issues are particularly important for the nonperturbative
functions, but they apply equally to corresponding perturbatively
calculable quantities.  Some of the universality properties were not
fully explicitly derived prior to Ref.\ \cite{Collins:2011qcdbook}.

First, we examine the function $\tilde{K}(\Tsc{b};\mu)$ that controls TMD
evolution (and its corresponding anomalous dimension $\gamma_K$ and
``nonperturbative function'' $g_K(\Tsc{b};\bmax)$).  Since $\tilde{K}$ is a
property of Wilson line operators, it depends on the color
representation of the two partons entering the hard scattering, but
not on quark flavor, hadron flavor, $Q$, and parton fractional momentum.
Thus, for all processes involving quarks, the same $\tilde{K}$ is used.
It was also proved not to depend on whether the quarks are
initial-state or final-state, so the same $\tilde{K}$ applies to all
versions of the Drell-Yan process, SIDIS, and $e^+e^-$-annihilation.
An immediate implication is that in Eq.\ (\ref{eq:soln.1}) etc, the
part of the evolution factor that involves $\tilde{K}$ (and hence
$\gamma_K$) is a common ($\Tsc{b}$-dependent) factor in the
$\Tsc{b}$-integrand, independent of which term it applies to in the
sum over parton flavors $j$, $j_A$, and $j_B$.

The most important real-world situation in which a different value of
$\tilde{K}$ is needed is in the Drell-Yan-like process of Higgs
production by gluon-gluon annihilation in hadron-hadron collisions,
since gluons are color octet.

The anomalous dimension $\gamma_j$, for the TMD quark densities,
arises from a mass-independent calculation with spin-$1/2$ quarks.  It
is therefore independent of quark flavor (and also of hadron flavor
and parton momentum fraction).  However, in hypothetical extensions of
QCD, there could be scalar color-triplet quarks (like squarks in a
supersymmetric theory), and these could have different anomalous
dimensions.
Gluons (not to mention gluinos) have a different
anomalous dimension.

The hard scattering depends on the process, but only on the
variables available for the partons initiating the hard scattering.
It is of course perturbatively calculable, and has well-known
dependence on partonic flavor.  There is an interesting partial
universality of higher order corrections.  For example, in the
electromagnetic hard scattering for Drell-Yan, the ratios of the one-
and two-loop corrections to the lowest-order term are quark-flavor
independent (when quark masses are neglected).  Flavor dependence 
first arises at order $\alpha_s^3$, where the virtual photon can
couple to a quark loop that has a different flavor than the quark and
antiquark initiating the hard scattering. However, the loop
corrections are generally different between ``time-like'' processes
(e.g., Drell-Yan) and ``space-like'' processes (e.g., SIDIS).  In 
addition, there are potential (and calculable) differences between the
hard scattering for unpolarized quarks and the parts that depend on
quark polarization.

As for the complete TMD functions (parton densities and fragmentation
functions), the nonperturbative parts that do not arise from $\tilde{K}$ are in general all different
and can depend on the flavors of the parton and hadron.  
The shape of the nonperturbative $\Tsc{b}$ dependence can depend on both
the values of $x$ and on the flavor.  It is therefore in general
incorrect to assume that the nonperturbative $\Tsc{b}$ dependence is
a universal factor times the corresponding integrated distribution.
The nonperturbative modeling by Schweitzer, Strikman, and Weiss
\cite{Schweitzer:2012hh} is important in suggesting a large difference
between the $\Tsc{b}$ dependence for sea and valence quarks.

Each particular TMD parton density is the same in all processes where it is used,
aside from the effects of evolution, and aside from the predicted
reversal of sign \cite{Collins:2002kn} of ``time-reversal-odd''
functions (Sivers function, etc) between Drell-Yan and SIDIS.

The universality properties of the nonperturbative functions
$g_{j/A}(x,\Tsc{b};\bmax)$ match those of the corresponding TMD functions.
(However, as pointed out above Eq.\ (\ref{eq:soln.2a}), the use of
functions like $g_{j/A}$
applies, in its simplest form,
only to the TMD densities that
correspond to the standard integrated densities, i.e., the
unpolarized, helicity, and transversity densities.)

The expansion coefficients $\tilde{C}$ in the OPE for TMD parton
densities depend on the color of the partons involved.  To the extent
that quark masses are neglected, the dependence of $\tilde{C}$ on
flavor is governed by exact flavor symmetry.  Beyond lowest order,
they do depend on the polarization type of the TMD functions: e.g.,
unpolarized TMD parton density as compared with the coefficients for
the corresponding transversity TMD parton densities.  They can be
different between the expansions for TMD parton densities and for TMD
fragmentation functions.

\section{A single master function for CS evolution of TMD densities}
\label{sec:master}

\subsection{Definition and properties}

In this section, we show how to gain a more unified view of TMD
evolution. The starting point is the first form of solution
(\ref{eq:soln.1}) of the evolution equations.  There, the TMD
densities are all independent of $Q$, and the $Q$-dependence, for each
combination of flavors of quark entering the hard scattering in the
first two lines, arises from three sources:
\begin{itemize}
\item The $Q^{2\tilde{K}(\Tsc{b},\mu_0)}$ factor.
\item The exponential of anomalous dimensions.
\item The coupling $\alpha_s(\mu_Q)$ in the hard scattering
  $\diff{\hat{\sigma}_{j\bar{\jmath}}}$. 
\end{itemize}
\begin{widetext}
We first observe that only the first item gives dependence on
$\Tsc{b}$, and therefore only this item gives a $Q$-dependent change
in the shape of $\Tsc{b}$ distribution, which would then be reflected
in the distribution of the cross section in transverse momentum.  
This statement is valid for the contribution of a particular quark
flavor.  If different quark flavors have different intrinsic
transverse-momentum distributions, then a change in the relative
normalization of the different flavor terms would be a source of
$Q$-dependence in the shape of the transverse-momentum dependence of
the cross section. However, the first two items in the list are flavor
independent.  Moreover, as regards the hard scattering, the ratios of
one- and two-loop corrections relative to the lowest graph are flavor
independent, as we observed in Sec.\ \ref{sec:univ}.  Thus flavor
dependence occurs only in the third item in the above list and only at
the rather high order $\alpha_s^3(Q)$.

Hence to a good approximation, the $Q$-dependence in the cross section
is merely an overall factor in the summed integrand, $\tilde{W}$, as in
(\ref{eq:W.sigma}).  This factor is a $Q$-dependent normalization
times the $Q^{2\tilde{K}(\Tsc{b},\mu_0)}$ factor that affects the
shape.  This implies that a measurement of the cross section alone is,
in principle, sufficient to test the evolution in $Q$, and to give a
measurement of $\tilde{K}$.  Hence, for dealing with evolution, there
is essentially no need to do a decomposition by parton
flavor, even though the evolution kernel $\tilde{K}$ is defined as a
property of the individual TMD parton densities.  In this sense, the
situation is quite different from the one of testing the evolution of
ordinary integrated parton densities.

What is also striking is that the same flavor independence and
evolution factor apply to all cases involving triplet quarks: not only
to unpolarized Drell-Yan, but also to polarized cases, e.g., with the
Sivers function, to SIDIS, and to back-to-back hadron production in
$e^+e^-$ annihilation.

Now consider the contribution of a particular flavor.  We defined
$\tilde{W}_j$ in Eq.~\eqref{eq:Wj.def}.  Differentiating it with
respect to $Q^2$ (or $s$) at fixed $x_A$ and $x_B$ gives
\begin{align}
\label{eq:xsect.evol}
    \frac{ \partial \ln \tilde{W}_j(\T{b},Q,x_A,x_B) }
         { \partial \ln Q^2 }
    = \frac{ \partial \ln \tilde{W}_j(\T{b},Q,x_A,x_B) }
           { \partial \ln s }
    = {}&
      \tilde{K}(\Tsc{b};\mu) + G^{\rm DY}_{j\bar{\jmath}}(\alpha_s(\mu),Q/\mu)
\nonumber\\
    & \hspace*{-3cm}
    =
      \tilde{K}(\Tsc{b};\mu_0) 
      + G^{\rm DY}_{j\bar{\jmath}}( \alpha_s(\mu_Q),Q/\muQ )
      - \int_{\mu_0}^{\muQ}  \frac{ \diff{\mu'} }{ \mu' }
             \gamma_K(\alpha_s(\mu'))
\nonumber\\
    & \hspace*{-3cm}
    =
       - g_K(\Tsc{b};\bmax) + \tilde{K}(\bstarsc;\mubstar)
      + G^{\rm DY}_{j\bar{\jmath}}( \alpha_s(\mu_Q),Q/\muQ )
      - \int_{\mubstar}^{\muQ}  \frac{ \diff{\mu'} }{ \mu' }
             \gamma_K(\alpha_s(\mu')),
\end{align}
where $G^{\rm DY}_{j\bar{\jmath}}$ is defined by 
\begin{equation}
\label{eq:G.def}
  G^{\rm DY}_{j\bar{\jmath}}(\alpha_s(\mu),Q/\mu)
  =
  \frac{ \partial{} }{ \partial{ \ln Q^2 } }
  \left[ \ln \frac{ Q^2\diff{\hat{\sigma}_{j\bar{\jmath}}}(Q,\mu,\alpha_s(\mu)) }
                  { \diff{\Omega} }
  \right].
\end{equation}
If $\mu$ is set to $\muQ=C_2Q$, proportional to $Q$, we can write 
\begin{equation}
\label{eq:G.other}
  G^{\rm DY}_{j\bar{\jmath}}(\alpha_s(\muQ),Q/\muQ)
  =
  \frac{ \diff{} }{ \diff{ \ln Q^2 } }
     \left[ \ln \frac{ Q^2\diff{\hat{\sigma}_{j\bar{\jmath}}}(Q,\muQ,\alpha_s(\muQ)) }
                     { \diff{\Omega} }
     \right]
  + \gamma_j(\alpha_s(\muQ);1) - \gamma_K(\alpha_s(\muQ))\ln\frac{Q}{\muQ}.
\end{equation}
\end{widetext}
Notice that the derivative with respect to $\ln Q^2$ in Eq.\
(\ref{eq:G.other}) is a total derivative, unlike the partial
derivative in Eq.\ (\ref{eq:G.def}).  That is, it acts on both the $Q$
and the $\mu_Q$ arguments of $\hat{\sigma}$, and so the first term in
the derivative of the hard cross section in Eq.\ (\ref{eq:G.other}) is
of order $\alpha_s(\muQ)^2$.  The notation on the right-hand side of
Eq.\ (\ref{eq:xsect.evol}) corresponds to a notation in the CSS
papers.  All the derivatives are at fixed $x_A$ and $x_B$.

Notice also that from our previous argument about flavor dependence
of the hard scattering, the lowest order in which $G^{\rm
  DY}_{j\bar{\jmath}}$ is flavor-dependent is $\alpha_s(Q)^4$: a
flavor dependence of $G$ arises from the $Q$-derivative of the
coupling in a 3-loop graph (in the electromagnetic case).
Process-dependence of $G$ (e.g., between Drell-Yan and SIDIS) would in
contrast arise at $\alpha_s(Q)^2$, from the process-dependence of the
one-loop hard scattering.  Given this (small) process-dependence, in
contrast to the flavor- and process-independence of the other terms,
we have notated $G$ with ``DY'' for the process under discussion.

The three forms on the right-hand side of Eq.\ (\ref{eq:xsect.evol})
have the following significance: In the first line all quantities are
at a fixed common renormalization scale $\mu$.  In the second line, RG
improvement is made on $G$, so that it can be calculated
perturbatively with small errors, while $\tilde{K}$ is kept at a fixed
scale.  In the last line a RG improvement and the CSS prescription are
applied to $\tilde{K}$, to allow perturbative calculations of
$\tilde{K}$ supplemented by a parametrization of any nonperturbative
part, by $g_K(\Tsc{b};\bmax)$.

Everything on the right-hand side of Eq.\ (\ref{eq:xsect.evol}) is
the same as the evolution of the standard exponent $-S$ defined by
Eq.\ (\ref{eq:S.def}), except for the addition of a term
\begin{equation}
\label{eq:GH.def}
  \frac{ \diff{} }{ \diff{ \ln Q^2 } }
     \left[ \ln \frac{ Q^2\diff{\hat{\sigma}_{j\bar{\jmath}}}(Q,\muQ,\alpha_s(\muQ)) }
                     { \diff{\Omega} }
     \right]
\end{equation}
that is associated with the running coupling in higher-order
corrections to the hard scattering.

The change with $Q$ of the shape of the $\Tsc{b}$ dependence is
governed solely by the $\tilde{K}(\Tsc{b};\mu_0)$ term.  All the
remaining terms on the right of Eq.\ (\ref{eq:xsect.evol}) give only a
change in the normalization, since they are independent of $\Tsc{b}$.
There appears to be an arbitrariness, because of the choice of
$\mu_0$.  However, because of the RG equation (\ref{eq:RG.K}),
a change of $\mu_0$ in $\tilde{K}(\Tsc{b};\mu_0)$ only adds a
constant, independent of $\Tsc{b}$, and 
therefore only affects the normalization of the cross-section but not
its shape.  Of course, the $\mu_0$-dependence in
$\tilde{K}(\Tsc{b};\mu_0)$ is canceled by $\mu_0$ dependence of the
integral over $\gamma_K$ in the first line of Eq.\
(\ref{eq:xsect.evol}).

Note that to the good extent that $G^{\rm DY}_{j\bar{\jmath}}$ can be
approximated as flavor-independent, Eq.\ (\ref{eq:xsect.evol}) applies
equally to the full $\Tsc{b}$-space integrand
$\tilde{W}=\sum_j\tilde{W}_j$ as well as to its individual flavor
components. 

It can be useful to discuss the evolution of the shape of the
$\Tsc{q}$ dependence of the cross section, and hence of the $\Tsc{b}$
dependence of $\tilde{W}_j$, separately from the evolution of the
normalization.  Therefore it would be useful to describe it by a
function of $\Tsc{b}$ that does not have an arbitrary additive
constant defined only by an abstract renormalization scheme, 
unlike $\tilde{K}$. 

One possibility would be to subtract the value of $\tilde{K}$ at one
fixed value $b_c$ of $\Tsc{b}$, by defining
\begin{align}
  \label{eq:Khat.def}
  \widehat{K}(\Tsc{b})
  ={}&
  \frac{ \partial \ln \tilde{W}_j(\Tsc{b},Q,x_A,x_B) }
       { \partial \ln Q^2 }
  - \mbox{value with $\Tsc{b} \mapsto b_c$}
\nonumber\\
  = {}&
    \tilde{K}(\Tsc{b};\mu_0) - \tilde{K}(b_c;\mu_0).
\end{align}
Although this subtracted quantity still has an arbitrary parameter,
the parameter is, so to speak, a data-related quantity.  Note that the
dependence on the RG scale $\mu_0$ cancels.  The significance of the
definition in the first line of (\ref{eq:Khat.def}) is
that it can be applied in any formalism for working with a TMD cross
section.  This is in contrast to $\tilde{K}$, whose definition is
within a specific CSS-like formalism.  Sometimes presenting results in
terms of $\widehat{K}$ will allow a convenient comparison of different
calculations.

Instead, we now propose what may be a better definition of a universal
measure of
the evolution of the shape of TMD functions,
with no arbitrary scale at all.  The definition is
motivated by observing that any $Q$-dependent change in the
$\Tsc{b}$-shape of $\tilde{W}_j$ arises from $\tilde{K}(\Tsc{b})$ not
being a constant.  So the derivative of $\tilde{K}(\Tsc{b})$ gives the
relevant information.  Therefore, let us define
\begin{equation}
  \label{eq:A.def}
  A(\Tsc{b})
  =
    - \frac{ \partial }{ \partial\ln \Tsc{b}^2 }
      \frac{ \partial }{ \partial\ln Q^2 }
      \ln \tilde{W}_j(\Tsc{b},Q,x_A,x_B)
\end{equation}
with the derivatives again being at fixed $x_A$ and $x_B$.
Logarithmic derivatives are used so that $A(\Tsc{b})$ is
dimensionless.  We have chosen a convention where $A$ is defined to be
the negative of a derivative of $\tilde{W}_j$, so that $A$ will be a
generally positive function.  If we make the good approximation of
neglecting the flavor dependence of the hard-scattering, then we can
apply definition (\ref{eq:A.def}) to the flavor summed integrand
$\tilde{W}$ instead of its flavor components $\tilde{W}_j$.

We have constructed the definition of $A$ so that it applies not only
in the formulation of TMD factorization that we have presented, but in
any other similar formalism where the cross section is given as a
Fourier transform of a quantity like $\tilde{W}_j$ (summed over $j$).
These are typically some kind of TMD factorization or of resummation
formalism, together with various approximations.  Therefore, in
principle, it could depend on all the kinematic parameters and on
flavor as well as having expected dependence on $\Tsc{b}$.  By use of
CSS-style TMD factorization, we will show that the true $A$ in QCD
does not depend on any of these other variables.

From the solution (\ref{eq:soln.1}) of the evolution equations, from
the definitions (\ref{eq:bstar}) and (\ref{eq:gK.def}) for the
$\bstarsc$ prescription, and from the RG equation (\ref{eq:RG.K}) for
$\tilde{K}$, we obtain
\begin{align}
  \label{eq:A.bstar}
  A(\Tsc{b})
  = {}&
    - \frac{ \partial \tilde{K}(\Tsc{b};\mu_0) }{ \partial \ln \Tsc{b}^2 }
\nonumber\\
  = {}&
    - \frac{ \partial \tilde{K}(\Tsc{b};\mu) }{ \partial \ln \Tsc{b}^2 },
\nonumber\\
  = {}&
      \frac{ \partial }{ \partial \ln \Tsc{b}^2 }
      \left[
         g_K(\Tsc{b};\bmax) - \tilde{K}(\bstarsc;\mu)
      \right]
\nonumber\\
  = {}&
      \frac{ \partial g_K(\Tsc{b};\bmax) }{ \partial \ln \Tsc{b}^2 }
      + \frac{ \bstarsc^2 }{ \Tsc{b}^2 }
        A(\bstarsc;\mu)
\nonumber\\
  = {}&
      \frac{ \partial g_K(\Tsc{b};\bmax) }{ \partial \ln \Tsc{b}^2 }
      + \frac{ \bstarsc^2 }{ \Tsc{b}^2 }
        A(\bstarsc;\mubstar)
\nonumber\\
  = {}&
    \left.
      \frac{ \partial }{ \partial \ln \Tsc{b}^2 }
      \left[
         g_K(\Tsc{b};\bmax) - \tilde{K}(\bstarsc;\mu)
      \right]
    \right|_{\mu \mapsto \mubstar}.
\end{align}
In the second line, we have changed the value of the renormalization
scale.  Since a RG transformation of $\tilde{K}$ adds to it a
constant, independent of $\Tsc{b}$, we have RG-invariance of its
derivative with respect to $\Tsc{b}$, and hence of $A$:
\begin{equation}
\label{eq:A.RG}
  \frac{ \diff{ A(\Tsc{b}; \mu) } }{\diff{\ln \mu}} = 0,
\end{equation}

In the last two lines of (\ref{eq:A.bstar}), we have chosen to set the
value of $\mu$ to $\mubstar$ in order give a suitable form for the use
of finite-order perturbation theory, without large logarithms of
$\mu\bstarsc$.  As usual in these situations, although the exact value
of $A$ does not depend on $\mu$, a finite-order truncation of its
perturbative expansion does, by an amount of order the first omitted
term.  The notation on the last line of (\ref{eq:A.bstar}) is to
emphasize that the derivative of $\tilde{K}$ is to be taken at fixed
renormalization scale.  Only after that is the renormalization-scale
set to its final value of order $1/\bstarsc$.  The factors of
$\bstarsc^2/\Tsc{b}^2$ arise from the following calculation:
\begin{equation}
  \frac{ \partial \ln \bstarsc^2 }{ \partial \ln \Tsc{b}^2 }
  = \frac{ 1 }{ 1 + \Tsc{b}^2/\bmax^2 }
  = \frac{\bstarsc^2}{\Tsc{b}^2}.
\end{equation}

We will review the two-loop calculation of $A$ below, where we also
show the equality of our master function $A$ with the function of the
same name defined by CSS.

In the later parts of (\ref{eq:A.bstar}), we have notated $A$ with an
additional scale argument.  This indicates that when perturbative
calculations are performed, a RG scale must be chosen and used with a
corresponding value of the running coupling (and in principle running
masses).  More correctly, we should remember that the full set of
arguments of $A$ include all the parameters of QCD.  To indicate, when
necessary, the relevant arguments and parameters, we write $A(\Tsc{b})
= A(\Tsc{b};\mu) = A(\Tsc{b}; \mu, \alpha_s(\mu), m(\mu))$, where the
number of arguments that we actually choose to write depends on the
context.  In the situations where we actually use perturbation theory,
the (light) quark masses will normally be approximated by zero.  But
in a bigger context, including nonperturbative physics, the arguments
must also include the $\mu$-dependent masses of the
quarks. Since $A$ is defined from a RG-invariant quantity
$\tilde{W}_j$, RG invariance applies also to $A$, as in Eq.\
(\ref{eq:A.RG}). 

Given its definition (\ref{eq:A.def}), we can
regard the function $A(\Tsc{b})$ as a fundamental property of parton
physics in QCD, independently of any particular factorization scheme
and of particular techniques for its calculation.  Therefore, we
propose that it is a good master function for analyzing the evolution
of the shape of transverse-momentum distributions.  The function is
directly related to experimental data, 
from Eq.\
(\ref{eq:A.def}).  But on the theoretical side, it is a property of
certain Wilson loops; this arises via TMD factorization
from Eq.\ (\ref{eq:A.bstar}) and the definition of $\tilde{K}$
in the recent formulation of TMD factorization in
\cite{Collins:2011qcdbook}.

Both the perturbative calculation of $A$ and the fits of the
large-$\Tsc{b}$ behavior of $\tilde{K}$ agree that $A(\Tsc{b})$ is
generally positive.  This implies that $\tilde{K}(\Tsc{b})$ decreases
when $\Tsc{b}$ increases.  Hence when $Q$ is increased, the
$\T{b}$-space integrand $\tilde{W}$ undergoes a larger fractional
decrease at larger $\Tsc{b}$ than at smaller $\Tsc{b}$.  Thus the
shape in $\Tsc{b}$ of $\tilde{W}$ undergoes a shift to where it is
dominated by ever smaller $\Tsc{b}$ as $Q$ increases, as is well
known, from, e.g., Ref.\ \cite{Konychev:2005iy}.  Correspondingly, the
transverse-momentum distribution broadens.  The function $A$ codes
these properties.

An important prediction of QCD is that $A$ is independent of the
process, of $Q$, and of the kinematic variables $x_A$ and $x_B$; it is
also independent of quark and hadron flavor.  These
are highly nontrivial predictions of QCD dynamics,
since the lack of dependence on
these variables is not guaranteed merely by the definition of $A$ in the
first line of Eq.\ (\ref{eq:A.def}).

Without the CSS $\bstarsc$ prescription, a purely perturbative
calculation from Eq.\ (\ref{eq:K}) gives the one-loop term in $A$
\begin{equation}
\label{eq:A.PT.1}
A(\Tsc{b}) = 
\frac{ \alpha_s(C_1/\Tsc{b})C_F }{ \pi } 
   + O\xleft( \alpha_s(C_1/\Tsc{b})^2 \right),
\end{equation}
where we have made the standard choice of scale $\mu=C_1/\Tsc{b}$. 

\begin{widetext}
\subsection{Equality with CSS's $A$; presentation of two-loop value}

Now in Ref.\ \cite{Collins:1984kg}, CSS transformed their TMD
factorization formula to a form where 
\begin{equation}
  \label{eq:CSS.A.B}
  \tilde{W}_j
  = (\text{$Q$-independent factor}) \times 
  \exp\left\{
    - \int_{C_1^2/\Tsc{b}^2}^{C_2^2Q^2}
      \frac{ \diff{ {\mu'}^2 } }{ {\mu'}^2 }
      \left[
             A_{\rm CSS}(\alpha_s(\mu'); C_1 ) \ln\left( \frac{C_2^2 Q^2}{{\mu'}^2} \right)
             + B_{\rm CSS}(\alpha_s(\mu');C_1,C_2 )
      \right]
  \right\},
\end{equation}
with the quantities $A_{\rm CSS}$ and $B_{\rm CSS}$ being those defined
by CSS. See, for example, Eq.~(5.1) of Ref.~\cite{Collins:1984kg}.
Applying our definition of $A$, Eq.\ \eqref{eq:A.def} gives
\begin{equation}
\label{eq:A.v.ACSS}
  A(\Tsc{b};\mu) = A_{\rm CSS}(\alpha_s(C_1/\Tsc{b}); C_1 ).
\end{equation}
Thus, we must numerically identify $A$, as we defined it, with CSS's
quantity of the same name.\footnote{\emph{Warning:} Moch, Vermaseren,
  and Vogt \cite{Moch:2005id} gave a calculation at $O(\alpha_s^3)$ of
  a quantity they called $A$.  What they calculate is in fact
  $\frac{1}{2}\gamma_K$ rather than $A$ as defined by CSS (and us).
  This can be checked from the definition (2.4) that
  \cite{Moch:2005id} uses for $A$.  
  Essentially the same observation was made by Becher and Neubert in
  Ref.\ \cite{Becher:2010tm}.
}
This is not an accident, of course.  CSS's construction involved a
transformation of their original TMD factorization formula so that the
quantities involved could be related to properties of cross sections.

However, despite the numerical equality of our $A$ with $A_{\rm CSS}$,
differences exist in how the arguments of the function are presented.
Our $A$ is treated primarily as a function of $\Tsc{b}$; but it has as
extra parameters the renormalization scale $\mu$ and the parameters of
QCD, $A = A(\Tsc{b}; \mu, \alpha_s(\mu), m(\mu))$, subject to RG
invariance. It is defined for all $\Tsc{b}$.  
By its definition, $A(\Tsc{b})$ is scale independent for all $\Tsc{b}$, as 
expressed by Eq.~\eqref{eq:A.RG}.
Therefore despite the extra arguments, $A$ should be treated as a
function of the single variable 
$\Tsc{b}$, when we make plots as in later sections.  

In contrast, CSS \cite{Collins:1984kg} presented $A$ as a function of $\alpha_s$ alone.
This is because they focused on the perturbative region for $\Tsc{b}$,
and $A_{\rm CSS}$ arose in the construction of a particular
transformation of their solution of the evolution equations --- see
their (3.11)--(3.15).  Moreover, because of the transformation they
made of their formula, they did not actually use
$A_{\rm CSS}(\alpha_s)$ with the coupling evaluated at the scale
$C_1/\Tsc{b}$ as it appears in Eq.\ (\ref{eq:A.v.ACSS}).  Instead,
after the transformation to obtain (\ref{eq:CSS.A.B}), they have
$A_{\rm CSS}(\alpha_s(\mu))$ evaluated at a general scale inside an
integral over $\mu$ in a purely perturbative context.  Thus it is
completely unobvious that $A_{\rm CSS}(\alpha_s)$ is to be regarded as
a function of $\Tsc{b}$ as we do.  Indeed, $A_{\rm CSS}$ gives the
appearance in (\ref{eq:CSS.A.B}) of being a modified version of
$\gamma_K$. 

Our definition is in a sense more general, and
the use of $\Tsc{b}$ as the primary argument implies that there is a
definite measurable functional form for $A$ in the nonperturbative
region, without the need to transform to a function of coupling, which
would appear to be entailed by the CSS definition if taken as it
stands. A consequence is that, for any methods and approximations that
are proposed for some kind of TMD factorization, the function
$A(\Tsc{b})$ can be obtained, and compared with the same function
obtained in another method.  It is not restricted to perturbative
uses.

This equality of our new definition $A$ with that of CSS allows us to
use the results of its calculation by Kodaira and Trentadue
\cite{Kodaira:1981nh} and by Davies and Stirling
\cite{Davies:1984hs}.  We find
\begin{equation}
\label{eq:A.pure.PT}
A(\Tsc{b}) = \frac{ \alpha_s(\mu)C_F }{ \pi } 
   + \left( \frac{ \alpha_s(\mu) }{ \pi } \right)^2
     C_F \left[
             \left( \frac{67}{36}-\frac{\pi^2}{12} \right) C_A 
              -\frac{5}{9} T_R n_f  
           + \left( \frac{11}{12}C_A - \frac{1}{3} T_R n_f \right)
             \ln\xleft( \frac{\mu^2\Tsc{b}^2}{4e^{-2\gamma_E}} \right)
     \right]
   + O\xleft( \alpha_s(\mu)^3 \right).
\end{equation}
\end{widetext}
This differs from the Davies-Stirling result by the addition of the
logarithmic term to give invariance under a change of $\mu$.  For the
Davies-Stirling calculation, the CSS definition of $A$ was used, 
which, via Eq.\ (\ref{eq:A.v.ACSS}), can be interpreted as our $A$
when $\mu$ is set to $2e^{-\gamma_E}/\Tsc{b}$. 

From this we can verify the known 2-loop value of $\gamma_K$, given in Eq.\
\eqref{eq:gammaK}, by the following manipulations:
\begin{align}
  \gamma_K(\alpha_s) & = - \frac{ \diff{\tilde{K}} }{ \diff{\ln \mu } }
\nonumber\\
   & \hspace*{-10mm}
 = - \frac{ \partial \tilde{K}(\Tsc{b};\mu,\alpha_s(\mu))  }{ \partial\ln \mu }
       - \frac{ \diff{\ \alpha_s(\mu)} }{ \diff{\ln \mu } }
        \frac{ \partial\tilde{K}(\Tsc{b};\mu,\alpha_s(\mu)) }{ \partial\alpha_s(\mu)}
\nonumber\\
   & \hspace*{-10mm}
 = - \frac{ \partial\tilde{K}(\Tsc{b};\mu,\alpha_s(\mu)) }{ \partial \ln \Tsc{b} }
       - \frac{ \diff{\ \alpha_s(\mu)} }{ \diff{\ln \mu } }
        \frac{ \partial \tilde{K}(\Tsc{b};\mu,\alpha_s(\mu)) }{ \partial\alpha_s(\mu) }
\nonumber\\
   & \hspace*{-10mm}
 = 2 A(\Tsc{b};\mu,\alpha_s(\mu))
       - \frac{ \diff{\ \alpha_s(\mu)} }{ \diff{\ln \mu } }
        \frac{ \partial\tilde{K}(\Tsc{b};\mu,\alpha_s(\mu)) }{ \partial\alpha_s(\mu) },
\end{align}
given that quark masses are neglected.
Then $\gamma_K$ to two loops, Eq.\ \eqref{eq:gammaK}, is obtained from the
value of $A$ to two loops together with the one-loop values of
$\tilde{K}$ and of the evolution of $\alpha_s$.
This also agrees with the result given in Moch, Vermaseren, and Vogt
\cite{Moch:2005id} in their Eq.\ (3.8), which is for what they call
$A$, but which is $\frac{1}{2}\gamma_K$ in our notation.  They give
the 3-loop value in their Eq.\ (3.9).

The only dependence on the specific method for implementing perturbation theory is in the 
choice of $\mu$, and, in particular, in the way of optimizing the relationship to $\Tsc{b}$.
If $\mu$ is set to $C_1/\Tsc{b}$, then $\alpha_s(C_1/\Tsc{b})$ is
small at small $\Tsc{b}$, with  
finite perturbative coefficients in the expansion in
Eq.~\eqref{eq:A.pure.PT}. If $C_1$ is excessively large or small, 
the coefficients beyond order $\alpha_s$ 
become large and the validity of a truncated perturbative expansion becomes suspect.
The common choice is $C_1= 2 e^{-\gamma_E}$.

\section{Assessment of approximations and fits}
\label{sec:assess}

As we already mentioned, there appears to be incompatibility between
different articles that have compared TMD factorization with data.  In
this section we will first review a selection of these works.  Then we
will present the results in terms of the master function $A(\Tsc{b})$,
defined in Sec.\ \ref{sec:master}, to assess their actual
compatibility and validity, or lack thereof.

\subsection{Review of fits, etc}

Some authors (e.g., \cite{Landry:2002ix, Konychev:2005iy, Qiu:2000ga,
  Qiu:2000hf}) have used the CSS formalism with the form of solution
given in Eq.\ (\ref{eq:soln.2}) (or some variant).  
These fits use
data to make fits for the functions $g_K(\Tsc{b};\bmax)$ and
$g_{j/H}(x,\T{b})$ that were intended to parametrize nonperturbative
$\Tsc{b}$ dependence and that were defined in Eqs.\ (\ref{eq:gK.def})
and (\ref{eq:gjH.def}).  
However, different analyses obtain quite
different numerical values.  Others argue (e.g.,
\cite{Echevarria:2012pw, Becher:2010tm, Sun:2013dya, Sun:2013hua,
  Boer:2008fr}) that they can successfully completely avoid the use of
the $\bstarsc$ method 
(or any similar matching method) 
and the function $g_K(\Tsc{b};\bmax)$.

\subsubsection{Fits to Drell-Yan data with CSS formalism}
\label{sec:BLNY.KN}

Landry et al.\ \cite{Landry:2002ix} made fits to Drell-Yan data using
the CSS formalism, including the CSS treatment of nonperturbative
regions by the $\bstarsc$ method.  They chose $\bmax$ to
be\footnote{To allow an easy comparison with the size of a proton, we
  give numerical distances in units of $\unit{fm}$ as well as
  $\unit{GeV^{-1}}$.}
$\unit[0.5]{GeV^{-1}} =  \unit[0.1]{fm}$.  Their
best fit used quadratic functions for $g_{j/A}$ and $g_K$, without
flavor dependence or hadron species dependence for $g_{j/A}$:
\begin{align}
\label{eq:BLNY.form}
 & \exp\left[ -g_{j/A}(x_A,\Tsc{b};\bmax) -g_{\bar{\jmath}/B}(x_B,\Tsc{b};\bmax) - \right. \nonumber \\ &  \left.  \qquad -  g_K(\Tsc{b};\bmax) \ln(Q^2/Q_0^2) 
       \right]
\nonumber\\
  & = \exp\xleft\{ 
           - \left[ g_1 + g_2\ln \frac{Q}{2Q_0} 
                    + g_1g_3\ln (100x_Ax_B)
             \right] 
           \Tsc{b}^2
      \right\} ,
\end{align}
with $Q_0=\unit[1.6]{GeV}$.  This factor alone, when
Fourier-transformed to transverse-momentum space, would give a
Gaussian shape to the transverse momentum distribution, and the
$\Tsc{q}$ distribution would broaden as $Q$ increases.  The measured
values of the parameters were:
\begin{subequations}
\label{eq:BLNY_g}
\begin{align}
    g_1 &= \unit[0.21_{-0.01}^{+0.01}]{GeV^2}, \label{eq:BLNY_ga}
\\
    g_2 &= \unit[0.68_{-0.02}^{+0.01}]{GeV^2}, \label{eq:BLNY_gb}
\\
    g_3 &= -0.6_{-0.04}^{+0.05}. \label{eq:BLNY_gc}
\\
&\mbox{(BLNY, $\bmax=\unit[0.5]{GeV^{-1}} = \unit[0.1]{fm}$)}
\nonumber
\end{align}
\end{subequations}
It is the $g_2$ coefficient that corresponds to the function
$g_K(\Tsc{b};\bmax)$ that embodies the nonperturbative
behavior of the CSS evolution kernel $\tilde{K}$.  The fitted form of
the function is then
\begin{multline}
\label{eq:BLNY.fit}
  g_K(\Tsc{b};\bmax) = \frac{g_2}{2} \Tsc{b}^2
              = \frac{0.68\pORm{0.01}{0.02}}{2} \Tsc{b}^2.
\\
\mbox{(BLNY, $\bmax=\unit[0.5]{GeV^{-1}} = \unit[0.1]{fm}$)}
\end{multline}

Updated fits were made by Konychev and Nadolsky (KN)
\cite{Konychev:2005iy}.  They quantified the effect of changing
$\bmax$ on the fits.  At $\unit[0.5]{GeV^{-1}} =
\unit[0.1]{fm}$, their value for $g_K$ was compatible with
(\ref{eq:BLNY.fit}).  But when $\bmax$ was increased, not only
did the fit become somewhat better (with an optimum near $b_{\rm
  max}=\unit[1.5]{GeV^{-1}} = \unit[0.3]{fm}$), but the coefficient
$g_2$ was substantially smaller, giving:
\begin{multline}
\label{eq:KN.fit}
  g_K(\Tsc{b};\bmax) = \frac{0.184\pm{0.018}}{2} \Tsc{b}^2.
\\
\mbox{(KN, $\bmax=\unit[1.5]{GeV^{-1}} = \unit[0.3]{fm}$, $C_1=2e^{-\gamma_E}$)}
\end{multline}

\begin{figure*}
  \centering
  \begin{tabular}{c@{\hspace*{1cm}}c}
     \includegraphics[width=6cm]{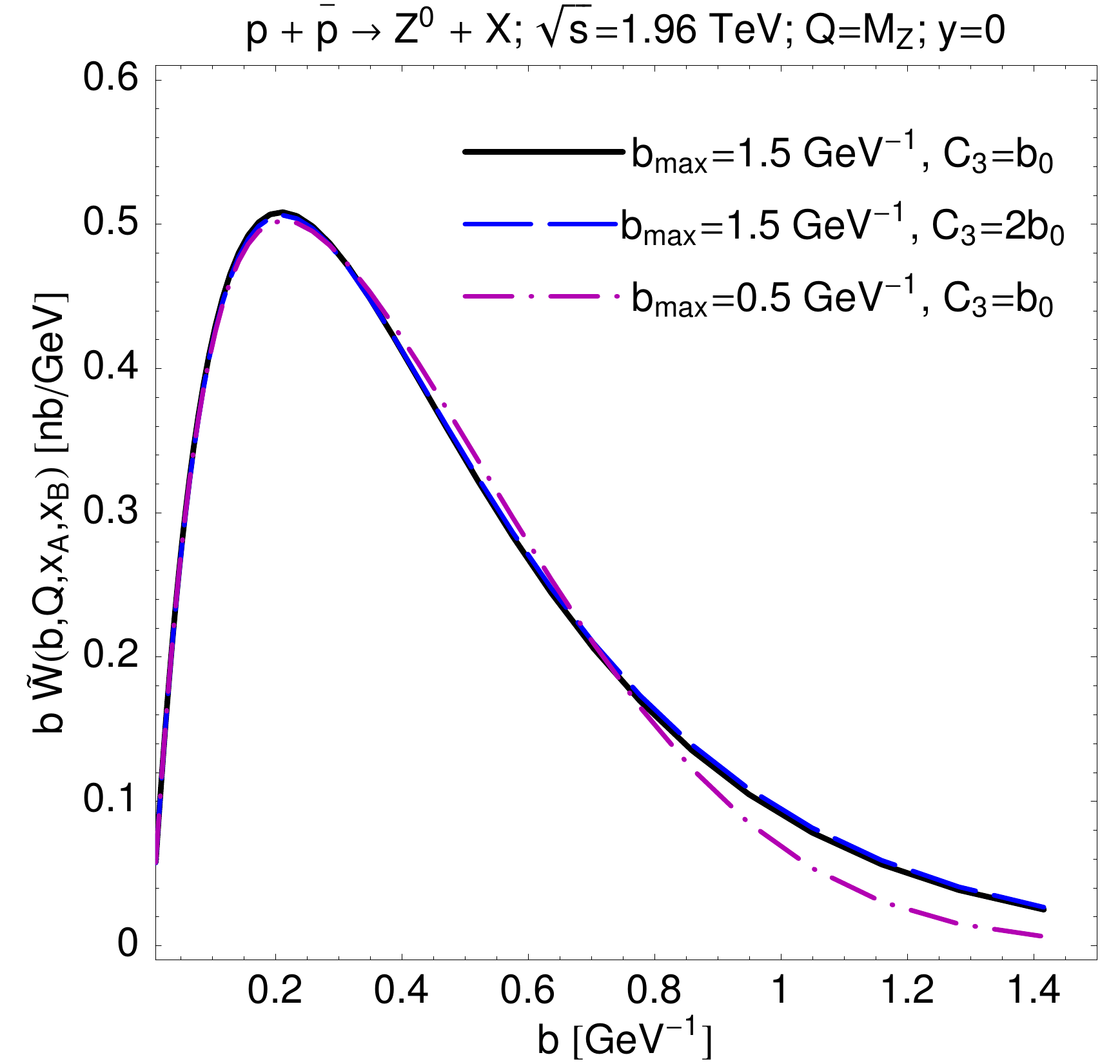}
  &
     \includegraphics[width=6cm]{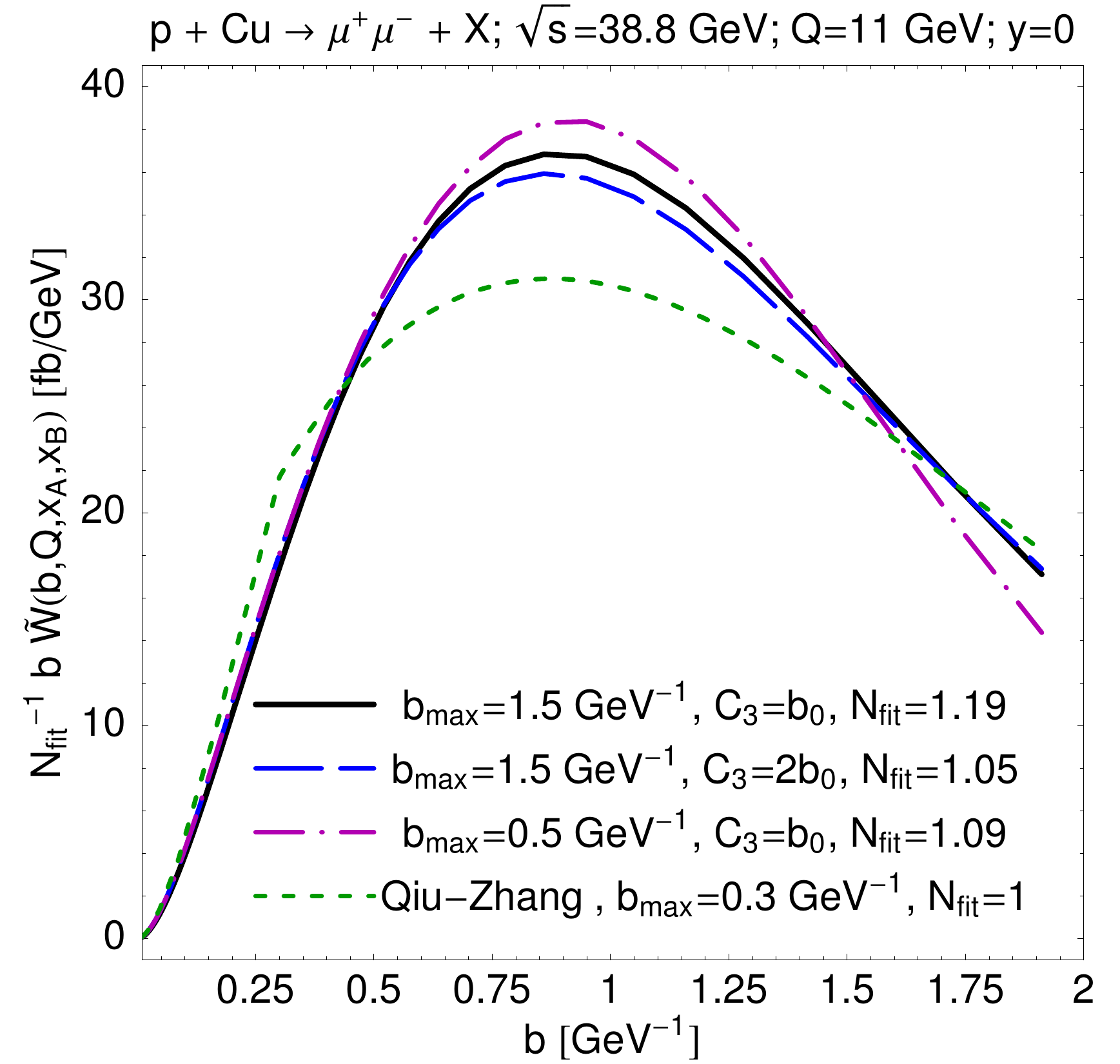}
  \\
     (a) & (b)
  \end{tabular}  
  \caption{$\Tsc{b}$-space integrand for the Drell-Yan process.  These
    plots are from Fig.\ 4 of \cite{Konychev:2005iy} and show the
    results of using different values of $\bmax$: (a) for $Z$
    production $\sqrt{s}=\unit[1.96]{TeV}$; (b) for
    $\sqrt{s}=\unit[38.8]{GeV}$ and $Q=\unit[11]{GeV}$. The two fits
    with $\bmax=\unit[1.5]{GeV^{-1}}$ correspond to two
    different choices for the ratio $C_3=\mubstar\bstarsc$, which
    gives a measure of sensitivity to truncation of perturbative
    expansions. The curve labeled ``Qiu-Zhang'', with $b_{\rm
      max}=\unit[0.3]{GeV^{-1}}$, uses the Qiu-Zhang \cite{Qiu:2000ga,
      Qiu:2000hf} parametrization.
    The normalization of $\tilde{W}$
    differs from that defined in Eq.\ \eqref{eq:Wj.def}.
    }
  \label{fig:KN.b.plots}
\end{figure*}

If one treated these fits as giving the true large-$\Tsc{b}$ behavior
of $g_K$, and hence of $\tilde{K}$, the two fits would be strongly
incompatible.  However, this is not a legitimate deduction, as can be
seen from Fig.\ \ref{fig:KN.b.plots} (which is Fig.\ 4 of Ref.\
\cite{Konychev:2005iy}).  There the integrand in TMD factorization is
plotted for several different fits, including the ones mentioned
above.  The integrand is the factor $\Tsc{b}\tilde{W}(\Tsc{b})$ in an
integral of the form
\begin{equation}
  \int_0^\infty \Tsc{b} \tilde{W}(\Tsc{b}) J_0(\Tsc{q}\Tsc{b})
  \diff{\Tsc{b}},
\end{equation}
which is obtained from the TMD factorization formula by performing the
integral over the azimuth of $\T{b}$ to obtain a Bessel function.

In Fig.\ \ref{fig:KN.b.plots}(a) is shown the situation at large $Q$,
specifically the integrand for production of the $Z$ boson.  At the
right-hand edge, at $\Tsc{b}=\unit[1.4]{GeV^{-1}}$, the 
difference between the fits for $g_K$, such as those in Eqs.\
(\ref{eq:BLNY.fit}) and 
(\ref{eq:KN.fit}), manifests itself as a difference between the curves
of more than a factor of two.
However although this is a large
\emph{relative} difference, it is in a
region where the integrand is small, so that the absolute difference is
quite small.  Thus the factor-of-two change in
the tail of the integrand only has a small effect on the cross
section, and does not greatly  
affect fitting within uncertainties at these large values of $Q$.  The
dominant values of $\Tsc{b}$ that contribute to the cross section are
in a perturbative region, around 0.1 to 0.5 $\unit{GeV^{-1}}$, i.e.,
0.02 to 0.1 $\unit{fm}$.  Here, the curves are very close to each
other.  This makes quantitatively manifest a phenomenon originally
found by Parisi and Petronzio \cite{Parisi:1979se}: When $Q$ is
sufficiently large, the whole of the transverse-momentum distribution,
even down to $\Tsc{q}=0$ is
determined by a range of $\Tsc{b}$ appropriate to perturbative
phenomena. That
is, the only important nonperturbative information is in the use of
ordinary \emph{integrated} parton densities, and no significant extra
nonperturbative information is needed for the $\Tsc{q}$ distribution.
Thus, the sensitivity to nonperturbative parameters for the
large $\Tsc{b}$ behavior becomes very weak, provided only that the
large $\Tsc{b}$ behavior is parametrized in a physically reasonable
way.  One can therefore assert that at large enough $Q$, the
nonperturbative information is only used qualitatively, to ensure
that the integrand is so small in the nonperturbative region that its
detailed behavior is unimportant.

However, if one examines lower energy data, the suppression of the
large $\Tsc{b}$ region is weaker, and correspondingly the relevant
values of $\Tsc{b}$ are larger. For example, Fig.\
\ref{fig:KN.b.plots}(b) shows that when $Q$ is around
$\unit[10]{GeV}$, the \emph{peak} of the integrand is near
$\Tsc{b}\simeq\unit[1]{GeV^{-1}} = \unit[0.2]{fm}$, closer to
nonperturbative phenomena.  There are noticeable differences between
the curves for different values of $\bmax$, but the relative
differences are much smaller than the factor of two differences seen
at the edge of graph (a).

Current discussions involve experiments and data at even lower $Q$,
while still having $Q$ large enough that it seems reasonable to use
TMD factorization.  The effects of evolution imply that these
experiments probe substantially larger values of $\Tsc{b}$ that are in
a more clearly nonperturbative region.  
In these contexts, one is often specifically interested in accessing 
the large $\Tsc{b}$ region in efforts to study nonperturbative properties of 
hadron structure.
However, there is a clear
danger that the fits resulting in Eqs.\ (\ref{eq:BLNY.fit}) and
(\ref{eq:KN.fit}) were primarily sensitive to smaller values of
$\Tsc{b}$ than are needed for the low $Q$ data.  
Hence the
extrapolation to larger $\Tsc{b}$ of the quadratic fit beyond the
region where it was determined by the fitted data could well
misrepresent the true consequences of QCD. 

In fact a difficulty immediately manifests itself, as particularly
emphasized by Sun and Yuan \cite{Sun:2013hua}:
There are regions of $Q$, $x_A$ and
$x_B$ where one might reasonably expect
factorization to continue to work, but where the coefficient of $\Tsc{b}^2$ in
the exponent in Eq.\ (\ref{eq:BLNY.form}), with the BLNY values, reverses sign to become
positive, given the values in Eq.\ (\ref{eq:BLNY_g}).  An example is
with $Q=\unit[3]{GeV}$ and $x_A=x_B=0.3$. With this reversed sign, the
integral over $\Tsc{b}$ for the cross section is badly divergent at
$\Tsc{b}=\infty$.  Even the Konychev-Nadolsky fit with its lower value
of $g_2$ does not \cite{Sun:2013hua} adequately account for the data
at lower $Q$ than where the fit was made.  (See also Ref.\
\cite{Aidala:2014hva} for a confirming analysis.)

This suggests the following
hypothesis: The fits in Refs.\ \cite{Landry:2002ix, Konychev:2005iy}
determine $g_K$ and $g_{j/H}$ in a range of relatively low $\Tsc{b}$,
say between about 0.5 and $\unit[2]{GeV^{-1}}$, 
but their extrapolation
to larger $\Tsc{b}$ is wrong.  The function $g_K$ that concerns the
$Q$ dependence should not continue to rise like $\Tsc{b}^2$; instead
it should flatten.  Then the evolution at low $Q$ is substantially
slower than the extrapolation of the old fits would give.  This is
without necessarily invalidating the earlier fits.

\subsubsection{Work without nonperturbative evolution function}

Several references, e.g., \cite{Echevarria:2012pw, Becher:2010tm,
  Sun:2013dya, Sun:2013hua, Boer:2008fr}, use a CSS-like formalism,
but without a fitted function to parametrize nonperturbative
large-$\Tsc{b}$ evolution.  (However, all agree in using fitted
nonperturbative function(s) for the $Q$-independent part of the TMD
parton densities.)

Echevarr\'\i a et al.\
\cite{Echevarria:2012pw} argued that one can avoid the
use of $\bmax$ and the corresponding parametrized function
$g_K$.  At the most fundamental level, they have TMD factorization and
evolution results that are equivalent
\cite{Collins:2012uy, Collins:2012ss} to a version of the CSS
formalism.  So the differences really only concern how the formalism
is exploited.  They apply a resummation procedure to the function
$\tilde{K}(\Tsc{b})$ (which is $-2 D$ in their notation).  They argue
that the resummed formula applies in a range of $\Tsc{b}$ up to about
a limiting value they term $b_C$.  The value of this limit is given
as $b_C=\unit[6]{GeV^{-1}} = \unit[1.2]{fm}$, 
when the scale $\mu$ is $\unit[5]{GeV}$.

Beyond this distance scale, they agree that nonperturbative
information in $\tilde{K}$ is needed.  However, they argue that since
the coordinate-space parton density is already small at the limiting
value $\Tsc{b}=b_C$, the nonperturbative information in
$\tilde{K}$ is irrelevant.  If this argument is really valid, it
increases the predictive power of TMD factorization. 

Unfortunately, this argument is not consistent with known
nonperturbative scales.  First, the distance scale quoted,
$\unit[1.2]{fm}$, is clearly in a region where nonperturbative
physics is important.  For example it is larger than the confinement
scale found in \cite{Schweitzer:2012hh} and a factor of 4 larger than
the chiral scale in the same reference.  Moreover, to obtain a small
parton density at this value of $\Tsc{b}$, Echevarr\'\i a et al.\ use
a Gaussian ansatz for the large-$\Tsc{b}$ behavior:
\begin{equation}
  e^{- \Tsc{b}^2 \langle \Tsc{p}^2 \rangle /4},
\end{equation}
where the value of $\langle \Tsc{p}^2 \rangle = \unit[0.38]{GeV^2}$ is
taken from a fit in Ref.\ \cite{Schweitzer:2010tt}.  The Gaussian
factor is evidently describing nonperturbative effects, which are not
in any of the Feynman graphs used, even with resummation.  The
distance scale associated with this factor is
\begin{equation}
\label{eq:non-pert.b}
  \frac{2}{ \sqrt{ \langle \Tsc{p}^2 \rangle } }
  = \unit[3.2]{GeV^{-1}}
  = \unit[0.65]{fm}.
\end{equation}
Quite reasonably this is roughly midway between the chiral and
confinement scales determined from very different theoretical
considerations in Ref.\ \cite{Schweitzer:2012hh}.

It follows then that Green functions already have substantial
nonperturbative contributions when transverse 
distances reach the
value in (\ref{eq:non-pert.b}).  Hence, a perturbative calculation,
even a resummed calculation, cannot be expected to be accurate at or beyond
this scale.  Wherever in $\Tsc{b}$ nonperturbative contributions are
important in a TMD parton density $\tilde{f}_{j/H}(x,\Tsc{b})$, one
should also expect them to be important for the evolution kernel
$\tilde{K}$, and one has not at all evaded the need to use a
nonperturbative contribution to it, either extracted by fitting to
data or by nonperturbative theoretical methods in QCD theory (or,
better, both).
In Sec.\ \ref{sec:EISS}, we will give a further analysis of the
argument given in Ref.\ \cite{Echevarria:2012pw}.

Given the above quantitative estimates of the onset of
nonperturbative physics, the previously used value of
$\bmax=\unit[1.5]{GeV^{-1}} = \unit[0.3]{fm}$ is reasonable.  A
substantially smaller value is excessively conservative, while
increasing it by more than about a factor of two goes too far into the
nonperturbative region.  

We conclude that Echevarr\'\i a et al.\ \cite{Echevarria:2012pw}
use perturbatively based calculations for $\tilde{K}$ (admittedly with
resummation) in a region where nonperturbative effects are important,
and that in other parts of their calculation, nonperturbative effects
are important in the same region.  

Becher and Neubert \cite{Becher:2010tm} also use a related formalism
without including nonperturbative effects at large $\Tsc{b}$.  But
they only claim that their formalism is valid when transverse momentum
is much larger than the QCD scale, i.e., $\Tsc{q} \gg \Lambda_{\rm
  QCD}$.  In that situation, the Fourier transform
probes variations of the integrand on a scale $1/\Tsc{q}$ and it
is dominated by
relatively small distances of order $1/\Tsc{q} \ll 1/\Lambda_{\rm
  QCD}$, and thus the nonperturbative contributions can be
numerically unimportant.

In contrast, the full TMD factorization method described above is
valid 
for transverse momenta down to zero, and the nonperturbative
large-$\Tsc{b}$ region is important, if $Q$ is not too large.  The
fact that the nonperturbative region is important is established by
the essentially universal use of a Gaussian form for TMD parton
densities at large $\Tsc{b}$ in fitting data.

Further work without a nonperturbative function for evolution is by
Sun and Yuan \cite{Sun:2013dya,Sun:2013hua}.
They use a certain
perturbative approximation to the exponent $S$ of the full evolution
factor, with $S$ being defined as in Eq.\ (\ref{eq:S.def}).
The approximation is essentially the same as
one used by Boer in \cite[Eq.\ (144)]{Boer:2008fr} and \cite[Eq.\
(34)]{Boer:2013zca}.  
We will refer to this as the Boer-Sun-Yuan (BSY) approximation.
The approximation was devised with the aim of expressing the TMD
factorization formula in terms of TMD densities,
$\tilde{f}_{j/H}\big( x_A, \T{b}; x_A^2Q_0^2,\mu_0 \bigr)$, with a
fixed scale.
(More recent work by Boer in Refs.~\cite{Boer:2013pya,Boer:2014tka} and by 
Sun-Yuan in Ref.~\cite{Su:2014wpa} no longer uses the BSY method.)

First, given the legitimate choices that $\mu_0=Q_0$ and $\mu_Q=Q$, the full
exponent (\ref{eq:S.def}) can be written as
\begin{widetext}
\begin{equation}
  \label{eq:S.variant}
  S(\Tsc{b},Q,Q_0,Q_0)
  =
  \int_{Q_0}^{Q}  \frac{ \diff{\mu'} }{ \mu' }
         \biggl[ 
                - 2 \tilde{K}(\Tsc{b};\alpha_s(Q_0),Q_0) 
                - 2 \gamma_j(\alpha_s(\mu'); 1) 
                + \ln\frac{Q^2}{ (\mu')^2 } \gamma_K(\alpha_s(\mu'))
         \biggr].
\end{equation}
Here, we have made explicit the coupling argument $\alpha_s(Q_0)$ of the
function $\tilde{K}$, since its scale does not match the scale in the
coupling elsewhere in the integrand.
This equation is equivalent to the exponent in \cite[Eq.\ (86)]{Ji:2004wu}.  
The approximations made in
Refs.\ \cite{Boer:2008fr, Sun:2013dya, Sun:2013hua} are to replace all
quantities 
in Eq.\ (\ref{eq:S.variant}) by their one-loop approximations and to
replace the coupling $\alpha_s(Q_0)$ in $\tilde{K}$ by $\alpha_s(\mu')$, to give
\begin{equation}
  \label{eq:S.B09}
  S_{\rm BSY}(\Tsc{b},Q,Q_0,Q_0)
  \equiv
  2C_F \int_{Q_0}^{Q}  \frac{ \diff{\mu'} }{ \mu' }
        \, \frac{\alpha_s(\mu')}{\pi}
        \biggl[
            \ln\frac{Q^2}{(\mu')^2}
            + \ln\frac{\Tsc{b}^2Q_0^2}{4} + 2\gamma_E 
            - \frac{3}{2}
        \biggr].
\end{equation}
\end{widetext}
This approximation is questionable for three reasons.  
First, the
actual value of the coupling $\alpha_s(Q_0)$ in $\tilde{K}$ has been replaced
by $\alpha_s(\mu')$.  Second, the perturbation series for $\tilde{K}$ has
been truncated at one loop, despite the fact that higher orders will
have higher powers of $\ln(\Tsc{b}Q_0)$, so it is not obvious that the higher order
corrections are small compared with lowest order.  
The exponent is used in an integral over all $\Tsc{b}$, so the integral
includes regions of $\Tsc{b}$ where the logarithms are arbitrarily
large.  Finally, we recall that in analyzing HERMES and COMPASS data,
important regions of $\Tsc{b}$ are in the domain of nonperturbative
phenomena, and therefore where a perturbative approximation is
generally invalid.

Observe that the $\Tsc{b}$ dependence in Eq.\ (\ref{eq:S.B09}) results
in a power-law $\Tsc{b}$ dependence in $e^{-S}$ proportional to
\begin{equation}
  \label{eq:B.09.b.factor}
  \left( \frac{1}{\Tsc{b}^2} \right)^{a(Q,Q_0)},
\end{equation}
where
\begin{equation}
  a(Q,Q_0)
  = 2C_F \int_{Q_0}^{Q}  \frac{ \diff{\mu'} }{ \mu' }
        \, \frac{\alpha_s(\mu')}{\pi}
\end{equation}
We get a divergence at $\Tsc{b}=0$ that is worse than what is present
in a correct formula, and we get a suppression at large $\Tsc{b}$.

Of course, the in-principle inadequacy of the approximation does not
in itself show that the approximation is actually numerically inadequate in its
application to data over a limited range of $Q$.  To see its effects
in practice, we need to evaluate the size of the error, which we will
do later.

As already mentioned, the reasoning leading to the BSY approximation
was aimed at expressing the TMD factorization formula in terms of
TMD densities at a fixed scale.  This is better achieved by our Eq.\
(\ref{eq:soln.1a}), where a RG transformation has been applied to
$\tilde{K}$ to remove large logarithms.  However, that formula also
shows that the need to parameterize the nonperturbative contribution to
evolution has not at all been evaded.

\subsubsection{Comparisons with single-spin asymmetry data at fairly low $Q$}

In contrast to the established good fits for Drell-Yan at relatively
large $Q$, the situation at the lower values of $Q$ for the HERMES and
COMPASS data is rather confusing, especially for the spin-dependent
TMD functions.

Aybat, Prokudin and Rogers \cite{Aybat:2011ta} (APR) used the results
of the Landry et al.\ \cite{Landry:2002ix} fit given in Eq.\
(\ref{eq:BLNY.fit}) for evolution, and used a fit by Anselmino et al.\
\cite{Anselmino:2011gs} for the 
Sivers function 
at HERMES at $\langle
Q^2\rangle_{\rm HERMES} \simeq \unit[2.4]{GeV^2}$.  (Average values of kinematic variables 
were used in the calculations.) The result seemed to be
a successful prediction/postdiction of COMPASS data at $\langle Q^2\rangle_{\rm COMPASS} \simeq \unit[3.8]{GeV^2}$. However, the Sivers function 
extracted from COMPASS data was sensitive to significantly smaller $x$ values than 
the HERMES data.
Hence, while the apparent consistency with the BLNY $g_K(\Tsc{b};\bmax)$ seen by APR is compelling, 
the relatively large variations between HERMES and COMPASS data 
were most likely due to a combination of variations in both $x$-dependence and evolution in $Q$, rather 
than a test of $Q$ evolution alone. 
See, for example, the discussion 
of this on page six of Ref.~\cite{Bradamante:2011xu} where it is explained how 
restricting to smaller $y$ likely enhances the influence of very small $Q$ and 
larger $x$, leading to a larger Sivers effect and greater direct agreement with HERMES data without explicit evolution.

Anselmino et al.\ \cite{Anselmino:2012aa} made a comprehensive
analysis of both the HERMES and COMPASS data.  They found improved
fits with TMD evolution than without, but the effect does not seem as
dramatic as in the results of APR,
even though they used same BLNY form (\ref{eq:BLNY.fit}) for TMD
evolution. 

A related group of authors \cite{Anselmino:2013vqa} fit the
transversity and Collins functions to data from HERMES, COMPASS and
Belle, covering a substantial range of $Q$, but did not need to use
TMD evolution to get a good fit.

Finally, Sun and Yuan \cite{Sun:2013dya, Sun:2013hua} were able to get
agreement 
with much data with their approximation (\ref{eq:S.B09}) for the
evolution exponent.  This exponent gives much less rapid evolution
than is implied from the Landry et al.\ \cite{Landry:2002ix} fit.  It
should be noted that the \emph{lowest} $Q$ used by Landry et al.\ is
about the \emph{highest} $Q$ considered by Sun and Yuan.  This is the
clue that we will exploit to motivate the idea that
compatibility between the results might be obtained once one allows
for the fact that they probe different ranges of $\Tsc{b}$.

\subsection{Comparison using master function $A$}
\label{sec:compL}

We now use $A(\Tsc{b})$ to analyze some of the phenomenological
approaches.

\subsubsection{BLNY and KN}

\begin{figure*}
\centering
  \begin{tabular}{c@{\hspace*{10mm}}c}
    \includegraphics[angle=90,scale=0.5]{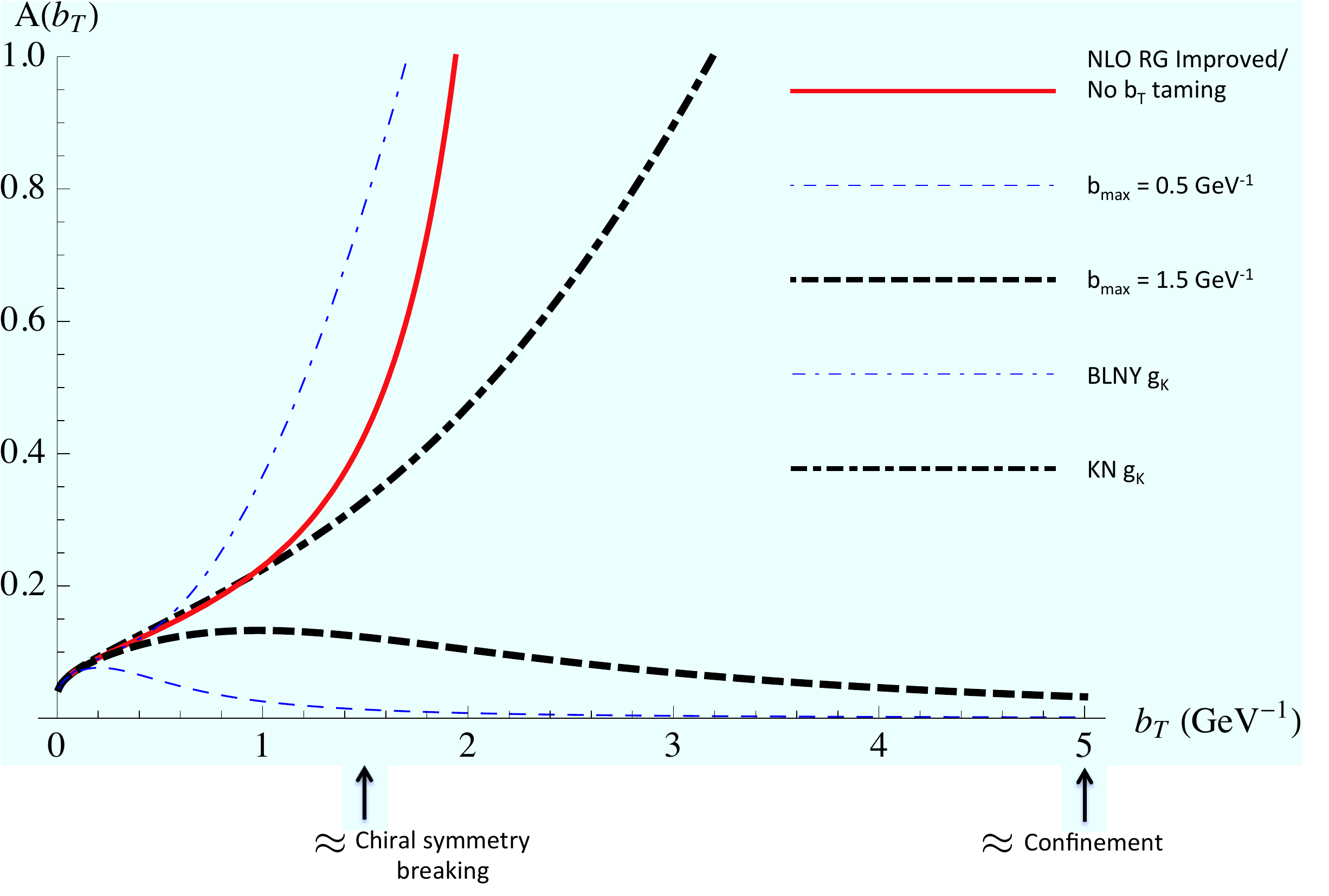} \\
    (a) \\
    \vspace{10mm} \\
    \includegraphics[angle=90,scale=0.5]{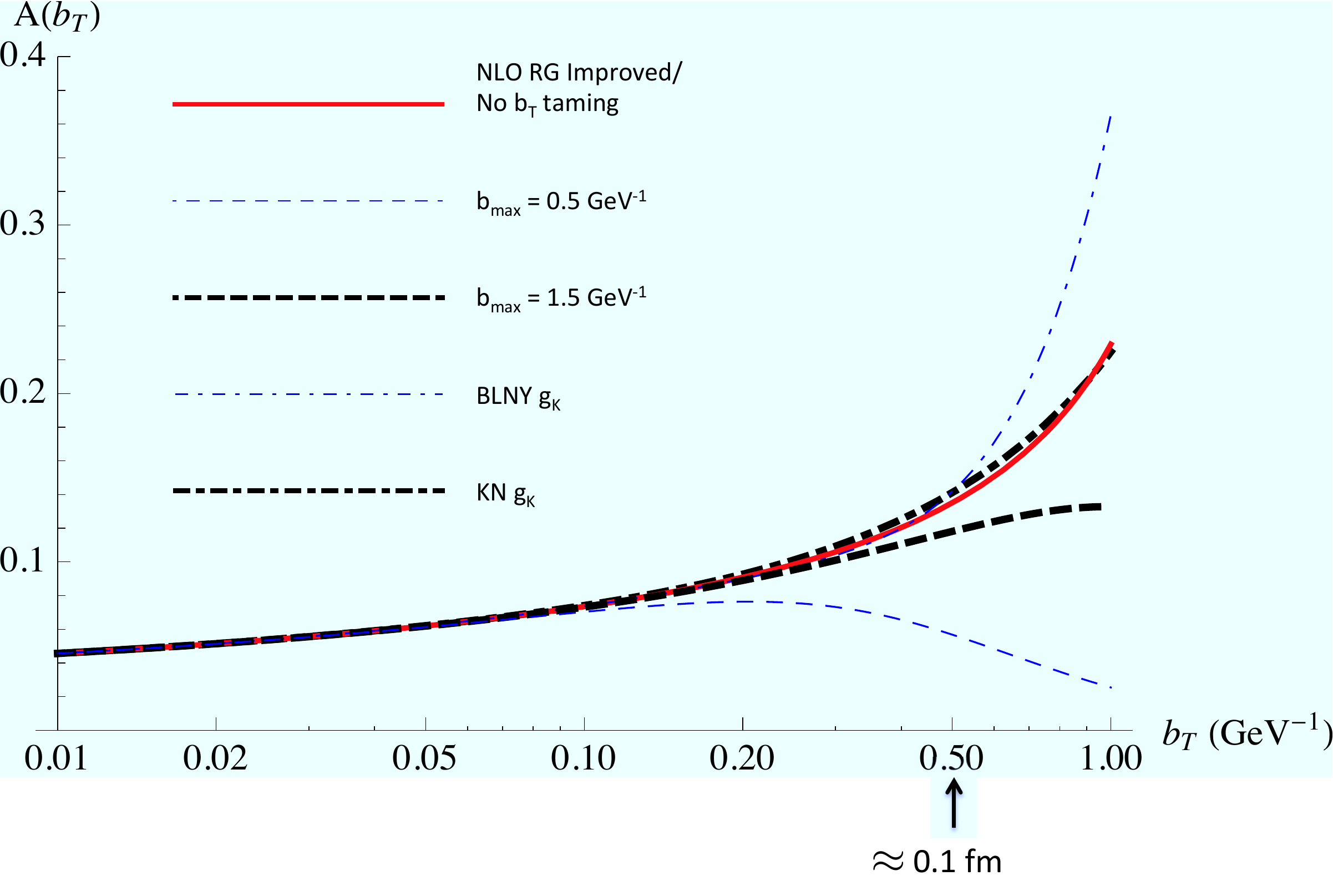}
    \\
    (b)
    \\[5mm]
   \end{tabular}
\caption{The $A(\Tsc{b})$ function defined in Eq.~\eqref{eq:A.def} for
  the BLNY and KN fits, together with some of its components.  
  The thick, solid, red curve shows a purely perturbative, but
  RG-improved, calculation, using the LO and NLO terms in Eq.\
  (\ref{eq:A.pure.PT}).
  The thin, blue, dashed curve shows the same quantity with the CS
  taming procedure, i.e., the second line in \eqref{eq:A.BLNY}, with
  the value $\bmax = 0.5$~GeV$^{-1}$ for the BLNY fit. 
  The thin, blue, dot-dashed curve shows the result after also adding
  in the $g_K(\Tsc{b};\bmax = 0.5 \, {\rm GeV}^{-1})$ function from BLNY\
  \cite{Landry:2002ix}, i.e., it is the actual $A(\Tsc{b})$ for the BLNY
  fit.
  The thick, black, dashed and dot-dashed curves similarly show the
  cut-off perturbative component and the full $A$ for the KN fit
  \cite{Konychev:2005iy} with $\bmax = \unit[1.5]{GeV^{-1}}$.
\\
  In (a), we plot the results with a linear scale out to 
  $\unit[5]{GeV^{-1}} = \unit[1]{fm}$. 
  In (b), we show the same functions, but with a logarithmic
  horizontal scale extending to $\unit[1]{GeV^{-1}} = \unit[0.2]{fm}$,
  to show better the region of small $\Tsc{b}$, to which large $Q$
  interactions are most sensitive.
}
\label{fig:quadplots}
\end{figure*}

First, in Fig.\ \ref{fig:quadplots} we show comparisons between the KN
and BLNY fits, described in Sec.\ \ref{sec:BLNY.KN}, and a purely
perturbative calculation.  The two sets of plots in the figure
differ only in their horizontal scales for $\Tsc{b}$.  Graph
(a) has a linear scale out to
$\Tsc{b}=\unit[5]{GeV^{-1}} = \unit[1]{fm}$, and thus emphasizes the
nonperturbative region. Graph (b) has a logarithmic scale, and goes
only up to $\Tsc{b}=\unit[1]{GeV^{-1}} = \unit[0.2]{fm}$; it thus shows a
primarily perturbative region for $\Tsc{b}$.  In graph (a), we have
also indicated some significant scales for nonperturbative physics,
from Ref.\ \cite{Schweitzer:2012hh}.

A baseline for all the comparisons is the perturbative value in Eq.\
(\ref{eq:A.pure.PT}).  With the two-loop approximation and the
indicated choice of scale for the strong coupling, we will call this
the ``purely perturbative, but RG-improved calculation'' of $A$.  It
is plotted in red in the figures.  We calculate the strong coupling
with three active flavors, the standard 2-loop formula in terms of
$\Lambda_3=\unit[339]{MeV}$, with the value of $\Lambda_3$ being from a
recent summary of world data by Bethke \cite{Bethke:2012jm}.

At low $\Tsc{b}$, the coupling is small, and the QCD prediction for
$A$ is accurate up to higher-order corrections.  When $\Tsc{b}$ is
decreased to zero, $A(\Tsc{b})$ slowly decreases to zero, like $1/\ln
(1/\Lambda\Tsc{b})$.  But when $\Tsc{b}$ is increased sufficiently, a
fixed-order approximation diverges at the Landau pole, as illustrated
by the red line in Fig.\ \ref{fig:quadplots}(a); a perturbative
approximation is evidently not trustworthy there.

The BLNY and KN fits use the CSS $\bstarsc$ prescription and a
quadratic form for $g_K(\Tsc{b};\bmax)$.  From 
Eq.~\eqref{eq:A.bstar} we get 
\begin{align}
\label{eq:A.BLNY}
A(\Tsc{b})_{\text{BLNY, KN}} = {}&  \frac{g_2}{2}\Tsc{b}^2
\nonumber \\ &\hspace*{-1cm}
  + \frac{1}{1 + \Tsc{b}^2/\bmax^2} 
    (\mbox{Two-loop $A(\bstarsc)$ from \eqref{eq:A.pure.PT}})
\nonumber \\ &\hspace*{-1cm}
  + O\xleft( \alpha_s(\mubstar)^3 \right).
\end{align}

For the BLNY fit, the blue dashed curves in each of the plots in
Fig.\ \ref{fig:quadplots} give the perturbative part of $A$ , i.e.,
the second line of Eq.\ \eqref{eq:A.BLNY}.  When $\Tsc{b}$ is well
below the cutoff $b_{\rm max, BLNY}=\unit[0.5]{GeV^{-1}}$, this
perturbative part matches the purely perturbative value in red.  But
at larger $\Tsc{b}$ it decreases to zero; the corresponding
contribution to $\tilde{K}$ goes to a constant.

When the fitted quadratic term for $g_K$ is added --- see Eq.\
\eqref{eq:BLNY.fit} for the coefficient --- we get the blue
dash-dotted curve.  This starts by compensating the effect of the
$\bmax$ cut-off on the perturbative term, and then rapidly gets large.

For the KN fit, with $\bmax=\unit[1.5]{GeV^{-1}}$, the corresponding
results are shown by the black curves.  The black dashed curve is the
cut-off perturbative part, and the black dot-dashed curve is the
result of adding in the fitted function for $g_K$, Eq.\
\eqref{eq:KN.fit}.

Observe first that the full BLNY curve matches the purely perturbative
term very closely up to $\Tsc{b}=\unit[0.5]{GeV^{-1}}$, even though
the perturbative term in Eq.\ \eqref{eq:A.BLNY} has already been
substantially reduced by the $\bstarsc$ cut off.  The difference
between the full BLNY value for $A$ and the cut-off perturbative term
is the $g_K$ function, which BLNY fitted to data.  Thus one can
reasonably say that BLNY verifies a perturbative prediction.

The KN fit matches the purely perturbative calculation slightly less
well at $\unit[0.5]{GeV^{-1}}$.  But then it matches its trend
substantially further out than does the BLNY fit, to a bit beyond
$\unit[1.5]{GeV^{-1}}$, admittedly in a region where the accurate
applicability of perturbation theory might be debatable.  Beyond that
it falls below the perturbative curve.  KN's fit has a notably better
fit to the data, as measured by $\chi^2$, so we should regard the KN
fit as more correctly corresponding to reality.

Recall from Fig.\ \ref{fig:KN.b.plots} that the $\Tsc{b}$-space
integrand decreases rapidly beyond around $\unit[1.5]{GeV^{-1}}$, for
the lower energy data, and that the decrease starts even earlier for
collider data.  Thus, although there are large differences at large
$\Tsc{b}$ between the $A$ functions in the two fits, the data used are
not sensitive to the higher values of $\Tsc{b}$, say beyond
$\unit[2]{GeV^{-1}}$; this insensitivity is even more marked at higher
energy, as is seen in Fig.\ \ref{fig:KN.b.plots}(a).  

For a more incisive measurement at large $\Tsc{b}$ one must use data
at lower $Q$, such as is provided by HERMES and COMPASS.

\subsubsection{Boer-Sun-Yuan}

We now examine the consequences of the Boer-Sun-Yuan (BSY)
approximation \cite{Boer:2008fr, Sun:2013dya, Sun:2013hua}, which gave
Eq.~\eqref{eq:S.B09} for the exponent in the $\Tsc{b}$ integrand.  We
apply the definition of $A$, the first line of Eq.~\eqref{eq:A.def},
to the value of $\tilde{W}$ that corresponds to Eq.~\eqref{eq:S.B09}.
This gives
\begin{equation}
A(\Tsc{b};Q)_{\rm BSY} \approx \frac{ \alpha_s(Q)C_F }{ \pi } \, .
\label{eq:appL}
\end{equation}
Now a highly nontrivial prediction of TMD factorization in QCD is
that $A(\Tsc{b})$ is independent of other kinematic variables, notably
$Q$, when TMD factorization is applied to leading power in
$\Tsc{q}/Q$.  So the $Q$-dependence in \eqref{eq:appL} is already in
contradiction with the prediction.  

\begin{figure*}
\centering
  \begin{tabular}{c@{\hspace*{10mm}}c}
    \includegraphics[angle=90,scale=0.5]{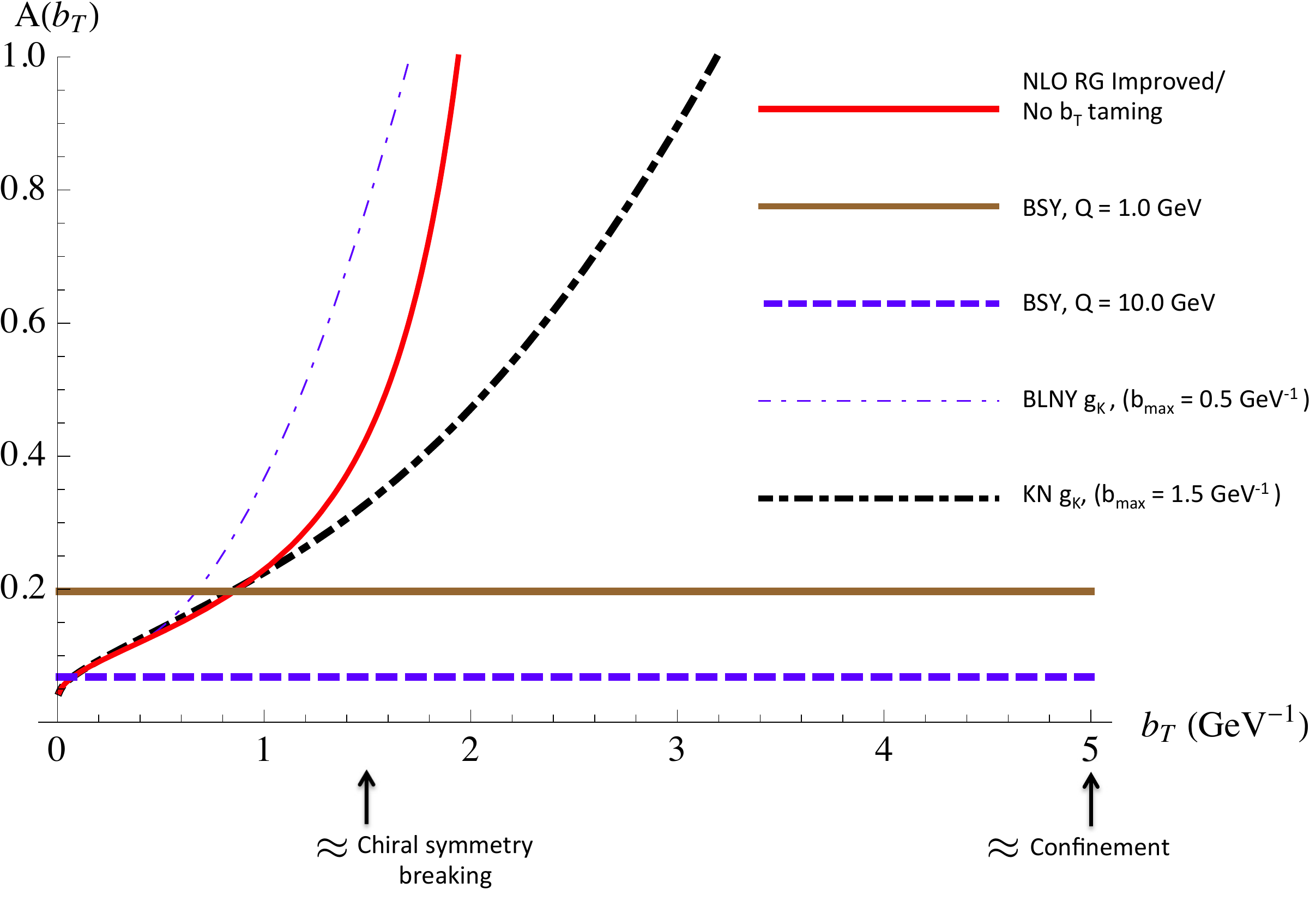} \\
    (a) \\
    \vspace{20mm} \\
    \includegraphics[angle=90,scale=0.5]{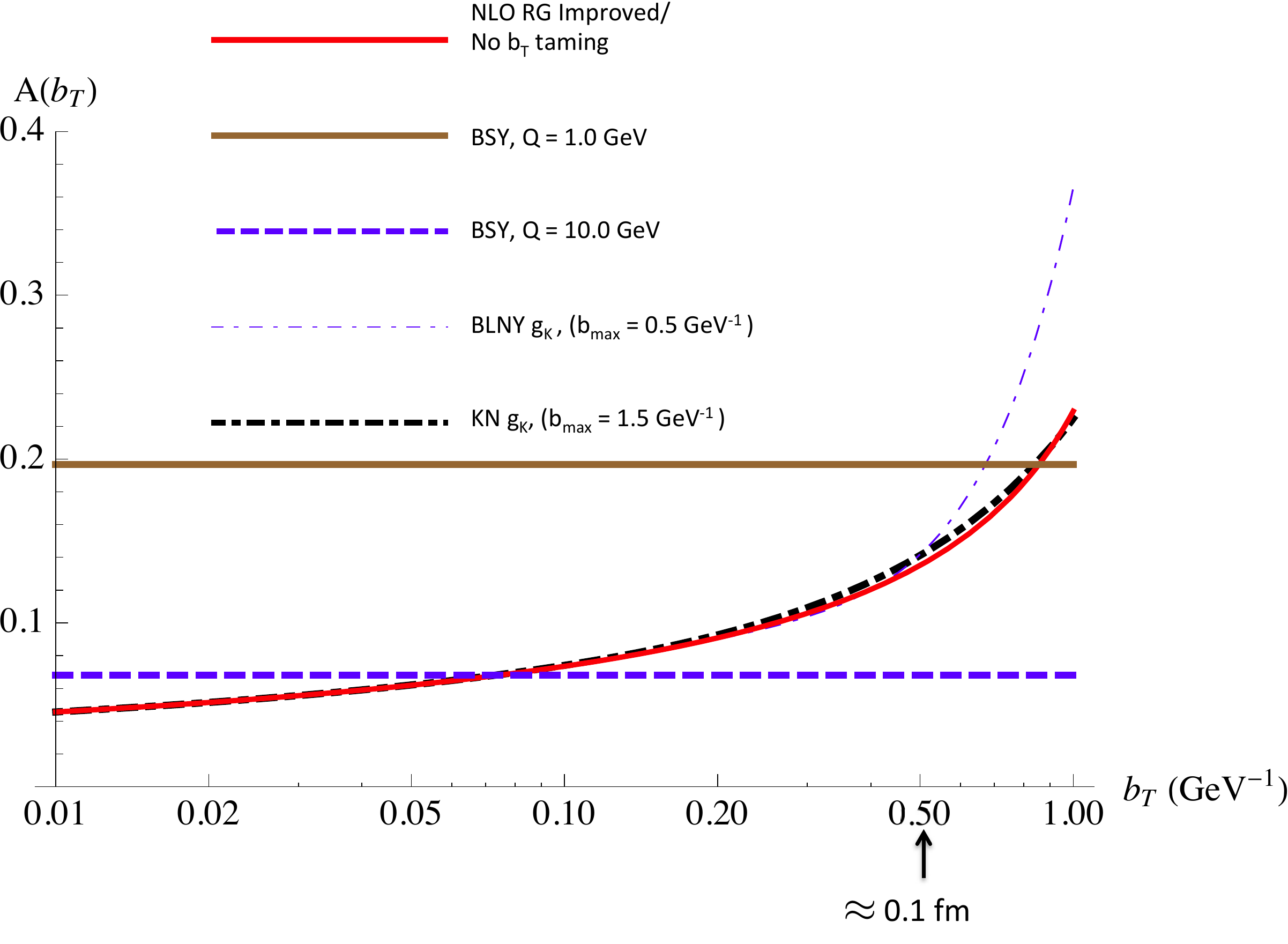}
    \\
    (b)
    \\[5mm]
   \end{tabular}
\caption{Like Fig.\ \ref{fig:quadplots}, but showing a comparison
  between the Boer-Sun-Yuan (BSY) approximation, the KN fit, and the 
  purely perturbative, RG-improved, calculation.  
  The thick, red, solid curve is the NLO perturbative calculation. The thick, black, dot-dashed, and thin, blue dot-dashed curves
  are the same KN and BLNY curves as in Fig.\
  \ref{fig:quadplots}.  The solid brown curve is for $A$ at $Q =
  1.0$~GeV in the BSY \cite{Sun:2013hua} method, from Eq.\
  \eqref{eq:appL}. The dashed blue curve is the BSY calculation for $Q =
  10.0$~GeV. 
    In (a), we plot the results with a linear scale out to 
  $\unit[5]{GeV^{-1}} = \unit[1]{fm}$. 
  In (b), we show the same functions, but with a logarithmic
  horizontal scale extending to $\unit[1]{GeV^{-1}} = \unit[0.2]{fm}$,
  to show better the region of small $\Tsc{b}$, to which large $Q$
  interactions are most sensitive.
}
\label{fig:syplots}
\end{figure*}

To show some of its implications, we have plotted in Fig.\
\ref{fig:syplots} values for the BSY $A$ at $Q=\unit[1]{GeV}$ (roughly
appropriate for HERMES) and at $Q=\unit[10]{GeV}$ (roughly appropriate
for the lower energy Drell-Yan data used in the KN and BLNY fits).
These are respectively the brown solid and blue dashed curves.  In
addition, we show the purely perturbative term (red solid line) and
the full KN and BLNY fitted values (black and blue dot-dashed lines). 

To assess the effect of $A$ on the cross sections under different
kinematic conditions, we need to know the typical values of $\Tsc{b}$
that are important.  We measure these by the point at which the
integrand $\Tsc{b}\tilde{W}(\Tsc{b})$ falls to half its peak value.
(We note that at asymptotically large $Q$, the peak occurs at around
$\Tsc{b} \sim 1/Q$, but the width falls as a slower power of $Q$
\cite{Parisi:1979se}; this is quite different than for a Gaussian.)
At the $Z$, the left-hand graph in Fig.\ \ref{fig:KN.b.plots} gives
a width of about $\unit[0.5]{GeV^{-1}}$.  For $Q$ around
$\unit[11]{GeV}$, the right-hand plot gives a width of about
$\unit[1.8]{GeV^{-1}}$.

In contrast, transverse momentum distributions at HERMES and COMPASS
are fit reasonably well by Gaussians:
\begin{equation}
  e^{-\Tsc{p}^2/\langle \Tsc{p}^2 \rangle },
\end{equation}
with the mean-square transverse momentum $\langle \Tsc{p}^2 \rangle$
being around $\unit[0.2]{GeV^2}$ to $\unit[0.3]{GeV^2}$.  Fourier
transformed, this gives 
\begin{equation}
  e^{-\Tsc{b}^2 \langle \Tsc{p}^2 \rangle /4 },
\end{equation}
which gives a width in position space of very roughly 
$2 / \sqrt{\langle \Tsc{p}^2 \rangle}$, perhaps
$\unit[4]{GeV^{-1}}$.

In Fig.\ \ref{fig:syplots}, the brown curve, for $Q=\unit[1]{GeV}$,
which is appropriate for low energy data, intersects the KN and
perturbative curves at $\Tsc{b} \simeq \unit[1]{GeV^{-1}}$, and falls
below them at higher $\Tsc{b}$.  Thus, the BSY approximation gives
slower
evolution than KN at the lowest values of $Q$.

When the energy is increased, the BSY form of $A$ becomes the blue
dashed curve.  This matches the QCD-predicted form at around $\Tsc{b}
\simeq \unit[0.1]{GeV^{-1}}$, and is well below the QCD perturbative
prediction in a region of $\Tsc{b}$ that is relevant for the
successful KN and BLNY fits, where the fits themselves confirmed the
accuracy of the perturbative predictions for $\Tsc{b} \lesssim
\unit[1]{GeV^{-1}}$.  As Sun and Yuan themselves acknowledge, their
approximation is not adequate for the higher energy data.  The best
that their approximation can manage is that the fall of the resulting
function $A$ with $Q$ roughly matches the values of the true $A$ at
the values of $\Tsc{b}$ probed at that $Q$; these distances decrease
as $Q$ increases.

From the success of the BLNY and KN fits, together with the knowledge
that perturbative QCD predictions are valid in the perturbative region
of $\Tsc{b}$, we should regard the values of $A(\Tsc{b})$ given by the
KN fit as reliable in the region $\Tsc{b} \lesssim
\unit[2]{GeV^{-1}}$, roughly.  Larger values of $\Tsc{b}$ are
unimportant for the Drell-Yan data fit by KN.  But it is larger values
of $\Tsc{b}$ that dominate for the conditions of the HERMES and
COMPASS experiments.  Therefore we should expect that the true $A$
turns down above about $\unit[2]{GeV^{-1}}$.  This would also match
the conclusions of Aidala et al.\ \cite{Aidala:2014hva}.  Most
importantly, we should be able to achieve this while preserving the
goodness of fit for the high-energy Drell-Yan data, not to mention
predictions for cross sections at high-energy hadron-hadron
colliders. 

In the future, it may become practical to analyze data directly in 
coordinate space, as proposed in Ref.~\cite{Boer:2011xd}. This would 
circumvent the complications associated with Fourier transforms or 
convolution integrals.

\subsubsection{Qiu-Zhang}

\begin{figure*}
\centering
    \includegraphics[angle=90,scale=0.5]{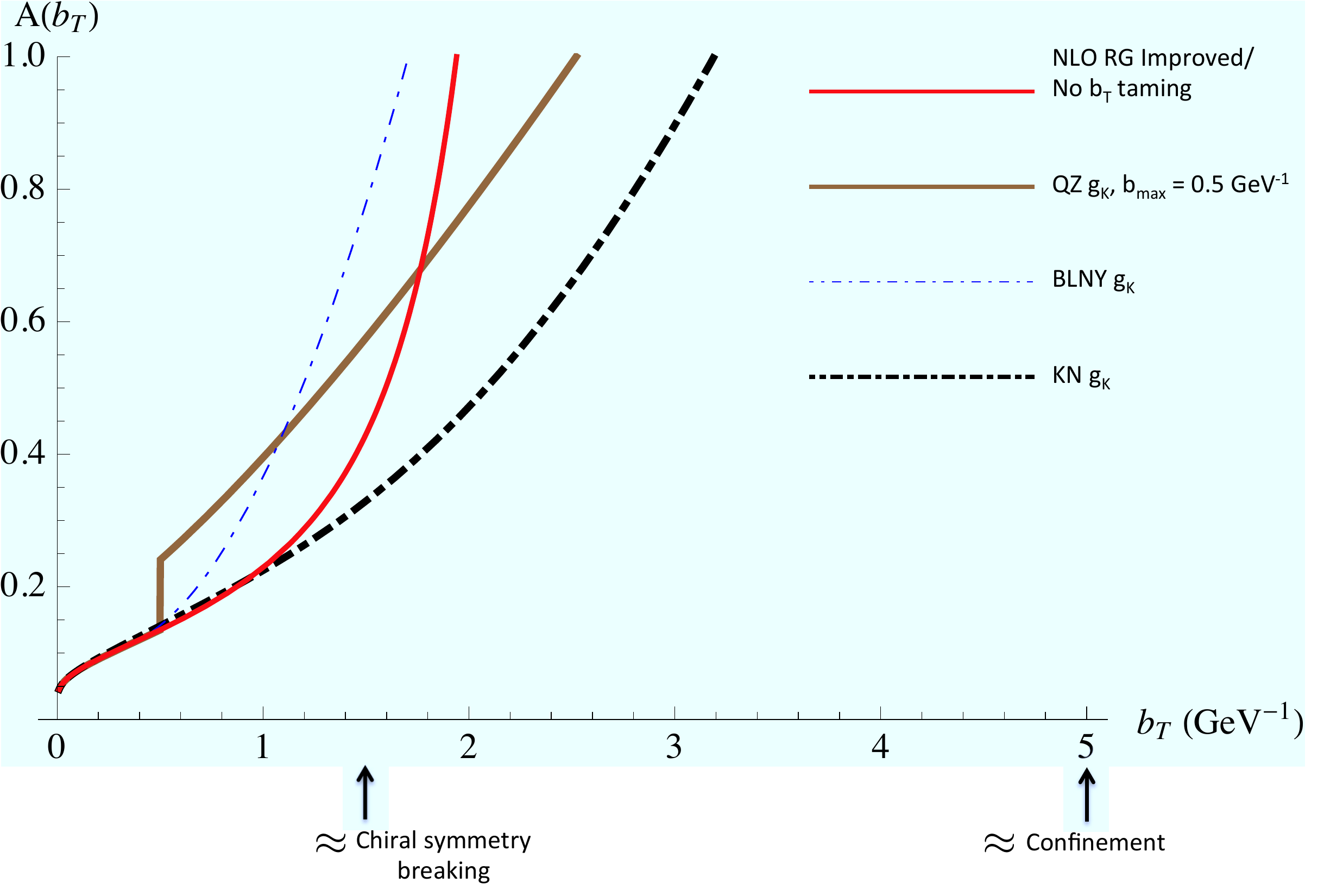}
\caption{Like Fig.\ \ref{fig:quadplots}(a), but now we show the
  comparison between $A(\Tsc{b})$ for the Qiu-Zhang model of
  Refs.~\cite{Qiu:2000ga,Qiu:2000hf} (brown line with step), the
  purely perturbative RG-improved calculation (red), and the full $A$
  functions from the BLNY fit (thin blue dot-dashed) and the KN fit
  (thick black dot-dashed).
  Since by construction, the Qiu-Zhang method agrees with the
  RG-improved perturbative calculation below the $\bmax$ cut off, we
  do not bother with the logarithmic plot for low $\Tsc{b}$.
}
\label{fig:QZplots}
\end{figure*}

In Fig.\ \ref{fig:QZplots}, we show comparisons for the result for
$A(\Tsc{b})$ in the Qiu-Zhang fit in
Refs.~\cite{Qiu:2000ga,Qiu:2000hf}.  The Qiu-Zhang result is the brown
line with a step at $\bmax=\unit[0.5]{GeV^{-1}}$.  They use a sharp
cut off at $\Tsc{b}=\bmax$ instead of CSS's smooth cut off.  Below
$\bmax$, they use exactly the perturbatively calculated formula.
Above $\bmax$, they freeze $\Tsc{b}$ at $\bmax$, and multiply the
$\Tsc{b}$-space integrand by a parametrized form corresponding to TMD
factorization and evolution:
\begin{widetext}
\begin{equation}
  \tilde{W}(\Tsc{b}, \dots)
  =
  \begin{cases}
    \tilde{W}^{(\text{pert})}(\Tsc{b}, \dots)
    & \mbox{if $\Tsc{b} < \bmax$},
  \\
    \tilde{W}^{(\text{pert})}(\bmax, \dots)
    F^{\text{NP}}_{\text{QZ}}(\text{b}, \dots)
    &\mbox{if $\Tsc{b} > \bmax$},
  \end{cases}
\end{equation}
where
\begin{equation}
  F^{\text{NP}}_{\text{QZ}}(\Tsc{b}, \dots)
=
  \exp\xleft\{ 
       -\ln\left(\frac{Q^2 \bmax^2}{c^2}\right) 
       \Bigl[
          g_1 \left( \Tsc{b}^{2\alpha} - \bmax^{2\alpha} \right)
          +g_2 \left( \Tsc{b}^2 - \bmax^2\right)
       \Bigr] 
       -\bar{g}_2 \left( \Tsc{b}^2 - \bmax^2\right) 
  \right\}\, ,
\end{equation}
and $\alpha$, $g_1$, $g_2$, and $\bar{g}_2$ are parameters to be fit
to data.  
\end{widetext}

The perturbative formula is used for small
$\Tsc{b}$.  By construction, $\tilde{W}$ is continuous at
$\Tsc{b}=\bmax$. The result seen in Fig.~\ref{fig:QZplots} is that 
$A(\Tsc{b})$ contains a discontinuity at $\Tsc{b} = \bmax$.

\subsection{Echevarr\'\i a-Idilbi-Sch\"afer-Scimemi}
\label{sec:EISS}

These authors (EISS) \cite{Echevarria:2012pw} argue that the practical
effect of the nonperturbative part of $\tilde{K}$ is small.
This conclusion is obtained from an argument
that the range of $\Tsc{b}$ in which perturbative
calculations are valid is a factor of two or more larger that has
normally been considered appropriate.  
Although underlying their derivations are
TMD factorization and evolution formulae that are exactly
equivalent to those we presented in Sec.\ \ref{sec:CSS.eqs},
their method of solution is very different.

In our solutions (following CSS), we used the RG equation
(\ref{eq:RG.K}) to evolve $\tilde{K}$ to a scale $\mu$ of order
$1/\Tsc{b}$, in order to avoid large logarithms of $\mu\Tsc{b}$ in the
perturbative expansion of $\tilde{K}$ in powers of $\alpha_s(\mu)$.
Expressed in terms of the quantity
$D^R(\Tsc{b};\mu)=-\frac12\tilde{K}(\Tsc{b};\mu)$ used by EISS, this
gives
\begin{equation}
  \label{eq:D.RG.soln}
  D^R(\Tsc{b};\mu) 
  = D^R(\Tsc{b};C_1/\Tsc{b}) 
    + \frac12 \int_{C_1/\Tsc{b}}^\mu \frac{\diff{\mu'}}{\mu'}
      \gamma_K(\alpha_s(\mu')).
\end{equation}

For a numerical estimate of $D^R(\Tsc{b};\mu)$, our method is to use
the right-hand-side of Eq.\ (\ref{eq:D.RG.soln}) with truncated
perturbative calculations for $D^R(\Tsc{b};C_1/\Tsc{b})$ and for
$\gamma_K(\alpha_s(\mu'))$, and with a truncated perturbation
expansion of the $\beta$ function that controls the running of the
coupling by
\begin{equation}
  \label{eq:beta}
  \frac{ \diff{ \alpha_s(\mu)/4\pi } }{ \diff{\ln{\mu}} }
  = \beta(\alpha_s(\mu))
  = -2 \sum_{n=0}^\infty 
    \beta_n \left( \frac{\alpha_s(\mu)}{4\pi} \right)^{n+2}.
\end{equation}
Errors in the resulting approximation can be estimated from the
truncation errors in the three perturbative expansions used.  When
$\mu$ is fixed in a perturbative region and $\Tsc{b}$ is increased
beyond $1/\mu$, the largest of the errors is controlled by the size of
$\alpha_s(C_1/\Tsc{b})$.  When $\Tsc{b}$ is large enough, the
inapplicability of perturbation theory is signaled by a Landau
singularity in the value used for $\alpha_s(C_1/\Tsc{b})$, and a
corresponding singularity in the \emph{approximated}
$D^R(\Tsc{b};\mu)$.
Strictly speaking, a full argument for the inapplicability of
perturbation theory at large distances arises not only from the large size
of the effective coupling, by itself, but also from the knowledge that
perturbation series in quantum field theory are often asymptotic
series rather than actually convergent series.  This results in an
ultimate limit to the accuracy of perturbative approximations, no
matter how many terms in the series are used.

Instead, EISS use an explicit resummation of logarithms for the
perturbative expansion of $D^R(\Tsc{b};\mu)$, i.e., for the
left-hand-side of (\ref{eq:D.RG.soln}).  The logarithms are powers of
$L_\perp(\mu \Tsc{b})=\ln(\mu^2\Tsc{b}^2/4e^{-2\gamma_E})$.  In massless
perturbation theory for $D^R$ the coefficient of $(\alpha_s(\mu))^n$
is a polynomial of degree $n$ in $L_\perp$:
\begin{equation}
  D^R(\Tsc{b};\mu)
  = \sum_{n=1}^\infty \sum_{p=0}^n D_{np} a^n L_\perp^p,
\end{equation}
where $a(\mu) =\alpha_s(\mu)/4\pi$.  The leading-logarithm (LL) resummation
is the sum of the terms with the maximal number of logarithms, the
next-to-leading-logarithm (NLL) term is the sum of the terms with one
fewer logarithm, etc.  To implement resummation, EISS define
$X(\mu \Tsc{b},\Tsc{b})=a(\mu) \beta_0L_\perp(\mu \Tsc{b})$, and then reorganize the perturbation series as
\begin{equation}
\label{ref:D.resum}
  D^R(\Tsc{b};\mu) 
  = \sum_{n=0}^\infty a^n D_n(X),
\end{equation}
where
\begin{equation}
  D_n(X) = \sum_{j=j_{\rm min}(n)}^\infty \frac{D_{j+n,j}}{\beta_0^j} X^j.
\end{equation}
Here $j_{\rm min}(0)=1$, and $j_{\rm min}(n)=0$ for $n\geq1$. The term
with $D_n$ sums the $n$th-to-leading logarithms.  The LL approximation
is $D_0(X)$, the NLL approximation is $D_0(X)+ aD_1(X)$, etc.  EISS
provide formulae for resummations at the LL, NLL, and
next-to-next-to-leading-logarithm (NNLL) approximations.  They give a
derivation in their App.\ B from the RG equation for $D^R$, and its
solution (\ref{eq:D.RG.soln}).

If one-loop approximations are
used for $\gamma_K$, for the $\beta$-function that controls the
running of the $\alpha_s$, and for $D^R(\Tsc{b};C_1/\Tsc{b})$ with
$C_1=2e^{-\gamma_E}$, then the right-hand-side of (\ref{eq:D.RG.soln})
gives the same results as the LL resummation.  

But beyond that level, there are generally differences between using
the resummation formula (\ref{ref:D.resum}) truncated to some order
in $a$, and directly using the right-hand-side of the solution
(\ref{eq:D.RG.soln}) with corresponding truncations in the quantities
that appear in it.

Ultimately what matters for making predictions for a cross sections is
the size (or expected size) of the errors in whatever approximation is
used to estimate $D^R(\Tsc{b};\mu)$.

A symptom of one source of error
is visible in EISS's Fig.\ 1 in Ref.\ \cite{Echevarria:2012pw}.  Here
are shown plots of the results of their resummations, for
$\mu=\unit[\sqrt{2.4}]{GeV}$ and for $\mu=\unit[5]{GeV}$.  (In the
plot, the symbol $Q_i$ is used instead of our $\mu$.)
If the exact $D^R(\Tsc{b};\mu)$ were plotted as a function of
$\Tsc{b}$ for two different values of $\mu$, then the two curves would
differ only by a simple vertical shift.  This follows directly from
the right-hand-side of (\ref{eq:D.RG.soln}), which implies that
\begin{equation}
\label{eq:D.mu1.mu2}
  D^R(\Tsc{b};\mu_2) - D^R(\Tsc{b};\mu_1)
  = \frac12 \int_{\mu_1}^{\mu_2} \frac{\diff{\mu'}}{\mu'}
      \gamma_K(\alpha_s(\mu')).
\end{equation}
These results follow from the RG equation for $D^R$:
\begin{equation}
  \label{eq:D.RGE}
  \frac{ \diff{D^R(\Tsc{b};\mu)} }{ \diff{\ln{\mu}} }
  = \frac12 \gamma_K(\alpha_s(\mu)).
\end{equation}

In going from $\mu=\unit[\sqrt{2.4}]{GeV}$ to $\mu=\unit[5]{GeV}$
the shift is upward, since $\gamma_K$ is dominated by its positive LO
term.  We now examine \cite[Fig.\ 1]{Echevarria:2012pw}.  At small
$\Tsc{b}$, below about $\unit[2]{GeV^{-1}}$ or even
$\unit[3]{GeV^{-1}}$, the upward shift is easily seen.  But at larger
$\Tsc{b}$, the plots disagree with the prediction: The calculated
$D^R$ decreases substantially instead of increasing as $\mu$
increases.

Of course, an approximate calculation of $D^R$ need not exactly obey a
property known to hold for the exact quantity.  But the deviations
from the predicted $\Tsc{b}$-independent shift give a lower bound on
the error in the approximation.  From the plots, we deduce that the
calculation is no longer trustworthy as $\Tsc{b}$ approaches about
$\unit[3]{GeV^{-1}}$.  This therefore falsifies EISS's assertion that
their resummation is valid up to around $b_c=\unit[6]{GeV^{-1}}$ when
$\mu=\unit[5]{GeV}$.

That the NLL and higher approximations do not give a simple vertical
shift can be seen by differentiating $D_0+aD_1$ with respect to $\mu$
(or $Q_i$).  From the formulae given by EISS, it is found that the
value of the derivative depends on $L_\perp$ and hence on $\Tsc{b}$.
The result is of order $a^2$ (or higher) times a function of $X$.
Thus it is of a form that would be compensated by including higher
order terms in the expansion in nonleading logarithms.

We next obtain the expected size of the truncation errors for
(\ref{ref:D.resum}) both from the explicit formulae given by EISS for
$D_0$, $D_1$ and $D_2$, and from their derivation in App.~B of
Ref.~\cite{Echevarria:2012pw}.  The properties we will list below
can be seen explicitly in EISS's formulae up to $n=2$.  They can be
proved to hold for all $n$, which is not too hard to do starting from
the derivation in their App.\ B.

Each higher term has one factor more of $a(\mu)$, i.e., of $\alpha_s(\mu)$.
But it is not this factor alone that is the true expansion parameter
relevant for the size of the higher terms.  Each term can be written
as a polynomial in $1/(1-X(\Tsc{b} \mu,\mu))$ and $\ln(1-X(\Tsc{b} \mu,\mu))$.  When $\Tsc{b}\mu$ gets
large enough, $X(\Tsc{b} \mu,\mu)$ approaches 1, so that the terms exhibit a Landau
singularity.  The leading singularity for the $a(\mu)^n$ term is
proportional to $\left[\dfrac{\ln(1-X(\Tsc{b} \mu,\mu))}{1-X(\Tsc{b} \mu,\mu)}\right]^n$, when $n\geq1$.
That is, each extra power of $a(\mu)$ is accompanied by one extra power of
$\dfrac{\ln(1-X(\Tsc{b} \mu,\mu))}{1-X(\Tsc{b} \mu,\mu)}$. Hence the effective expansion parameter, when
$X(\Tsc{b} \mu,\mu)$ gets near to 1, is not $a(\mu)$, but $\dfrac{a(\mu)\ln(1-X(\Tsc{b} \mu,\mu))}{1-X(\Tsc{b} \mu,\mu)}$.  This is
roughly the effective coupling (divided by $4\pi$) at a scale
proportional to $1/\Tsc{b}$, i.e., $a(C_1/\Tsc{b})$.  [Note that if the 1-loop approximation
is used for $\beta$, then the corresponding approximation to the
effective coupling is $\alpha_s(C_1/\Tsc{b}) = \alpha_s(\mu)/(1-X(\Tsc{b} \mu,\mu))$.]

We therefore deduce that the series in leading and nonleading
logarithms becomes inapplicable when $a/(1-X(\Tsc{b} \mu,\mu))$ is of order unity,
i.e., when $1-X(\Tsc{b} \mu,\mu)$ is of order $a(\mu)$.  This implies that the other terms in the
polynomial in $1/(1-X(\Tsc{b} \mu,\mu))$ and $\ln(1-X(\Tsc{b} \mu,\mu))$ are a smaller relative correction since they
are smaller in size than the term with the leading singularity.  

The result is that the error in a truncation of the expansion in
leading and nonleading logarithms is governed by the same parameter
$a(C_1/\Tsc{b})$ that determines the sizes of truncation errors in the
direct use of the solution of the evolution equation, i.e., the
truncation errors for the right-hand side of Eq.\
(\ref{eq:D.RG.soln}).

So nothing in terms of accuracy has been gained by a use of literal
resummation instead of the use of the right-hand-side of
(\ref{eq:D.RG.soln}) with truncations in perturbative expansions of
quantities that have no logarithms.  What is lost is relative
simplicity in the formulae and a transparency in the nature of the
errors. 

Another important issue is EISS's treatment of heavy quarks.  Their
resummation is obtained from calculations in Ref.\ \cite{Moch:2005id}
of massless Feynman graphs. Now, we know that propagators of heavy
particles decay exponentially at large distances, like $e^{-m\Tsc{b}}$
(aside from a power law that is irrelevant here).  The distances we
are concerned with are for potential nonperturbative contributions to
$\tilde{K}$ or $D^R$, i.e., for $\Tsc{b}$ above about
$\unit[1]{GeV^{-1}}$.  In this region of $\Tsc{b}$, the exponential
$e^{-m\Tsc{b}}$ is very important for both the charm and bottom
quarks.  For an approximation using massless quarks to be suitable,
this region of $\Tsc{b}$ is such that one must decouple the heavy
quarks first.  That is, the appropriate number of active quark flavors
is 3.

\begin{figure}
  \centering
  \includegraphics[scale=0.6]{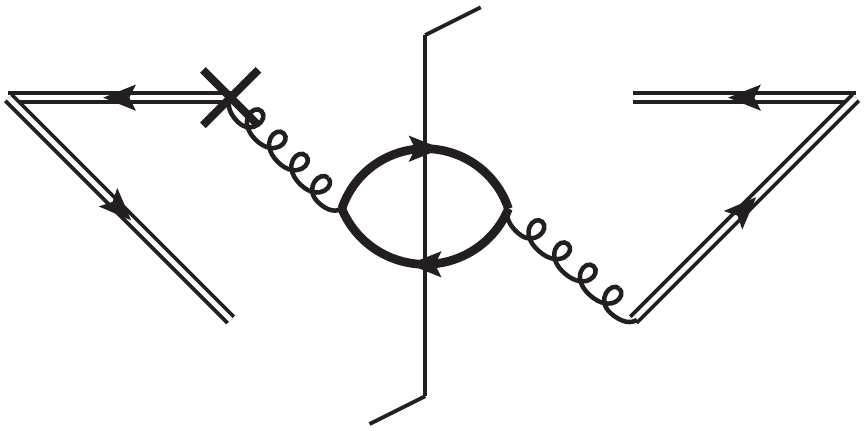}
  \caption{Typical two-loop graph with a heavy quark loop that
    contributes to $\tilde{K}$.  The notation is as in Ref.\
    \cite{Collins:2011qcdbook}: The cross denotes a derivative of a
    space-like Wilson line with respect to its rapidity.} 
  \label{fig:K.heavy.quark}
\end{figure}

Furthermore, when $m_q\Tsc{b}$ is not small for some quark of mass
$m_q$, the dependence on $\Tsc{b}$ is nontrivial; one cannot expect a
simple logarithmic dependence.  An example of a two-loop graph with a
heavy quark that contributes to $\tilde{K}$ is shown in Fig.\
\ref{fig:K.heavy.quark}.  This and similar graphs have a heavy quark
loop correcting a gluon propagator in any of the Wilson-line matrix
elements that occur in the definition of $\tilde{K}$.

EISS first estimate the location of the Landau singularity by setting
the number of active flavors to 5, and using the value\footnote{Note
  that the value of $\Lambda_5$ used by EISS is somewhat 
  smaller than the measured value \cite{Bethke:2012jm}.}
$\Lambda_5=\unit[157]{MeV}$ for the scale parameterizing the running
coupling.  They calculate a corresponding distance scale
$b_{\Lambda_{\rm QCD}}\simeq \unit[7]{GeV^{-1}}$; this calculation
corresponds to $b_{\Lambda_{\rm QCD}} = 2e^{-\gamma_E}/\Lambda_5$.

But, as we observed, the appropriate number of active flavors is 3.
Then one should use \cite{Bethke:2012jm} $\Lambda_3=\unit[339]{GeV}$,
more than a factor of two larger.  This brings the corresponding
distance scale down to $b_{\Lambda_{\rm QCD}}\simeq \unit[3]{GeV^{-1}} =
\unit[0.6]{fm}$, which is much more in line with standard ideas of
distance scales where nonperturbative physics is important.  

Having obtained a form for $D^R$ (or $\tilde{K}$) at some low scale
$\mu_i$, one naturally wishes to evolve it to a much larger value of
$\mu$, whenever processes with large $Q$ are to be treated.  As is
usual with similar matrix elements (e.g., densities of light quarks),
one should then evolve it by 
Eq.\ (\ref{eq:RG.K}), used in regions
with different numbers of active quark flavors, together with the
relevant matching conditions.  This evolution preserves the property
that the difference (\ref{eq:D.mu1.mu2}) between $D^R$ at different
scales is independent of $\Tsc{b}$.  (As mentioned earlier, a full
treatment of the effects of heavy quarks on TMD factorization remains
to be worked out; we hope to deal with this in a future article.)

In any case, it is entirely incorrect to use a massless approximation
in perturbative calculations of $D^R$ when $m_q\Tsc{b}$ is not small.
So performing a resummation with 4 or 5 active flavors, as EISS do, is
incorrect when $\Tsc{b}$ is in the region of $\unit[1]{GeV^{-1}}$
upwards. 

We have one final concern with the EISS approach.  This is that it
implements the transition from the perturbative part of $D^R$ to the
nonperturbative part through a sharp cutoff at a certain position
$\Tsc{b}=b_c$. For $\Tsc{b}$ below $b_c$, EISS
calculate $D^R$ from resummed perturbation theory alone, and use a
parametrized form only for larger $\Tsc{b}$.  However, as explained in
the next section, such a sharp boundary between perturbative and
nonperturbative regions does not correspond to what happens in actual
quantum field theory.  CSS's motivation for a smooth cutoff remains
valid.  (Of course, there is the possibility that RG-improved or
resummed perturbation theory remains accurate at larger $\Tsc{b}$ than
has been assumed earlier.)

Hence, we conclude that, in contrast with the implication of the EISS results,
one does indeed need to allow for a nonperturbative parameterization in the
phenomenologically important range of $\Tsc{b}$ beginning above roughly 1 or 2
$\unit{GeV^{-1}}$.  A separate question, to be discussed below, is
whether changes are needed compared with previously accepted
parameterizations. 

We note that the more recent phenomenological work of Ref.~\cite{DAlesio:2014vja}, 
which utilizes the EISS method, does in fact consider a nonperturbative part for $D^R$.
However, the problems discussed above remain in their calculations.

\section{Small-$\Tsc{b}$ behavior of $g_K$}
\label{sec:small-b}

The function $g_K(\Tsc{b};\bmax)$ was defined in Eq.~\eqref{eq:gK.def}
as (the negative of) the difference between the exact $\tilde{K}$ and
a cut-off version restricted to a roughly perturbative region of
$\Tsc{b}$.  It is power suppressed when $\Tsc{b}$ is much less than
$\bmax$.  CSS deliberately chose a smooth cutoff, because a normal
calculation of $\tilde{K}$ uses RG-improved perturbation
theory with truncations of the RG functions $\beta(\alpha_s)$ and
$\gamma_K(\alpha_s)$ (or some resummation with similar accuracy).  The
accuracy of such a calculation gradually degrades as $\Tsc{b}$
increases.  nonperturbative phenomena are
not restricted to a sharp range of $\Tsc{b}$ but contribute at all
$\Tsc{b}$; but they are expected to be power-suppressed at small
$\Tsc{b}$.  

Hence, when Eq.\ \eqref{eq:soln.2} is actually  used to fit data, there is a
subtle shift in the meaning of $g_K(\Tsc{b};\bmax)$.  Instead of $g_K$
defined in Eq.~\eqref{eq:gK.def}, the fits should be regarded as
actually measuring the following quantity\footnote{It is not entirely
  clear which value $\mu_0$ of the renormalization scale should be used here;
  but that only affects an additive constant, with no dependence on
  $\Tsc{b}$.}
\begin{equation}
  \label{eq:gKhat.def}
  \hat{g}_K(\Tsc{b};\bmax) 
  = -\tilde{K}(\Tsc{b},\mu_0)_{\rm exact} 
    + \tilde{K}(\bstarsc,\mu_0)_{\rm approx.}.
\end{equation}
Here the perturbatively approximated $\tilde{K}$ is defined as
follows.  From the RG equation for $\tilde{K}$ we find
\begin{widetext}
\begin{equation}
\label{eq:K.RG.improved}
   \tilde{K}(\Tsc{b};\mu_0;\alpha_s(\mu_0)) 
  = \tilde{K}(\Tsc{b};\mubstar;\alpha_s(\mubstar)) 
    - \int_{\mubstar}^{\mu_0} \frac{ d \mu' }{ \mu' }
       \gamma_K(\alpha_s(\mu')),
\end{equation}
where the renormalization scale in $\tilde{K}$ on the right-hand side
is chosen to avoid large logarithms in its perturbation series.  Then
we define $\tilde{K}(\bstarsc,\mu_0)_{\rm approx}$ by applying to the
right-hand side of \eqref{eq:K.RG.improved} truncated perturbation
theory for $\tilde{K}$, $\gamma_K$, and for the function $\beta$
controlling evolution of the coupling.  (Note that this form of
approximation preserves the property of the exact
$\tilde{K}(\Tsc{b};\mu_0)$ that its dependence on $\Tsc{b}$ and
$\mu_0$ is the sum of a function of $\Tsc{b}$ and a function of
$\mu_0$.)
Thus a fit of $g_K$ can allow both for nonperturbative phenomena and
for uncalculated higher-order terms in the perturbative part of
$\tilde{K}$.

To better understand the changed $g_K$, we write it in two forms:
\begin{subequations}
\label{eq:gKhat.versions}
\begin{align}
\label{eq:gKhat.version.1}
  \hat{g}_K(\Tsc{b};\bmax) 
  ={}&
    \left[ 
       -\tilde{K}(\Tsc{b},\mu_0)_{\rm exact} 
       + \tilde{K}(\bstarsc,\mu_0)_{\rm exact}
    \right]
  +
    \left[ 
       -\tilde{K}(\bstarsc,\mu_0)_{\rm exact} 
       + \tilde{K}(\bstarsc,\mu_0)_{\rm approx.}
    \right]
\\
\label{eq:gKhat.version.2}
  ={}&
    \left[ 
       -\tilde{K}(\Tsc{b},\mu_0)_{\rm exact} 
       + \tilde{K}(\Tsc{b},\mu_0)_{\rm approx.}
    \right]
  +
    \left[ 
       -\tilde{K}(\Tsc{b},\mu_0)_{\rm approx.} 
       + \tilde{K}(\bstarsc,\mu_0)_{\rm approx.}
    \right].
\end{align}  
\end{subequations}
In (\ref{eq:gKhat.version.1}) the quantity in the first brackets is
$g_K$ as 
originally defined, and the second brackets contain the error in using
truncated perturbative methods to compute $\tilde{K}(\bstarsc)$.
In (\ref{eq:gKhat.version.2})
the first brackets show the difference between the
exact $\tilde{K}$ and a truncated perturbatively-based estimate, while
the second brackets give a perturbative approximation to $g_K$.

If $\bmax$ is chosen conservatively (as in the BLNY fits), then
perturbatively based calculations of $\tilde{K}$ are applicable for
the whole region of $\Tsc{b}$ less than $\bmax$, and even at somewhat
larger $\Tsc{b}$.  Actual fits for $g_K$, or rather $\hat{g}_K$,
particularly with a simple quadratic approximation, are a compromise,
between reproducing $g_K$ in a region where it is predicted, and
fitting $g_K$ at larger $\Tsc{b}$ where it is less perturbative.  Even
so, we expect to estimate, roughly, the small-$\Tsc{b}$ behavior of
$g_K$ from perturbative calculations.  We can regard such an estimate
as giving a property of the first term on the right-hand side of
\eqref{eq:gKhat.version.1}, i.e., of $g_K$ itself.  Alternatively it
gives a property of the second term on the right-hand side of
\eqref{eq:gKhat.version.2}.  The validity of perturbation theory when
$\Tsc{b}$ is small is coded in a small value for the other term on
each line, which is a difference between the exact $\tilde{K}$ and its
perturbative estimate.

Real nonperturbative physics is at larger
$\Tsc{b}$, and, as we will see in more detail in Sec.\
\ref{sec:large-b}, a simple extrapolation of $g_K$ from small
$\Tsc{b}$ is likely to be wrong.

Once a less conservative value of $\bmax$ is chosen, more of the
fitting is concerned with effects beyond those predicted by low-order
perturbation theory.  This is exhibited on the right-hand side of
\eqref{eq:gKhat.version.1}, where the first term is the exact $g_K$
and the second term gives the error in replacing the exact value of
$\tilde{K}(\bstarsc,\mu_0)$ by a perturbatively-based approximation.

We now show how to predict approximately the quadratic behavior of
$g_K$ when $\Tsc{b} \lesssim \bmax$.  This amounts to an examination of
the second term on the right-hand side of \eqref{eq:gKhat.version.2}.
We will find that the results
roughly reproduce the values of the coefficient $g_2$ in Eq.\
(\ref{eq:BLNY.form}) that were fitted by BLNY and KN.

If $\tilde{K}$ were an analytic function of $\Tsc{b}$ around
$\Tsc{b}=0$, then $g_K$ as defined by Eq.~\eqref{eq:gK.def} would be
correctly given by a quadratic in $\Tsc{b}$ at small $\Tsc{b}$.  But
in fact $\tilde{K}$ has a mild singularity at $\Tsc{b}=0$, as is
verified by doing a renormalization-group improvement, as in
\eqref{eq:K.RG.improved}. 
Because the effective coupling $\alpha_s(\mubstar)$ is not analytic at
$\Tsc{b}=0$, neither is $\tilde{K}$.  This mildly modifies the
quadratic small-$\Tsc{b}$ behavior of $g_K$.  But normally we are not
concerned with accurately approximating $g_K$ at very small $\Tsc{b}$,
precisely because $g_K$ is small there and has little effect on the
cross section.  What we need to obtain is an approximation that is
useful when $\Tsc{b}$ gets closer to $\bmax$.

To get a simple approximation, we first set $\mu=\mubstar$ in the
definition of $g_K$, to remove large logarithms:
\begin{equation}
 g_K(\Tsc{b};\bmax)
=   - \tilde{K}(\Tsc{b};\mubstar;\alpha_s(\mubstar))
    + \tilde{K}(\bstarsc;\mubstar;\alpha_s(\mubstar))  \, .
\label{eq:expandedgK}
\end{equation}
\end{widetext}
We assume that $\Tsc{b}$ is less than $\bmax$.  Then there are no
large logarithms involving $\Tsc{b}$ or $\bstarsc$.  Using the
lowest-order formula for $\tilde{K}$ gives
\begin{equation}
\label{eq:gK.pert.approx}
   g_K(\Tsc{b};\bmax)
\simeq
   \frac{ \alpha_s(C_1/\bstarsc) C_F }{ \pi }
   \ln\xleft( 1 + \Tsc{b}^2/\bmax^2 \right). 
\end{equation}
This has $\Tsc{b}^2$ behavior at small $\Tsc{b}$ but a slower rise
above $\bmax$. 
It is the form used in Ref.~\cite{Collins:1985xx} to 
optimize matching between the perturbative calculation and $g_K(\Tsc{b};\bmax)$ 
at moderate $\Tsc{b}$.

To compare with fitted values of $g_K$ with $g_K=\frac12
g_2\Tsc{b}^2$, we propose two methods.  One is to expand at small
$\Tsc{b}$:
\begin{equation}
\label{eq:LPgK}
   g_K(\Tsc{b};\bmax)
\simeq
   \frac{ \alpha_s(C_1/\bstarsc) C_F }{ \pi }
   \frac{ \Tsc{b}^2 }{ \bmax^2 },
\end{equation}
and then to replace $C_1/\bstarsc$ by $C_1/\bmax$, since fits for
$g_K$ concerns $\Tsc{b}$ not far from $\bmax$.  Then we equate the
coefficients of $\Tsc{b}^2$ in this formula and in the fitted $g_K$,
to obtain
\begin{equation}
  g_2 \simeq  \frac{ 2 \alpha_s(C_1/\bmax) C_F }{ \pi \bmax^2 }
  \quad \mbox{(by small $\Tsc{b}$ expansion).}
\end{equation}
The other method is to equate the derivatives with respect to
$\bmax^2$ at $\Tsc{b}=\bmax$; this may be more representative of how
$g_K$ affects the evolution of the cross section because 
this is where $g_K$ gives a substantial correction to the cut-off
$\tilde{K}$.
The result gives an
estimate that is a factor of two smaller:
\begin{equation}
  g_2 \simeq  \frac{ \alpha_s(C_1/\bmax) C_F }{ \pi \bmax^2 }
  \quad \mbox{(by derivative at $\bmax$)}.
\end{equation}
Neither method can exactly reproduce the fitted $g_K$, since the
perturbative estimate for $g_K$ has a different functional form than
the fitted $g_K$; the best we can do is an approximate match.

To obtain numerical values, we use the two-loop parametrization of
$\alpha_s(\mu)$ from Ref.~\cite{Bethke:2009jm} with 3 active 
flavors
of quark.  We make the standard choice $C_1 = 2e^{-\gamma_E}$.  For
the two standard values $\bmax = \unit[0.5]{GeV^{-1}}$ and $\bmax =
\unit[1.5]{GeV^{-1}}$, we find
\begin{align}
\left. \frac{C_F}{\pi } \frac{1}{\bmax^2}  \alpha_s(C_1/\bmax) \right|_{\bmax = 0.5 \, {\rm GeV}^{-1}} & \approx 0.45 \, {\rm GeV}^2, \label{eq:p5est} \\
\left. \frac{C_F}{\pi } \frac{1}{\bmax^2}  \alpha_s(C_1/\bmax) \right|_{\bmax = 1.5 \, {\rm GeV}^{-1}} & \approx 0.13 \, {\rm GeV}^2  \label{eq:1p5est} . 
\end{align}
We compare with the measured values in Table~\ref{table1}.
\begin{table}
\begin{center}
\renewcommand{\arraystretch}{1.4}
\begin{tabular}{|c|c|c|c|}
     \hline
\multicolumn{4}{|c|}{\rule[-3mm]{0mm}{8mm}  $g_2$ values in quadratic parametrizations: }   \\
  $\bmax$  & Fitted 
  & \renewcommand{\arraystretch}{0.1}
    \begin{tabular}{cc}Expansion\\Method\end{tabular} 
  & \renewcommand{\arraystretch}{0.1}
    \begin{tabular}{cc}Derivative\\Method\end{tabular} 
  \\
\hline
$\unit[0.5]{GeV^{-1}}$ & $\unit[0.68_{-0.02}^{+0.01}]{GeV^2}$ & $\unit[0.9]{GeV^2}$ &  $\unit[0.45]{GeV^2}$ \\
$\unit[1.5]{GeV^{-1}}$ & $\unit[0.18\pm{0.02}]{GeV^2}$ & $\unit[0.26]{GeV^2}$ &  $\unit[0.13]{GeV^2}$ \\
\hline
\end{tabular}
\end{center}
\caption{}
\label{table1}
\end{table}
We see a rough agreement, with the two methods of matching a value of
$g_2$ to (\ref{eq:gK.pert.approx}) giving results that bracket the
measured value.  
We deduce that some of the work in the fits simply
reproduces perturbative predictions in a region where the predictions
have a useful, if approximate validity.  We also deduce that the
values of $\bmax$ are 
conservative.  If one wants to genuinely measure the nonperturbative
part of $g_K$, one needs a more general parameterization and one needs
to ensure that data is used that is sensitive to higher values of
$\Tsc{b}$. We will address this issue in the next section.

Of course, the above estimates are crude and meant only to check for
general consistency.  
At large $\Tsc{b}$, Eq.~\eqref{eq:gK.pert.approx} is not expected to be an accurate 
parametrization of $g_K(\Tsc{b};\bmax)$. First, it is based on 
an extrapolation of a low order 
perturbative calculation. Second, at large $\Tsc{b}$ it depends strongly on 
$\bmax$. The complete TMD factorization formalism is $\bmax$ independent, and fully optimized fits should  approximately 
reflect this if they are to account for large $\Tsc{b}$ behavior. 

Notice that the arguments for approximately quadratic behavior for
$g_K(\Tsc{b})$ at small $\Tsc{b}$ equally apply to the functions
$g_{j/H}$ defined in Eq.\ \eqref{eq:gjH.def}.  This small $\Tsc{b}$
behavior corresponds, after exponentiation, to a Gaussian for a TMD
parton density.  

We should emphasize that our result that perturbation theory
approximately reproduces the fitted values of $g_2$ does not imply
that it should get them exactly correct: The fitted values have also
to allow for both uncalculated higher-order perturbative terms and for
genuinely nonperturbative effects.

\section{Large-$\Tsc{b}$ behavior of correlation function}
\label{sec:large-b}

\subsection{General properties}
\label{sec:large-b.general.properties}

Appropriate parameterizations for the nonperturbative large-$\Tsc{b}$
behavior of TMD parton densities and of the CSS kernel $\tilde{K}$
need to be informed by the expectations from the general principles of
quantum field theory.  All of these quantities are certain kinds of
Euclidean correlation function.  Therefore we generally expect them to decay
exponentially (supplemented by a power law):
\begin{equation}
\label{eq:exp.decay}
  \frac{1}{{\Tsc{b}}^\alpha} e^{-m\Tsc{b}}
\end{equation}
for large distance $\Tsc{b}$.  Here $m$ is the mass of the lowest mass
state that can be exchanged in the relevant channel.  The exponent
$\alpha$ depends on the dimensionality of the problem.  

Contributions to the correlation functions arise from quantities like
$\langle P| \text{op}(\T{b}) |X\rangle \langle X| \text{op}(0)
|P\rangle$. Exceptions to the property of exponential decay only arise
when the theory has massless particles, or when we have vacuum matrix
elements. nonperturbative QCD has only
massive states, so the exception of masslessness does not apply.  The
issue of vacuum matrix elements does not arise for
quark and gluon densities in a hadron, but it does happen for
$\tilde{K}$, at least in perturbation theory, as we will discuss
later.

The decaying exponential behavior of coordinate-space parton densities
at large $\Tsc{b}$ is illustrated by the calculations by Schweitzer et
al.\ \cite{Schweitzer:2012hh} in a chiral model.  It can also be
illustrated by simple perturbative calculations with massive
fields.\footnote{Parton densities are matrix elements of
  gauge-invariant quark and gluon operators in hadron states. What the
  appropriate nonperturbative final states $|X\rangle$ should be is
  not entirely obvious in a theory with color confinement.  This issue
  also gives a potential loophole in the argument for exponential
  decay at large $\Tsc{b}$.}

From many one-loop calculations of the TMD quantities of interest, we
know that a typical integral giving $\Tsc{b}$ dependence is of the
form:
\begin{equation}
\label{eq:FT}
  \int d^2\T{k}
  \frac{ e^{i\T{k}\cdot \T{b}} }{ \Tsc{k}^2 + m^2 }.
\end{equation}
Note that $m$ is generally an $x$-dependent function of actual
particle masses.  We can systematically obtain the large-$\Tsc{b}$
behavior by deforming the integral over $\T{k}$ into the space of
complex momenta, so as to make the real part of the exponent negative.
The dominant behavior is from the neighborhood of the pole at
$\Tsc{k}^2 = -m^2$.  Simple power counting gives the dominant region
of $\T{k}$ as
\begin{equation}
  \T{k} \sim \left( im + O(1/\Tsc{b}), O(\sqrt{m/\Tsc{b}}) ) \right),
\end{equation}
where Euclidean coordinates $\T{k}=(k_x,k_y)$ are used in a situation
where $\T{b}=(\Tsc{b},0)$ is in the $x$-direction.

Since the lowest mass state is a property of the theory, this suggests
(but does not strictly prove) that the exponential (and probably the
accompanying power of $\Tsc{b}$) is valid for the TMD parton density
independently of $\zeta$ (or $Q$).  
This implies that the nonperturbative part of $\tilde{K}$ goes to a constant at
  large $\Tsc{b}$, to preserve the exponential and the power-law behavior of
  $\Tsc{b}$ in Eq.~\eqref{eq:exp.decay}.
Naturally, the numerical
coefficient of the exponential decreases when $\zeta$ (and hence $Q$)
increases; this corresponds to the known qualitative behavior of TMD
parton densities.

For $\tilde{K}$ this matches what is obtained from perturbative
calculations with a massive gluon (with the mass mimicking the effect
of massive states in nonperturbative QCD).  Relevant formulae can be
found in Ref.\ \cite{Collins:2011qcdbook}.  Graphs with only emission
of virtual particles, i.e., with the vacuum for the final state, are
independent of $\Tsc{b}$.  Graphs with particle emission into the
final state decay exponentially.  The commonly assumed quadratic
behavior of $\tilde{K}$ and Gaussian behavior of TMD parton densities
can only be an approximation, valid at best only for moderate
$\Tsc{b}$.

We therefore propose that the following constraints be applied to
nonperturbative parameterizations:
\begin{enumerate}
\item The nonperturbative TMD parton density at a starting value of
  $\zeta$ has the above exponential behavior, as coded in a linear
  large-$\Tsc{b}$ behavior of the functions $g_{j/H}$ in Eq.\
  \eqref{eq:gjH.def}. 
\item For the functions $g_{j/H}$, there should 
  therefore be a transition from 
  approximately quadratic low-$\Tsc{b}$ behavior to linear high-$\Tsc{b}$ behavior. 
\item The nonperturbative part of $\tilde{K}$ goes to a constant at
  large $\Tsc{b}$.
\item This constant is negative, so that the large $\Tsc{b}$ tail is
  reduced as $\zeta$ increases.\footnote{The value of
    $\tilde{K}(\Tsc{b},\mu)$ does depend on the RG scale $\mu$.  At
    large scales, the RG evolution and the positivity of the LO term
    in the anomalous dimension $\gamma_K$ ensures that $\tilde{K}$ is
    strongly negative at large $\Tsc{b}$.  But the negativity should
    apply even at fairly low $\mu$.}
\item Correspondingly the master function $A$ goes to zero:
  $A(\infty)=0$. 
\end{enumerate}
Thus a key property of the dynamics of nonperturbative QCD is the
number $\tilde{K}(\infty;\mu_0)$ (at a chosen reference scale
$\mu_0$).  This should be extracted from fits to data.  The value at
other scales is given by the RG applied in a perturbative region.
Thus
\begin{equation}
\label{eq:K.infty.RG.improved}
   \tilde{K}(\infty;\muQ)
  = \tilde{K}(\infty;\mu_0) 
    - \int_{\mu_0}^{\muQ} \frac{ d \mu' }{ \mu' }
       \gamma_K(\alpha_s(\mu')).
\end{equation}

As for the function $g_K(\Tsc{b};\bmax)$, it follows from its
definition that at a \emph{minimum} it should obey:
\begin{itemize}
\item At large $\Tsc{b}$, $g_K(\Tsc{b};\bmax)$ goes to a constant,
  such that $\tilde{K}(\Tsc{b};Q)$ goes to a negative constant.
\item At small $\Tsc{b}$, $g_K(\Tsc{b};\bmax)$ is (approximately)
  quadratic, with roughly the coefficient found in Sec.\
  \ref{sec:small-b}.
\end{itemize}
Note that Tafat \cite{Tafat:2001in} argues that $\tilde{K}$ is
proportional to $\Tsc{b}$ at large $\Tsc{b}$.  We do not adequately
understand the justification of Tafat's argument.  In any case, all
such properties are subject to experimental test.

\subsection{Dependence of $g_K$ on $\bmax$}
\label{sec:bmaxdep}

Since the exact $\tilde{K}$ is independent of $\bmax$, it is useful to
devise methods of parameterizing $g_K$ to achieve this to a useful
accuracy, with the aid of perturbative calculations like those in
Sec.\ \ref{sec:small-b}.  This will add
(approximate) constraints on $g_K$ beyond
the two just listed.  In the next section we will show a
parameterization obeying the constraints.

The extra constraints are to arrange automatic $\bmax$ independence in
certain important regions, and that $g_K$ agrees with its perturbative
calculation when the perturbative calculation is valid:
\begin{widetext}
\begin{enumerate}
  \item The form of $g_K(\Tsc{b};\bmax)$ is such that
  \begin{equation}
    \asy_{\Tsc{b} \ll \bmax}
    \left. \frac{\diff{}}{\diff{\bmax}} g_K(\Tsc{b};\bmax)
     \right|_{\text{parametrized}}
 = 
    \asy_{\Tsc{b} \ll \bmax}
    \left. \frac{\diff{}}{\diff{\bmax}} g_K(\Tsc{b};\bmax)
    \right|_{\text{truncated PT}}.
  \end{equation}
  Here the notation $\asy$ (for ``asymptote'') denotes the extraction
  of the approximately quadratic behavior at small $\Tsc{b}$, with
  neglect of higher power corrections.  The notation ``truncated PT''
  on the right-hand side indicates an application of the calculational
  methods of Sec.\ \ref{sec:small-b}, but possibly taken to higher
  order.

\item When $\Tsc{b} \to \infty$, the value of $g_K(\infty;\bmax)$ is
  to be arranged so that $\tilde{K}(\infty)$ is independent of
  $\bmax$. 

  To implement this, consider the RG-improved form \eqref{eq:K.RG.improved} for
  $\tilde{K}$.
  We differentiate with respect to $\bmax$, and take the limit
  $\Tsc{b}\to\infty$.  We also replace the perturbative part of the
  right-hand side by its truncated approximation.  Since the full
  $\tilde{K}$ is independent of $\bmax$, we find
  \begin{equation}
     \frac{\diff{}}{\diff{ \ln \bmax} }  g_K(\Tsc{b} = \infty;\bmax)
   =
     \left[
          \frac{\diff{\tilde{K}(\bmax;C_1/\bmax)}}{\diff{ \ln \bmax} } 
            - \gamma_K (\alpha_s(C_1/\bmax))
     \right]_{\text{truncated PT}},
  \label{eq:dbmaxlargeb}
  \end{equation}
  which is our second requirement on the parametrization of $g_K$. It
  determines the $\bmax$ dependence of $g_K(\Tsc{b} {=} \infty;\bmax)$.
\end{enumerate}
\end{widetext}

\subsection{Simple Parametrization of $g_K(b_T;\bmax)$}
\label{sec:simpleparam}

Here we propose a very simple example of a single-parameter form for
$g_K(\Tsc{b};\bmax)$ that follows the strategy established by the above
conditions.
While not unique, it is useful for illustrating the basic
properties that are desired for a parametrization of $g_K(b_T;\bmax)$.
The general principle is that we enforce $\bmax$-independence in the
asymptotic small- and large-$\Tsc{b}$ limits, and that the
perturbatively predicted small-$\Tsc{b}$ behavior is reproduced.  The
functional form of $g_K$ interpolates smoothly between the small- and
large-$\Tsc{b}$ regions, and one parameter determines the
numerical value of $\tilde{K}$ at $\Tsc{b}=\infty$.  Modifications and
additions to this function can be proposed later if required to get a
better fit to data.

The proposed form is
\begin{multline}
 g_K(\Tsc{b};\bmax)
\\
    = g_0(b_{\rm max}) \left(1 - \exp \left[ - \frac{C_F
          \alpha_s(\mubstar) \Tsc{b}^2}{\pi g_0(b_{\rm max})\,
          \bmax^2} \right] \right) \,,
\label{eq:expmodel}
\end{multline}
where $g_0(b_{\rm max})$ is a function of $\bmax$ that we will determine.
Expanding in small $\Tsc{b}^2 \ll b_{\rm max}^2$ gives
\begin{multline}
 g_K(\Tsc{b};\bmax) \\
  = \frac{C_F}{\pi} \frac{\Tsc{b}^2}{\bmax^2} \alpha_s(\mubstar)
    + O \left( \frac{\Tsc{b}^4 C_F^2 \alpha_s(\mubstar)^2}{\bmax^4 \pi^2 g_0(b_{\rm max})} \right) \, , 
\label{eq:gKexpanded}
\end{multline}
which matches the $O(\alpha_s\Tsc{b}^2/\bmax^2)$ term in
Eq.~\eqref{eq:LPgK}, independently of what the function $g_0(b_{\rm
  max})$ is.  
Thus, $g_K$ both agrees with the predicted small
$\Tsc{b}$ behavior and gives $\bmax$ independence of the full
$\tilde{K}$ in this region. See also Ref.~\cite{Collins:2014loa} for more discussion of the 
motivation for Eq.~\eqref{eq:expmodel}.

Note that a (RG improved and truncated) perturbation expansion for
$\tilde{K}(\bstarsc)$ is only valid if $\bmax$ is not so large that one
encounters the Landau pole before the $\bstarsc$ taming sets in.  On the other hand, if $\bmax$ is too 
small, then an expansion of a perturbative treatment of 
$(-\tilde{K}(\Tsc{b},\mu) + \tilde{K}(\bstarsc,\mu))$ around small $\Tsc{b}^2/\bmax^2$ 
is sensitive to quartic $\Tsc{b}^4$ and higher power terms, even in regions of small $\Tsc{b}$ where 
a perturbative description is valid. Thus,
Eq.\ \eqref{eq:expmodel} is optimized for choices of $\bmax$ roughly in
the transition region between perturbative and nonperturbative
$\Tsc{b}$. 

To enforce that $\tilde{K}(\Tsc{b}{=}\infty)$ is independent of $\bmax$,
we solve Eq.~\eqref{eq:dbmaxlargeb} with the leading order expressions
for $\gamma_K$ and $\tilde{K}$ from
Eqs.~(\ref{eq:gammaK},~\ref{eq:K}).  This gives
\begin{equation}
  g_0(\bmax) = g_0({\bmax}_{,0}) + \frac{2 C_F}{\pi}
  \int_{C_1/{\bmax}_{,0}}^{C_1/\bmax} \frac{d \mu^\prime}{\mu^\prime}
    \alpha_s(\mu^\prime) \,.
 \label{eq:gmax}
\end{equation}
Here, ${\bmax}_{,0}$ is a reference value for $\Tsc{b}$ used to fix
$g_0$; it is effectively an integration constant.  If one chooses a
particular value of $\bmax$ when fitting the TMD factorization formula
to data, then the single numerical value $g_0(\bmax)$ is a parameter of
the fit.  If a second fit were made with a different value of $\bmax$,
then we expect the fitted value of $g_0(\bmax)$ at the new $\bmax$ to be
given (approximately) in terms of the old value by Eq.\
\eqref{eq:gmax}. 

At large $\Tsc{b}$, the resulting $\tilde{K}$ goes to a constant, as planned, and at 
small $\Tsc{b}$, $g_K$
is an approximate power series in $\Tsc{b}^2$. In principle, we may also include a treatment of  
$b_{\rm max}$ independence in higher powers of $\Tsc{b}$ in the region of intermediate $\Tsc{b}$. Since the resulting corrections 
are small, we regard the current approximation as adequate for the moment. 
If in Eq.\ \eqref{eq:expmodel} we take the limit $g_0 \to \infty$,
this is equivalent to setting $g_K(\Tsc{b};\bmax)$ to zero and
describing the entire range of $\tilde{K}(\Tsc{b};Q)$ using the cutoff
form $\tilde{K}(\bstarsc;Q)$.  If, in addition, we take the limit
$\bmax \to \infty$, and use the perturbative formula for $\tilde{K}$,
then the result for $\tilde{K}(\Tsc{b};\mu_Q)$ in Eq.\
\eqref{eq:K.RG.improved} is to use the perturbative RG-improved
version for all $\Tsc{b}$. 

Improvements can of course be made by using higher order terms in the
perturbative expansion of $\tilde{K}$ and $\gamma_K$.

\begin{figure*}
\centering
  \begin{tabular}{c@{\hspace*{10mm}}c}
    \includegraphics[angle=90,scale=0.5]{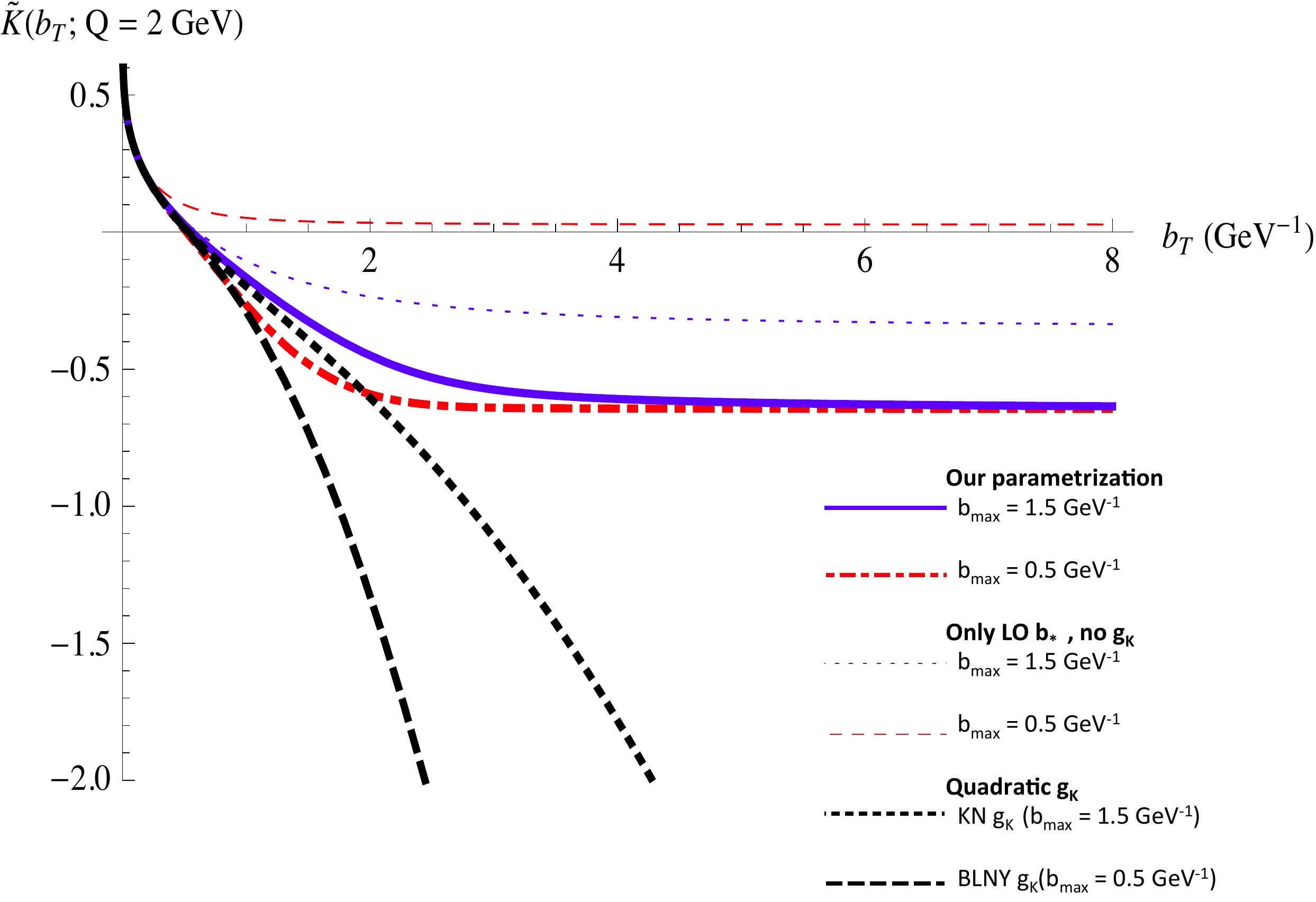} \\ \; \\
    (a) \\
    \vspace{5mm} \\
    \includegraphics[angle=90,scale=0.5]{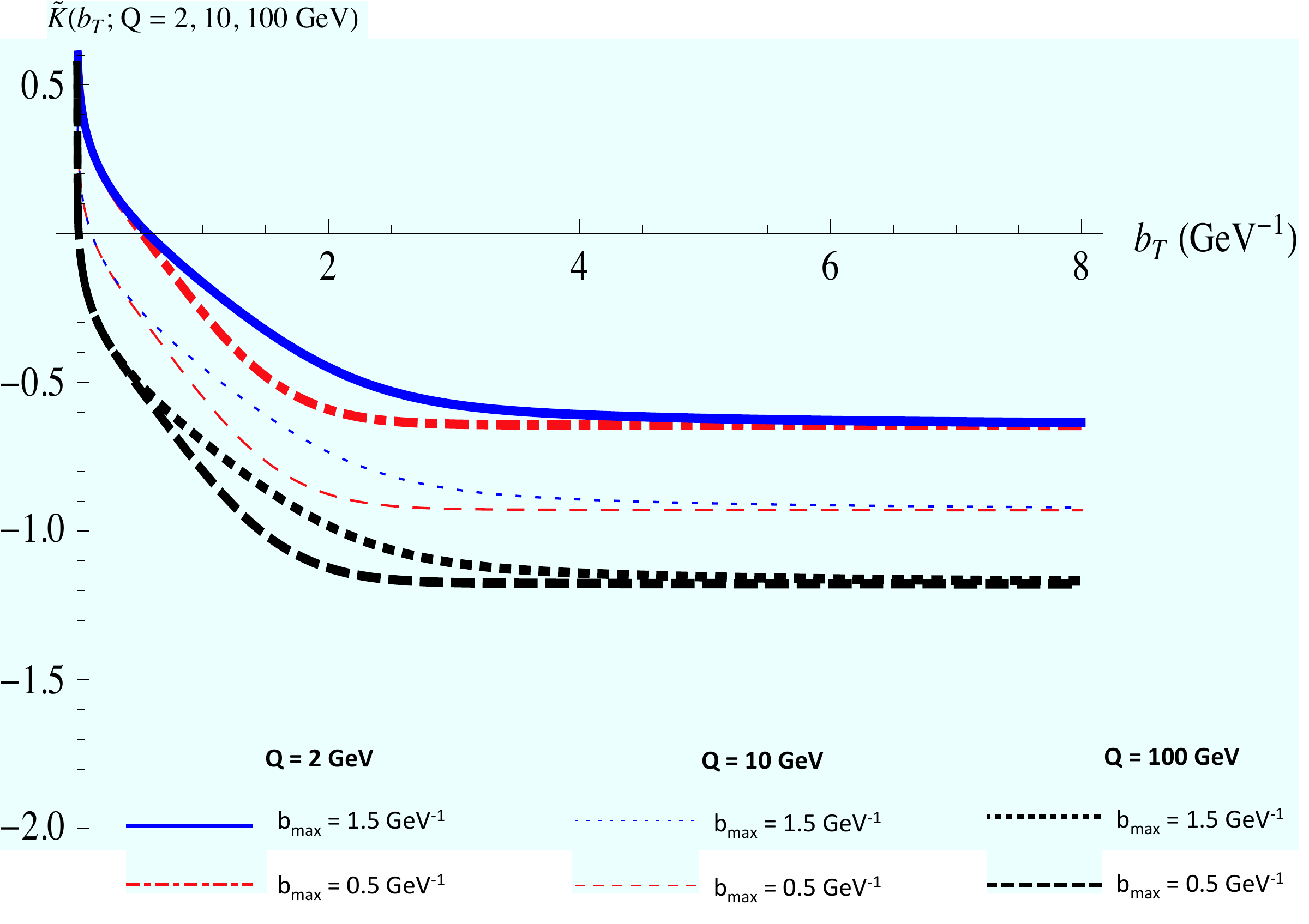}
   \\ \; \\
    (b)
    \\[5mm]
   \end{tabular}
\caption{(a) Comparison of different treatments of $g_K(\Tsc{b};\bmax)$ in calculation 
of $\tilde{K}(\Tsc{b};Q)$ for $Q = 2.0$~GeV. 
The solid blue and dot-dashed red curves are the calculation using the $g_K(\Tsc{b};\bmax)$ parametrization from 
Eq.~\eqref{eq:expmodel} with, respectively, $\bmax = 1.5$~GeV$^{-1}$ and
$0.5$~GeV$^{-1}$. The value at $\Tsc{b}=\infty$ is set by
$g(\bmax {=} \unit[1.5]{GeV^{-1}}) = 0.3$.
The thin dotted blue and dashed red curves are the LO RG improved calculations of $\tilde{K}(\bstarsc;Q)$ using 
$\bmax = 1.5$~GeV$^{-1}$ and $0.5$~GeV$^{-1}$ and \emph{no} $g_K(\Tsc{b};\bmax)$.
The black dotted and dashed curves are using the KN and BLNY fits for $g_K(\Tsc{b};\bmax)$ with 
$\bmax = 1.5$~GeV$^{-1}$ and $0.5$~GeV$^{-1}$ respectively.
\\
(b) Calculation of $\tilde{K}(\Tsc{b};Q)$ using 
Eq.~\eqref{eq:expmodel} with $\bmax = 1.5$~GeV$^{-1}$ and
$0.5$~GeV$^{-1}$ and several values of $Q$: $Q = 2$, $10$ and
$100$~GeV. \emph{Note the change in the meaning of the line types
  between graphs (a) and (b).}
}
\label{fig:gKNP}
\end{figure*}
\begin{figure*}
\centering
  \begin{tabular}{c@{\hspace*{10mm}}c}
    \includegraphics[angle=90,scale=0.5]{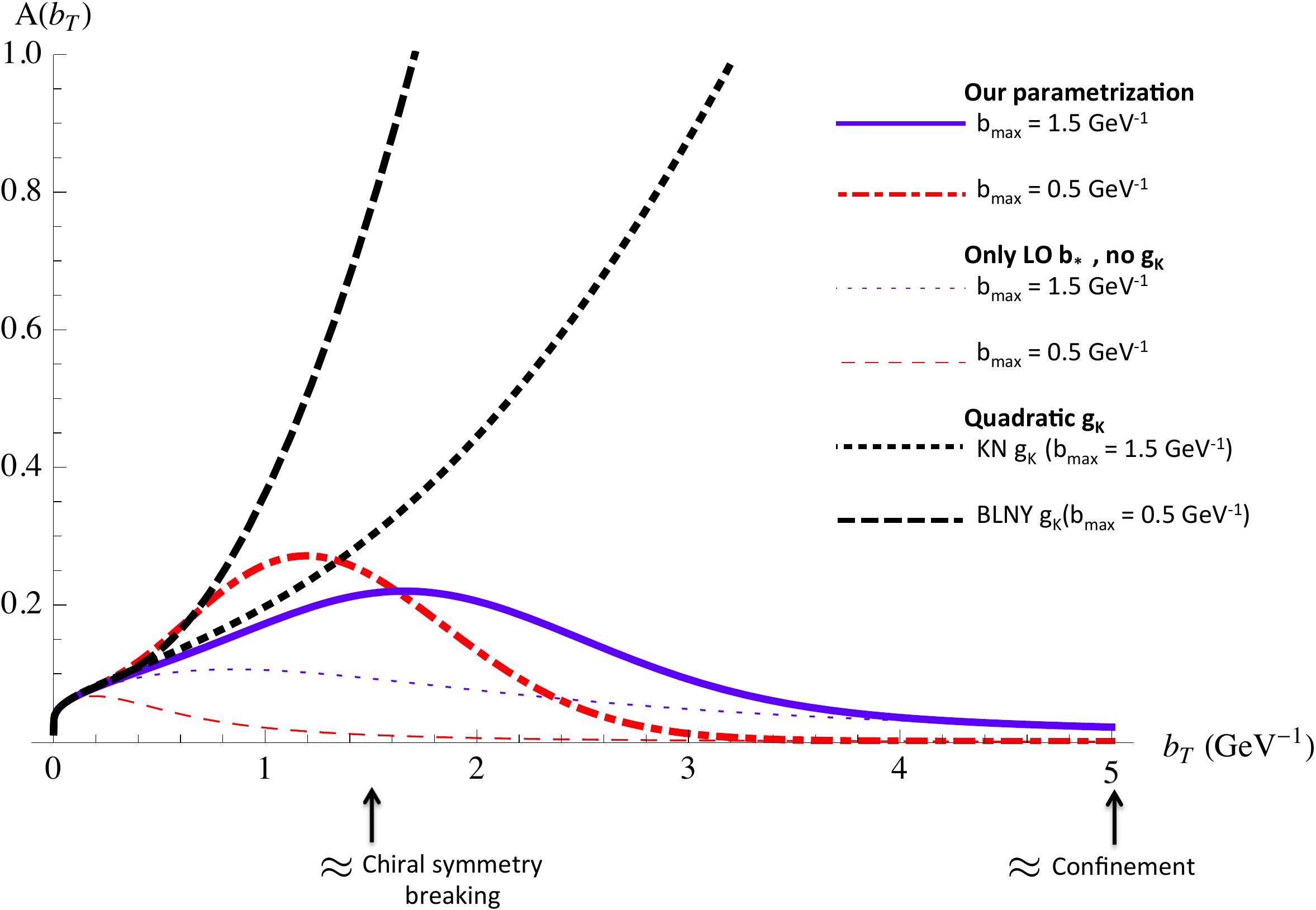} \\ \; \\
    (a) \\
    \vspace{5mm} \\
    \includegraphics[angle=90,scale=0.5]{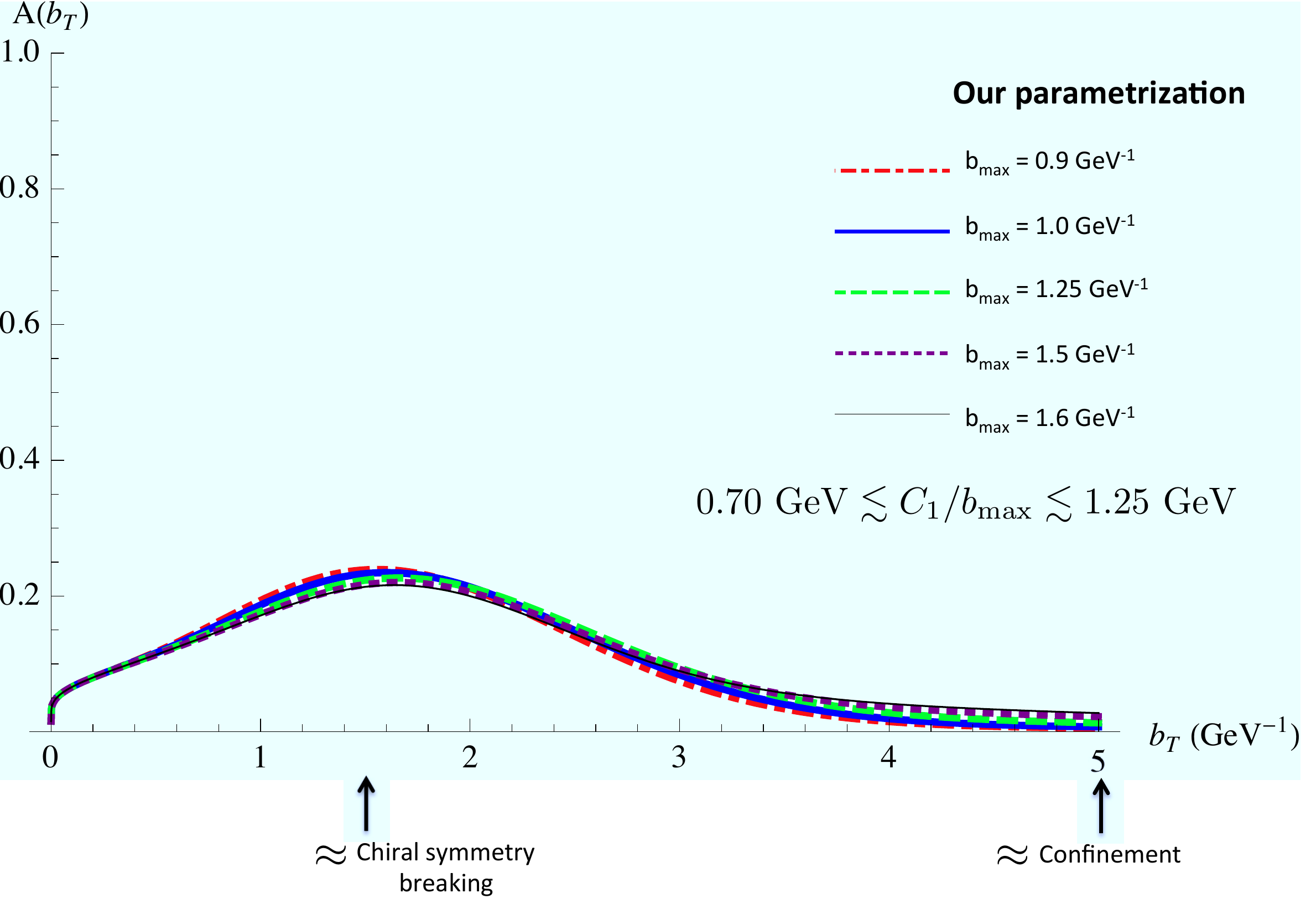}
   \\ \; \\
    (b)
    \\[5mm]
   \end{tabular}
\caption{(a) The values of the function $A(\Tsc{b})$ that corresponds to
  $\tilde{K}(\Tsc{b})$ for the curves in
  Fig.~\ref{fig:gKNP}(a). 
  \\
  (b) The function $A(\Tsc{b})$ for a selection of $\bmax$ values
  in the range
  $\unit[0.7]{GeV} \lesssim C_1/\bmax \lesssim \unit[1.25]{GeV}$.  They use
  \eqref{eq:expmodel} and \eqref{eq:gKexpanded}, with our parameter
  values.  
  For $b \lesssim 2.5 {\rm GeV}^{-1}$, we have checked 
  that $A(\Tsc{b})$ varies by less than $20 \%$ with the changes in the cutoff scale 
  corresponding to the range plot (b), so long as $0.2 \lesssim g_0(1.5 {\rm GeV}^{-1}) \lesssim 0.6$.}
\label{fig:ANP}
\end{figure*}

We now display an example of results obtained using the parametrization in
Eq.~\eqref{eq:expmodel}.  
In Fig.~\ref{fig:gKNP}, we have plotted the $\tilde{K}(\Tsc{b};Q)$ obtained using Eq.~\eqref{eq:expmodel} 
and Eq.~\eqref{eq:gmax}. 
For the reference value of $\bmax$, we choose ${\bmax}_{,0} = 1.5\,  {\rm GeV}^{-1}$.   
The goal of this paper is only to illustrate the general method, so we need only 
to estimate a reasonable value for $g_0(\bmax = 1.5 \,  {\rm GeV}^{-1})$.  To do this, we note that, 
if one assumes that perturbation theory is at least roughly applicable up to $\Tsc{b} \sim 1.5 \, {\rm GeV}^{-1}$, then 
the leading quadratic terms should dominate in the expansion in Eq.~\eqref{eq:gKexpanded} for the region of $\Tsc{b} \lesssim 1.5$~GeV$^{-1}$. 
Thus, for the quartic terms  in Eq.~\eqref{eq:gKexpanded} to be small at $\Tsc{b} \sim \bmax$ we must have 
\begin{equation}
\frac{ C_F \alpha_s(\mubstar) }{\pi g_0(b_{\rm max})} \lesssim 1 \, ,
\end{equation}
or, 
\begin{equation}
g_0(b_{\rm max} = 1.5 \, {\rm GeV}^{-1}) \gtrsim \frac{ C_F \alpha_s(C_1 / 1.5) }{\pi} \approx 0.3 \, .
\end{equation}
Therefore, for the sake of illustration, 
we will use $g_0(b_{\rm max} = 1.5 \, {\rm GeV}^{-1}) = 0.3$. In the future, this parameter should be extracted from actual fits.

We have chosen to plot $\tilde{K}(\Tsc{b};Q)$ in Fig.~\ref{fig:gKNP}(a) at the relatively low scale of $Q = \unit[2.0]{GeV}$.  
The result of using $\bmax = \unit[1.5]{GeV^{-1}}$ is shown as the solid blue curve while the result of switching to   
$\bmax = \unit[0.5]{GeV^{-1}}$ using Eq.~\eqref{eq:gmax} is shown as the red dot-dashed curve.
For comparison, Fig.~\ref{fig:gKNP}(a) also shows the calculation of $\tilde{K}(\bstarsc;Q=\unit[2]{GeV})$ in the LO RG improved approximation, 
but with no $g_K(\Tsc{b};\bmax)$ term at all (thin blue dotted and red dashed curves for  $\bmax = \unit[1.5]{GeV^{-1}}$ and 
$\bmax = \unit[0.5]{GeV^{-1}}$ respectively). Finally, we have shown 
the calculation with the KN ($\bmax = \unit[1.5]{GeV^{-1}}$) and BLNY
($\bmax = \unit[0.5]{GeV^{-1}}$)  
parametrizations (thick black dotted and dashed curves), which use the quadratic form for $g_K(\Tsc{b};\bmax)$ from 
Eqs.~(\ref{eq:BLNY.form},\ref{eq:KN.fit}).  Fig.~\ref{fig:gKNP}(a) illustrates how the parametrization from 
Eq.~\eqref{eq:expmodel} undergoes a comparatively small variation with $\bmax$, relative to methods that use a quadratic 
$g_K(\Tsc{b};\bmax)$ or no $g_K(\Tsc{b};\bmax)$ at all. 

Most of the residual variation in the parametrization from Eq.~\eqref{eq:expmodel}
is confined to the intermediate region between about $\Tsc{b} \sim 1.0$~GeV$^{-1}$ and $\Tsc{b} \sim 3.0$~GeV$^{-1}$ where 
one expects sensitivity to the precise details of the physics involved in the
transition between a perturbative and a purely nonperturbative description of the $\Tsc{b}$-dependence.
At both very large and very small $\Tsc{b}$ there is no sensitivity to
$\bmax$, as imposed by our construction.

The true $\tilde{K}(\bstarsc;\mu)$ is independent of $\bmax$
everywhere, so the gap between, for example, the blue and red 
thick curves in 
Fig.~\ref{fig:gKNP}(a) at around $\Tsc{b} \sim \unit[2.0]{GeV^{-1}}$ is
symptomatic of the limitations of the 
simplistic
parametrization of Eq.~\eqref{eq:expmodel}.  Our present aim is simply
to design a simple functional form that greatly reduces the $\bmax$
dependence. It is suggestive of improvements that can be made, by 
including higher order calculations and greater knowledge of the nonperturbative 
effects, that reduce the dependence even further.

In Fig.~\ref{fig:gKNP}(b), we have again plotted the $\tilde{K}(\Tsc{b};Q)$ obtained using Eq.~\eqref{eq:expmodel} 
and Eq.~\eqref{eq:gmax}, for $\bmax = 1.5$~GeV$^{-1}$ and $\bmax = 0.5$~GeV$^{-1}$, but now we show the 
result of using $Q= 2, 10$, and $100$~GeV. The variation in $Q$ results only in a vertical shift of $\tilde{K}(\Tsc{b};Q)$. 
This is an exact property of the CS kernel, and the vertical shift needs to be accounted for when addressing the evolution of 
the normalization of the cross section. 

As explained in Sect.~\ref{sec:master}, the physical consequences of 
$\tilde{K}(\Tsc{b};Q)$ for the evolution of the shape of transverse
distributions 
are conveniently illustrated through the master function $A(\Tsc{b})$ defined 
in Eq.~\eqref{eq:A.def}. Therefore, in Fig.~\ref{fig:ANP}(a) we have
plotted the versions of the function $A(\Tsc{b})$ that correspond to
the curves in Fig.~\ref{fig:gKNP}(a). 

In principle, like $\tilde{K}(\Tsc{b};Q)$, $A(\Tsc{b})$ should also be exactly independent of $\bmax$.
The problem of the instability of the quadratic parametrization for $g_K(\Tsc{b};\bmax)$ is 
made clear in Fig.~\ref{fig:ANP}(a) in the large $\Tsc{b}$ region.
There, we see dramatic differences between the results of the two
standard parameterizations, with different $\bmax$.  These differences
result in large changes in the evolution with $Q$ of the shape of the
integrand function $\tilde{W}(\Tsc{b})$.
The variation in $A(\Tsc{b})$ with $\bmax$ obtained with the parametrization in 
Eq.~\eqref{eq:expmodel} is noticeable but much smaller, and it approaches zero at very large $\Tsc{b}$.

Note that the variation of $\bmax$ in Fig.~\ref{fig:ANP}(a) corresponds to a rather large variation in the 
lower cutoff on the hard scale: $0.75~{\rm GeV} \lesssim C_1/\bmax \lesssim 2.25~{\rm GeV}$. 
For the smallest values of $\bmax$, the expansion of the perturbative expression for 
$(-\tilde{K}(\Tsc{b},\mu) + \tilde{K}(\bstarsc,\mu))$ is sensitive to quartic powers 
of $\Tsc{b}/\bmax$ (and higher powers) even for quite small values of $\Tsc{b}$. 

If we instead restrict to a slightly narrower window of 
$0.70~{\rm GeV} \lesssim C_1/\bmax \lesssim 1.25~{\rm GeV}$, sensitivity 
to $\bmax$ becomes essentially negligible, as illustrated by Fig.~\ref{fig:ANP}(b).

\subsection{Evolution of the  normalization}

Many discussions of TMD evolution focus on the variation with $Q$ of
the shape of transverse momentum distributions.  Among the reasons are
that this is a particularly recognizable effect of evolution, while
accurate measurements of normalizations are more difficult.  In addition,
the quadratic form commonly used for $g_K(\Tsc{b})$ predicts important
shape changes even at low $Q$.  However it is also useful to ask how
the normalization evolves.

Our proposal that $\tilde{K}(\Tsc{b})$ goes to a constant as
$\Tsc{b}\to\infty$ changes this situation in an interesting way.  The
effect of the nonperturbative part of $\tilde{K}$ is now to change
the normalization of the Fourier transformed cross section $\tilde{W}$
at large $\Tsc{b}$ instead of changing its shape dramatically. The
variation of normalization with $Q$ is approximately a power law as we
now show.  Measurement of this power law would be particularly useful
for determining the value of $\tilde{K}$ at $\Tsc{b}=\infty$. 

It follows from Eq.\ \eqref{eq:soln.2} that the shape of the
$\Tsc{b}$-space integrand at large $\Tsc{b}$ is determined by the
functions $g_{j/H}(\Tsc{b})$, which can be regarded as parameterizing
the intrinsic transverse momentum distribution of the quarks.

By replacing $\tilde{K}(\Tsc{b})$ by its asymptotic value
$\tilde{K}(\infty)$, we now show that the effect of evolution is to
give the normalization an approximate power dependence on $Q$.  
This can be seen from Eq.~\eqref{eq:xsect.evol}, where for large
$\Tsc{b}$
\begin{multline}
\label{eq:xsect.evol.large-b}
    \frac{ \partial \ln \tilde{W}(\T{b},Q,x_A,x_B) }
         { \partial \ln Q^2 }
\\     \simeq
      \tilde{K}(\infty;\mu_0) 
      + G( \alpha_s(\mu_Q),Q/\muQ )
      - \int_{\mu_0}^{\muQ}  \frac{ \diff{\mu'} }{ \mu' }
             \gamma_K(\alpha_s(\mu')).
\end{multline}
The variation of the right-hand-side with $\ln Q$ is an
order-$\alpha_s(Q)$ effect.  If to a first approximation we neglect
this variation, we find a power law:
\begin{equation}
 \tilde{W}(\Tsc{b},Q,x_A,x_B)
 \simeq
   \tilde{W}(\Tsc{b},Q_0,x_A,x_B) \left( \frac{Q_0^2}{Q^2} \right)^a \,,
\end{equation}
with
\begin{align}
   a ={}& -K(\infty;Q_0) - G(\alpha_s(Q_0), Q/Q_0)
\nonumber\\
    ={} &  g_K(\infty;\bmax) - \tilde{K}(\bmax;C_1/\bmax) 
\nonumber\\ 
     &   -G(\alpha_s(Q_0), Q/Q_0)
           + \int_{C_1/\bmax}^{Q_0}
          \frac{ \diff{\mu'} }{ \mu' } \gamma_K(\alpha_s(\mu')) \, .
\end{align}
As $Q$ increases, the right-hand side of \eqref{eq:xsect.evol.large-b}
becomes more negative, so the decrease of the normalization becomes
even stronger.  At small-$\Tsc{b}$, the situation is totally
different, of course, as can be read off the plots in Fig.\
\ref{fig:gKNP}, for example.  The power-law only applies at
large-$\Tsc{b}$.  

At low values of $Q$, around a GeV or two at the lower limit of TMD
factorization's applicability, the power law just derived gives a
corresponding decrease in the cross section at small transverse
momentum.  The approximate Bjorken scaling of the total cross section
is restored by a compensating change at larger transverse momentum,
which comes from relatively small $\Tsc{b}$. 

\section{Summary and Conclusion}
\label{sec:concl}

There is a wide variety of sometimes apparently contradictory methods
and results in the theoretical treatment and analysis of
transverse-momentum-dependent cross sections, especially as regards
the nature and importance of nonperturbative contributions from large
transverse distances.  We presented a systematic analysis of the
issues.  The basis of the logic is a TMD factorization theorem,
together with the associated evolution equations etc for the TMD
parton densities (and fragmentation functions).  Several different
forms of solution were presented, each emphasizing particular aspects
of the physics.  These have mostly been seen in the literature before,
but here they are unified by the link to scale-dependent TMD densities
as the foundation of the reasoning.

The evolution of the shape of TMD functions is primarily governed by a function
$\tilde{K}(\Tsc{b},\mu)$, which appears under different names in some
SCET-based formalisms.  Although this function is strongly universal,
it is nontrivial to gain a unified view of it.  

From an experimental point of view, the range of $\Tsc{b}$ dominantly
probed depends on the kinematic region of the data, so that no single
experiment can measure or test $\tilde{K}$ for all
$\Tsc{b}$. Moreover, $\tilde{K}$ is scale-dependent.  Although the
scale-dependence is just a perturbatively calculable upward shift as
$\mu$ increases, independent of $\Tsc{b}$, it does add an important
complication in measuring $\tilde{K}$ and in testing measured and
predicted values.

We proposed a master function $A(\Tsc{b})$ as a measure of the
evolution of the shape of TMD functions.  Unlike $\tilde{K}$, it is
independent of both scheme and scale (but is related to $\tilde{K}$ by
a derivative). We showed how the function $A(\Tsc{b})$ can be used to
diagnose disagreements between different methods and approximations
for TMD cross sections.  We suggest that the results of calculations
and fits should include a presentation of the resulting values of
$A(\Tsc{b})$, and one aim should be to find the values of $A(\Tsc{b})$, in addition 
to $\tilde{K}(\Tsc{b};Q)$,
for all $\Tsc{b}$, as an important property of QCD.

A further complication concerns the predictability of $\tilde{K}$.  At
low $\Tsc{b}$, perturbative calculations, supplemented by RG
improvement or by explicit resummation, are accurate.  As $\Tsc{b}$ is
increased, these predictions gradually become less accurate, and at
large enough $\Tsc{b}$ (beyond about 2 or $\unit[3]{GeV^{-1}}$),
$\tilde{K}$ becomes essentially nonperturbative.  We pointed out that
the use of resummation instead of RG improvement does not at all
change this situation; the accuracy of either kind of
perturbatively based calculation is governed by the value of
$\alpha_s(1/\Tsc{b})$.
 
Sensitivity to large $b_T$ in $A(\Tsc{b})$, for its part, opens new avenues 
of opportunity 
for probing nonperturbative properties of QCD. The relevant objects 
are the vacuum expectation values of Wilson loops, which are basic 
nonperturbative subjects of interest in, for example, lattice QCD methods and 
have already attracted interest in TMD studies~\cite{Musch:2010ka,Musch:2011er,Ji:2014hxa}.

When $Q$ is increased, the dominant range of $\Tsc{b}$ needed to
calculate the cross section shifts to ever smaller values.  Thus even
dramatic differences in the form of $\Tilde{K}$ at large $\Tsc{b}$
typically have little effect on TMD cross sections at large enough
$Q$, because at large $\Tsc{b}$ the $\Tsc{b}$ integrand is
exponentiated to a small value.  There is a stability to the evolution
towards larger $Q$.  In contrast, backward evolution in $Q$ is
unstable.  It is found that evolution with $Q$ is in fact considerably
less rapid at low $Q$ 
than is given by a backward evolution from
standard fits to Drell-Yan data.

(Even at relatively large $Q$, however, some knowledge 
of the large-$\Tsc{b}$ behavior may be desirable if 
very high degrees of precision are needed.)

From general principles about correlation functions at large
distances, we argued that $\tilde{K}(\Tsc{b},\mu)$ should go to a
constant as $\Tsc{b}\to\infty$.  This contrasts with the widely used
quadratic parameterizations.  We proposed a new form for a
one-parameter approximation for interpolating between the perturbative
result for $\tilde{K}$ at small $\Tsc{b}$ and a constant at large
$\Tsc{b}$.  
An important task is to fit the constant from data that
are sensitive to evolution with $\Tsc{b}$ in the range of 3 to
$\unit[4]{GeV^{-1}}$ upwards.  Our parameterization is intended to
approximately agree with standard Drell-Yan fits in the region of $\Tsc{b}$
to which they are sensitive, while giving slower evolution of the
shape of the cross section at lower $Q$.  It is also arranged to give
automatically weak sensitivity to CSS's parameter $\bmax$.  Of course,
our parameterization can be supplemented by higher order perturbative
calculations where available, and by extra parameterized functions for
the nonperturbative part.  But the new parameterization is designed
so that these corrections should be relatively weak.

It is worth emphasizing that
important and interesting physics is encoded in both the large and small $\Tsc{b}$ regions of cross sections like 
Eq.~\eqref{eq:kt.fact}. Depending on the specific underlying motivation for applying 
TMD factorization, different ranges of $\Tsc{b}$ may be of greater or lesser 
interest, and the relative importance of small and large $\Tsc{b}$ contributions depends on 
the size of $Q$. However,
a good TMD factorization formalism incorporates 
both types of behavior and smoothly relates a diverse range of different observables with 
different degrees of relative sensitivity to large and small $\Tsc{b}$. 

As a conclusion, we propose that one important aim of theoretical and
phenomenological work in QCD should be to obtain accurate values of
$A(\Tsc{b})$ and also of $\tilde{K}(\Tsc{b},\mu)$ over a wide range of
$\Tsc{b}$.  The results will have a similar significance to the
well-known global fits of collinear parton densities and fragmentation
functions, which provide definitive values, and uncertainties, for
these functions.\footnote{In the range of variables where they are
  actually determined, of course.}
We propose that $\tilde{K}$ goes to a constant at large $\Tsc{b}$
instead of being quadratic in $\Tsc{b}$.

One important benefit of presenting results directly for $A(\Tsc{b})$
and $\tilde{K}(\Tsc{b},\mu)$ should be a much better understanding of
how the transition occurs from perturbative behavior at small
$\Tsc{b}$ to nonperturbative behavior at large $\Tsc{b}$.  This
transition should be gradual rather than very sudden.  The strongly
universal nature of any nonperturbative contribution to $\tilde{K}$,
mentioned earlier, gives broad implications for these results.


\begin{acknowledgments}
  T. R. is supported in part by the National Science Foundation under
  Grant Nos.\ PHY-0969739 and \ PHY-1316617.  T.R. also acknowledges support 
  from the University of Michigan and the Lightner-Sams
  Foundation. J. C. and T.R. are supported by the
  U.S. Department of Energy under Grant No.\ DE-SC0008745. 
  This work was also supported by the DOE contract No. DE-AC05-06OR23177,
  under which Jefferson Science Associates, LLC operates Jefferson Lab.
  We acknowledge useful conversations with  D. Boer, L. Gamberg, A. Idilbi, P. Nadolsky, and 
  G. Sterman.
\end{acknowledgments}

\bibliography{jcc}

\end{document}